\documentclass[bj,authoryear]{imsart}

\RequirePackage{amsthm,amsmath,amsfonts,amssymb}
\RequirePackage[colorlinks,citecolor=blue,urlcolor=blue]{hyperref}
\RequirePackage{graphicx}
\RequirePackage{ulem}
\startlocaldefs
\theoremstyle{plain}

\newtheorem{theorem}{Theorem}[section]
\newtheorem{lemma}[theorem]{Lemma}
\theoremstyle{definition}
\newtheorem{definition}[theorem]{Definition}
\newtheorem{example}{Example}

\newtheorem{remark}{{\bf Remark}}
\newtheorem{assumption}{Assumption}

\theoremstyle{remark}

\newcommand{\setword}[2]{%
  \phantomsection
  #1\def\@currentlabel{\unexpanded{#1}}\label{#2}%
}
\newcommand{\indep}{\rotatebox[origin=c]{90}{$\models$}}

\newcommand{\dx}{\mathop{}\!\mathrm{d}}
\endlocaldefs

\begin{document}
\begin{frontmatter}
\title{SID: a novel class of nonparametric   tests of independence for censored outcomes}
\runtitle{Survival independence divergence}

\begin{aug}
\author[A]{\fnms{Jinhong}~\snm{Li}\ead[label=e1]{jinhongli0106@gmail.com}}
\author[B]{\fnms{Jicai}~\snm{Liu}\ead[label=e2]{liujicai1234@126.com}}
\author[C]{\fnms{Jinhong}~\snm{You}\ead[label=e3]{johnyou07@163.com}}
\author[D]{\fnms{Riquan}~\snm{Zhang}\ead[label=e4]{zhangriquan@163.com}}
\address[A]{School of Statistics and Mathematics, Zhejiang Gongshang University\printead[presep={,\ }]{e1}}
\address[B]{School of Statistics and Mathematics,
Shanghai Lixin University of Accounting and Finance\printead[presep={,\ }]{e2}}
\address[C]{School of Statistics and Management,
Shanghai University of Finance and Economics\printead[presep={,\ }]{e3}}
\address[D]{School of Statistics and Information,
Shanghai University of International Business and Economics\printead[presep={,\ }]{e4}}
\runauthor{Jinhong Li et al.}
\end{aug}

\begin{abstract}
We propose a new class of metrics, called the survival independence divergence (SID), to  test dependence between a right-censored outcome and covariates. A key technique for deriving the SIDs is to use a counting process strategy, which equivalently transforms the intractable independence test due to the presence of censoring into a test problem for complete observations.
The SIDs are equal to zero if and only if the right-censored response and covariates are independent,
and they are capable of detecting various types of nonlinear dependence. We propose  empirical estimates of the SIDs and establish their asymptotic properties. We further develop a wild bootstrap method to estimate the critical values   and show the consistency of the  bootstrap tests.   The   numerical  studies  demonstrate that our SID-based tests are highly competitive with existing methods in a wide range of settings.
\end{abstract}


\end{frontmatter}

\section{Introduction}\label{sec:intro}
\setcounter{page}{1}

Let $T$ be an event time and $\mathbf{X}$   a $p$-dimensional vector of covariates.
In this paper, we test the following hypotheses
\begin{equation}\label{hypthesis-independ}
H_{0}: T \text{ and }   \mathbf{X} \text{ are independent }  \quad \text{ versus } \quad H_{1}: \text{otherwise}.
 \end{equation}
Problem \eqref{hypthesis-independ} is   fundamental  in statistics with broad
applications. Among many applications, $T$ is frequently subject to censoring.  For instance, in clinical trials, patients' survival times to death   are often right censored due to the termination of follow-up study or patients' drop out; in employment studies, the unemployment duration may be censored by the cutoff nature of the sampling. When  censoring occurs, we only obtain partial information about the subjects'  survival times, but do not know
the exact time. Most existing  methods  for testing independence  are not designed to deal with censoring, and thus we need to adapt them to censored time-to-event data.

There is a long history of measuring dependence for uncensored
data. Pearson's correlation coefficient, Spearman's $\rho$, and Kendall's $\tau$ are probably the three most popular classical measures to quantify dependence between two univariate random variables. However, it is well known that these coefficients can detect only linear or monotone associations. Thus, many other flexible measures have been developed to overcome these difficulties, such as rank-based methods \citep{weihs2018symmetric,chatterjee2021new},  kernel-based methods \citep{ gretton2007kernel, ke2019expected}, and distance-based methods \citep{szekely2007measuring,sejdinovic2013equivalence}.

There have been  lots of studies on   testing indirectly the independence   between    $T$ and $\mathbf{X}$ for censored data in survival analysis.
The classical log-rank test is arguably the most popular approach
 for a two-sample test \citep{mentel1966evaluation} and is the most powerful test against local proportional hazards alternatives. However, the log-rank test cannot directly be used with general covariates, such as discrete or continuous covariates. Alternatively, one may test whether the regression coefficients corresponding to the covariates are equal to zero by fitting a semiparametric regression model, such as the proportional hazards models \citep{cox1972regression}, and  the accelerated failure time models \citep{wei1992accelerated}.
Notably, a model-based test may also suffer a severe loss of power if the models are violated.
 Thus, it is urgent to develop model-free  methods to problem \eqref{hypthesis-independ}.

 Recently, a few nonparametric test methods have been proposed. For example,
\citet{edelmann2021consistent} developed the inverse probability of censoring-weighted (IPCW) scheme and generalized the distance covariance \citep{szekely2007measuring} to right-censored data. \citet{Tamara} extended the classic weighted log-rank tests through kernel methods and proposed a kernel log-rank test (KLR).
The main challenge to test  problem \eqref{hypthesis-independ} for censored data is that
the problem depends on the relationship between the censoring time  $C$   and the covariates $\mathbf{X}$. To specify the relationship,   \citet{edelmann2021consistent}
assume  that    $C$ is independent of $(\mathbf{X}, T)$. Generally, the assumption is  strong   and may be violated in many applications. In contrast,  \citet{Tamara} adopted a mild and commonly used censoring mechanism that $C$ is independent of $T$ conditional on $\mathbf{X}$, called the so-called   independent censoring scheme in the literature on survival analysis \citep{fleming2011counting}.

In this paper, we introduce a novel  survival independence divergence (SID) to  test problem~\eqref{hypthesis-independ} under the framework of the independent censoring scheme. Our approach employs a counting process strategy to  overcome the challenges    caused by censoring.
Specifically, with the counting process technique, we equivalently transform the original test problem~\eqref{hypthesis-independ} into a tractable test problem~\eqref{hypthesis-condition-twosample}; see Theorem~\ref{theorem1.1}.
We then propose two types of SIDs to test the equivalent problem.
The first type is formulated by the discrepancy between two conditional characteristic functions, and the second is constructed using a Hilbert space distance between two distribution embeddings. We further build the connection between the two types of SIDs through the equivalence of distance-based and kernel-based statistics \citep{sejdinovic2013equivalence}.  The proposed SIDs have several appealing properties.  They equal zero if and only if right-censored  response  and covariates are independent,  they  do not require  strong  assumptions on   the censoring mechanism,  and  they are capable of detecting various  types of nonlinear dependence.

We emphasize that our approach is predicated on the independent censoring scheme, which is a mild assumption and   seems difficult to weaken further.  Under the assumption,    it is intriguing to note   that    $ T\indep \textbf{X}$  is equivalent to $T\indep (C,\textbf{X})$, and also  $C \indep \textbf{X}$  is equivalent to  $C\indep (T,\textbf{X})$, where
$\indep$ denotes statistical independence.
The assumption of $C\indep (T,\textbf{X})$ is common   in the literature, as seen in \citet{stute1993consistent}, \citet{chen1999dimension}, \citet{wang2007asymptotics}. Utilizing the above fact,  our approach can be effectively employed to test    for $C \indep \textbf{X}$  and thus  $C\indep (T,\textbf{X})$.

The idea of the SIDs is   different from  that of the KLR proposed by \citet{Tamara}.  The KLR can be viewed  as  a joint method,  which is based on the difference between the joint distribution of $T$ and $\mathbf{X}$,   denoted as $\nu_1(t,\mathbf{x})$,   and the product of their marginal distributions, denoted as $\nu_0(t,\mathbf{x})$, as detailed in Theorem~3.1 of \citet{Tamara}. In contrast, the SIDs are categorized as conditional methods, focusing on    the difference between a conditional distribution  and a marginal distribution, as described in our Theorem~\ref{theorem1.1}. In fact, the two approaches are  frequently  used  to test independence between two variables in the context of uncensored data.





The rest of the paper is organized as follows. In Section~\ref{Methodology-Identifiability}, we
  discuss identifiability of  problem \eqref{hypthesis-independ}.
 In Section~\ref{Methodology}, we introduce the SIDs and their properties. Sections~\ref{chapter-Estimation} and~\ref{chapter-asymptotic} develop their sample counterparts and establish their theoretical properties. Section~\ref{chapter-bootstrap} proposes a wild bootstrap method to approximate the null distribution and shows its asymptotic validity.
  Numerical studies are conducted in Sections~\ref{chapter-simulation} and \ref{chapter-Realdata}.
 We provide a brief discussion in Section~\ref{Discussion}.

Throughout this paper, we adopt the notation $i=\sqrt{-1}$ as  the imaginary unit. For
a complex-valued function $f(\cdot)$,  its complex conjugate is denoted  by $\bar{f}$,  and the modulus squared is defined as $\|f\|^2=f\bar{f}$.
For any function \( f(t, x) \) that takes real or complex values,  we define the limit \( \lim_{\Delta t \to 0^+} f(t + \Delta t, x) = g(t, x) \)  at specific \( t\) and \(x\)  by measuring the difference  \( f(t + \Delta t, x) \) and \( g(t, x) \) with an appropriate norm.   For real-valued functions, we employ the absolute value of the difference; for complex-valued functions, we use the modulus.

\section{Identifiability}\label{Methodology-Identifiability}

Problem \eqref{hypthesis-independ}  may be unidentifiable due to the presence of censoring, a significant distinction from tests of independence for uncensored data. For example,  assume that the maximum supports of  $C$ and  $T$ satisfy $\sup\{t: S_{C}(t)>0\}=c_0$ and $\sup\{t: S_{T}(t)>0\}=c_0+1$, for some positive $c_0$. If $T\cdot I(0\leq T\leq c_0)$ is independent of $\mathbf{X}$ but $T\cdot I(c_0<T\leq c_0+1)$  is dependent on $\mathbf{X}$, problem \eqref{hypthesis-independ} in the case is  not identified  because we only observe  $T$ before $c_0$.  We require specific identifiability constraints to address this issue.

Throughout the paper, we   focus on the  case that $T$
  is right-censored by  $C$.
Denote  \(Y = \min\{T, C\}\) and  \(\delta = I(T \leq C)\), where \(I(\cdot)\)  is the indicator function.  In practice, we only observe the triplet \((Y, \delta, \mathbf{X})\). Assume that $T$ is  a  continuous positive random variable,  and the covariates $\mathbf{X}$   take  values in a bounded subset  $\mathcal{X} \subseteq \mathbb{R}^p$. Let \(\lambda(t|\mathbf{x})\) be the conditional hazard function of \(T\) given \(\mathbf{X} = \mathbf{x}\), for \(t \geq 0\) and \(\mathbf{x} \in \mathcal{X}\). Note that \(\lambda(t|\mathbf{x})\) can be used to measure the dependence between \(T\) and \(\mathbf{X}\). Specifically, \(T\) is independent of \(\mathbf{X}\) if and only if \(\lambda(t|\mathbf{x}) = \lambda(t)\) for almost all \(t \geq 0\) and  \(\mathbf{x} \in \mathcal{X}\). Thus, the identifiability of the independence between \(T\) and \(\mathbf{X}\) can be transformed into the identifiability of \(\lambda(t|\mathbf{x})\). To characterize the latter, we utilize the joint probability distribution of  \((Y, \delta, \mathbf{X})\).  For this purpose, we introduce the following   independent censoring scheme
\begin{assumption}\label{assum1}
 The censoring time $C$  is conditionally independent of  $T$ given $\mathbf{X}$.
\end{assumption}

 Under Assumption \ref{assum1}, we can obtain the   the joint density function of $(Y,\delta, \mathbf{X})$, given by
 \begin{eqnarray}\label{liklihood-function}
		p_{_{Y,\delta,\mathbf{X}}(t,i,\mathbf{x};\lambda) }=
    \lambda(t|\mathbf{x})^i \exp\left\{-\int^t_0\lambda(s|\mathbf{x})ds\right\}S_{C|\mathbf{X}=\mathbf{x}}(t)^i f_{C|\mathbf{X}=\mathbf{x}}(t)^{1-i}p_{_{\mathbf{X}}}(\mathbf{x}),
\end{eqnarray}
for $t\in [0,+\infty)$, $i\in\{0,1\}$ and $\mathbf{x} \in\mathcal{X}$,
where
$S_{C|\mathbf{X}=\mathbf{x}}(t)$ and $f_{C|\mathbf{X}=\mathbf{x}}(t)$ are the conditional survival and density functions of  $C$ (for discrete  $C$, $f_{C|\mathbf{X}=\mathbf{x}}(t)=\operatorname{P}\{C=t|\mathbf{X}=\mathbf{x}\})$, and
$p_{_{\mathbf{X}}}(\mathbf{x})$  is  the    probability function of  $\mathbf{X}$. The derivation of \eqref{liklihood-function} can be referred to Chapter 3 of \citet{klein2003survival}.
Then,  we   formally define the identifiability of problem \eqref{hypthesis-independ}   as follows.

\begin{definition}\label{defin-identifiable}
Problem \eqref{hypthesis-independ}  is identifiable    if   $p_{_{Y,\delta,\mathbf{X}}}(t,i,\mathbf{x};\lambda_1)=p_{_{Y,i,\mathbf{X}}}(t,i,\mathbf{x};\lambda_2)$
 for any two hazard functions $\lambda_1(t|\mathbf{x})$ and $\lambda_2(t|\mathbf{x})$   implies $\lambda_1(t|\mathbf{x})=\lambda_2(t|\mathbf{x})$.
\end{definition}

Definition \ref{defin-identifiable} suggests that  $p_{_{Y,\delta,\mathbf{X}}}(t,i,\mathbf{x};\lambda)$ is identifiable if there do not exist two distinct conditional hazard  functions that correspond to the same distribution.
This definition extends the concept of identifiability within classical parametric maximum likelihood estimation \citep{wald1949note}.
By Definition \ref{defin-identifiable}, we   establish the identifiability condition  for problem \eqref{hypthesis-independ}  in the following theorem.

\begin{theorem}\label{identifiability}
Under  Assumption \ref {assum1},  Problem \eqref{hypthesis-independ} is identifiable  from the observed \((Y, \delta, \mathbf{X})\)
if $T\cdot I(T>\tau)$ is independent of $\mathbf{X}$, where   $\tau=\sup\{t:S_Y(t)>0\}$ and $S_Y(t)=P\{Y>t\}$.
\end{theorem}

Throughout the paper, we  refer to     the  condition  that   $T\cdot I(T>\tau)$ is independent of $\mathbf{X}$ in Theorem \ref{identifiability},    as the identifiability condition.
This condition encompasses    Assumption 3.1  from  \citet{Tamara}, which states that if $S_{C|\mathbf{X}=\mathbf{x}}(t)=0$, then $S_{T|\mathbf{X}=\mathbf{x}}(t)=0$  for almost all $\mathbf{x} \in \mathcal{X}$. In fact,  by  their Assumption 3.1, we have    that    $\tau_{_{T,\mathbf{x}}}\leq\tau_{_{C,\mathbf{x}}}$,
where  $\tau_{_{T,\mathbf{x}}}=\sup\{t: S_{T|\mathbf{X}=\mathbf{x}}(t)>0\}$ and   $\tau_{_{C,\mathbf{x}}}=\sup\{t: S_{C|\mathbf{X}=\mathbf{x}}(t)>0\}$. Furthermore, by
Proposition C.1 in \citet{Tamara}, we have that $\tau_{_{Y,\mathbf{x}}}=\tau_{_{T,\mathbf{x}}}\leq\tau$ for  almost all $\mathbf{x} \in \mathcal{X}$.  This  suggests  that $T\cdot I(T>\tau)$ is independent of $\mathbf{X}$.

\section{Methodology}\label{Methodology}
In the section, we equivalently transform   problem~\eqref{hypthesis-independ} into a tractable   problem by a counting process strategy. To this end, we define the observed counting process $N(t)=I(Y\leq t, \delta=1)$ and the at-risk process $Y(t)=I(Y \geq t)$.
Let $\Delta N(t)= N((t + \Delta t)^-)-N(t^-)$ be an increment of $N(\cdot)$ over $[t, t + \Delta t)$ with $\Delta t>0$.   Let  $f_{\mathbf{X},Y,\delta}(\mathbf{x},t,1)$  be  the subdensity of  $(\mathbf{X},Y,\delta=1)$,  defined as
 $f_{\mathbf{X},Y,\delta}(\mathbf{x},t,1)= {\partial^2 P\{\mathbf{X} \leq \mathbf{x}, Y \leq t, \delta=1\}}/{\partial \mathbf{x} \partial t}$. Hereafter, we need the following assumptions.

\begin{assumption}\label{condition1}
	Assume that $T\cdot I(T>\tau)$ is independent of $\mathbf{X}$.
\end{assumption}
\begin{assumption}\label{condition-3}
Assume that $T$ is  a  continuous positive  random variable,  and $f_{\mathbf{X},Y,\delta}(\mathbf{x},t,1)$  is bounded for   $(\mathbf{x}, t) \in \mathcal{X} \times [0, \tau)$  with a bounded and continuous second-order partial derivative with respect to $t$.
\end{assumption}

Assumption 2 is derived from the identifiability condition discussed in Section \ref{Methodology-Identifiability}. Assumption 3  is a common requirement in the survival analysis literature. It imposes smoothness conditions on the subdensities, as also considered in \citet{10.1214/aos/1017939141} and \citet{10.1214/21-AOS2153}.

\begin{theorem}\label{theorem1.1}
	Under  Assumptions 1-3, $\mathbf{X}$ is independent of $T$ if and only if
	\begin{equation}\label{conditequl}
		\lim_{ \Delta t\rightarrow 0^+}P\{\mathbf{X}\leq \mathbf{x} \mid \Delta N(t)=1,Y(t)=1\}=P\{\mathbf{X}\leq \mathbf{x} \mid Y(t)=1\},
	\end{equation}
	almost surely, for  any $\mathbf{x}\in \mathcal{X}$ and $t\in[0,\tau)$.
\end{theorem}

In Theorem \ref{theorem1.1},  the term  ``almost surely'' means that equality \eqref{conditequl} holds with probability 1. Note that  the  result  depends only on the observable counting processes  $N(t)$ and $Y(t)$. Thus,   it can be tested using the observed data, effectively addressing the challenge posed by censoring.  We   provide   additional  remarks on the theorem.
\begin{remark}
 If the identifiability condition  is invalid, i.e., $T$ and $\mathbf{X}$ may be dependent when $T>\tau$,  \eqref{conditequl} is  still useful in many clinical applications, which tests  equivalently the independence between   $\min\{T, \tau\}$ and $\mathbf{X}$.  In practice, we can  use \eqref{conditequl} to characterize the independence between the truncated  time $\min\{T, t_0\}$ and $\mathbf{X}$,   where $t_0 (<\tau)$  is a clinically meaningful endpoint.
 For example, in drug-efficacy studies as reported by 
  \citet{dugue2016deal} and \citet{rodrigues2018use},
researchers focus on the treatment effect during the trial period  $[0, t_0)$ (e.g., $t_0=2$ years  in oncology trials).
 In this case, Theorem \ref{theorem1.1}   can be used to test whether $T\cdot I(T\leq t_0)$ is independent of the treatment assignment $\mathbf{X}$   by replacing   $t\in[0,\tau)$ by $t\in[0, t_0)$ in Theorem \ref{theorem1.1}.

\end{remark}

\begin{remark}
It should be noted that the limit on the left-hand side of \eqref{conditequl} indeed exists. In fact,
 using $\{\Delta N(t)=1,Y(t)=1\}=\{ t \leq Y <t+\Delta t, \delta=1\}$ and Assumption 2,  we can show that
\begin{eqnarray}\label{eq:remark2}
\lim_{\Delta t\rightarrow 0^+}P\{\mathbf{X}\leq \mathbf{x} \mid \Delta N(t)=1,Y(t)=1\}=P\{\mathbf{X}\leq \mathbf{x} \mid Y=t,\delta=1\}.
\end{eqnarray}
See the detailed derivation on page 3  of  the   Supplementary Material.
\end{remark}

\begin{remark}
 According to  Remark 2 and the definition of $Y(t)$,   both  sides of  \eqref{conditequl} are equivalent to
\begin{align}
\lim_{\Delta t\rightarrow 0^+}P\{\mathbf{X}\leq \mathbf{x} \mid \Delta N(t)=1,Y(t)=1\} &= P\{\mathbf{X}\leq \mathbf{x} \mid T=t, C\geq t\},   \label{conditequl-noncensor-1} \\
P\{\mathbf{X}\leq \mathbf{x} \mid Y(t)=1\} &  =P\{\mathbf{X}\leq \mathbf{x} \mid T\geq t,C\geq t \}.\label{conditequl-noncensor-2}
\end{align}
Under    Assumption \ref{assum1},  that is, $T\indep C| \textbf{X}$,  we derive the following insights into \eqref{conditequl}.
\begin{description}
    \item[(i)] When $T\indep \textbf{X}$, by taking $X_1=T$, $X_2=C$ and $Y=\mathbf{X}$  in Proposition 1(ii)-(iii) of  \citet{YANG2019480}, we have that
   \begin{equation}\label{imply-relationship}
   T\indep C| \textbf{X}~ \text{ and }~~   T\indep \textbf{X}  \Longrightarrow  \textbf{X} \indep T|C,  \end{equation}
    where $\Longrightarrow$ denotes ``implies''.  This result suggests that   \eqref{conditequl-noncensor-1} and~\eqref{conditequl-noncensor-2}   are equal to $P\{\mathbf{X}\leq \mathbf{x} \mid  C\geq t \}$  for  any $\mathbf{x}\in \mathcal{X}$ and $t\in[0,\tau)$.

\item[(ii)] When   $T\not\!\!\indep  \textbf{X}$,  the result in \eqref{imply-relationship} does not hold. Thus,  \eqref{conditequl-noncensor-1} and~\eqref{conditequl-noncensor-2}  may be not equal for some   $t$ and  $\mathbf{x}$.

\item[(iii)] In the special case where  $C=+\infty$, implying  that  $T$ can be fully observed, the right-hand sides of \eqref{conditequl-noncensor-1} and~\eqref{conditequl-noncensor-2} become $P\{\mathbf{X}\leq \mathbf{x} \mid T=t\}$ and $P\{\mathbf{X}\leq \mathbf{x} \mid T\geq t\}$. Note that the events   $\{T=t \}$  and
$\{T\geq t\}$  are distinct for   $t \geq 0$. Thus,  $P\{\mathbf{X}\leq \mathbf{x} \mid T=t\}$  is not equal to $P\{\mathbf{X}\leq \mathbf{x} \mid T\geq t\}$ unless  $T\indep \textbf{X}$.
\end{description}
Form the   discussion above, it is evident that        \eqref{conditequl} can detect    the independence between      $T$ and $\mathbf{X}$.
\end{remark}

Let \( P_{\mathbf{X}|\Delta N_t,Y_t} \) and \( Q_{\mathbf{X}|Y_t} \) be the probability measures generated by  \( P\{\mathbf{X} \leq \mathbf{x} \mid \Delta N(t) = 1, Y(t) = 1\} \) and \( P\{\mathbf{X} \leq \mathbf{x} \mid Y(t) = 1\} \), respectively.
Intuitively,   Theorem~\ref{theorem1.1} implies that the cumulative functions  \( P\{\mathbf{X} \leq \mathbf{x} \mid \Delta N(t) = 1, Y(t) = 1\}\) converges  to \(P\{\mathbf{X} \leq \mathbf{x} \mid Y(t) = 1\} \) as \( \Delta t \rightarrow 0^+ \). Thus, their generated measures should satisfy \( P_{\mathbf{X}|\Delta N_t,Y_t} \rightarrow Q_{\mathbf{X}|Y_t} \) as \( \Delta t \rightarrow 0^+ \), denoted   as
\(\lim_{\Delta t \rightarrow 0^+} P_{\mathbf{X}|\Delta N_t,Y_t} = Q_{\mathbf{X}|Y_t}\). Here, the limit of the measures can be understood as
\( \sup_{\mathbf{x} \in \mathcal{X}} | P_{\mathbf{X}|\Delta N_t, Y_t}\{ (-\infty, \mathbf{x}]\} - Q_{\mathbf{X}|Y_t}\{ (-\infty, \mathbf{x}]\} | \to 0 \text{ as } \Delta t \to 0^+. \)
 In summary, by Theorem~\ref{theorem1.1}, the testing problem~\eqref{hypthesis-independ} is equivalent to testing
 \begin{equation}\label{hypthesis-condition-twosample}
	H_{0}: \lim_{ \Delta t\rightarrow 0^+}P_{\mathbf{X}|\Delta N_t ,Y_t}=Q_{\mathbf{X}|Y_t}, \text{ for any } t\in[0,\tau) \quad \text{ versus } \quad H_{1}: \text{otherwise}.
\end{equation}
In the following sections, we develop two new classes of measures
 to quantify the discrepancy between $\lim_{ \Delta t\rightarrow 0^+}P_{\mathbf{X}|\Delta N_t,Y_t}$ and
$Q_{\mathbf{X}|Y_t}$. One is based on the characteristic function approach, and the other is based on the kernel embedding technique. Using the general framework proposed by \cite{sejdinovic2013equivalence}, we further build the connection between the two
 classes of measures.

\subsection{Characteristic function-based approach}\label{Methodology-sub1}
In this section, we introduce a class of measures based on the discrepancy between the characteristic functions of  $\lim_{ \Delta t\rightarrow 0^+}P_{\mathbf{X}|\Delta N_t,Y_t}$  and  $Q_{\mathbf{X}|Y_t}$   to test problem~\eqref{hypthesis-condition-twosample}. For simplicity, we define
$
 \varphi_{\mathbf{X}|Y_t} (\mathbf{u},t) = E\{\exp\{i\mathbf{u}^T\mathbf{X}\} \mid Y(t) = 1 \}$ and
 $
\varphi_{\mathbf{X}| \Delta N_t,Y_t} (\mathbf{u}, t) = E\{\exp\{i\mathbf{u}^T\mathbf{X}\} \mid \Delta N(t) = 1, Y(t) =1\}$. Note that   $\varphi_{\mathbf{X}| \Delta N_t,Y_t} (\mathbf{u}, t)$ as  $\Delta t\rightarrow 0^+$ is well defined,    as derived in Remark 2.
 From the one-to-one relation between characteristic functions and distribution functions, under Assumption~\ref{condition1},   we have  $\lim_{ \Delta t\rightarrow 0^+}P_{\mathbf{X}|\Delta N_t,Y_t}=Q_{\mathbf{X}|Y_t}$ if and only if
\begin{equation}\label{eq:char-function-equal}
\lim_{ \Delta t\rightarrow 0^+}\varphi_{\mathbf{X} |\Delta N_t,Y_t}(\mathbf{u},t)=\varphi_{\mathbf{X} |Y_t}(\mathbf{u},t), \text{ for all }  \mathbf{u}\in \mathbb{R}^p  \text{ and } t\in[0,\tau).
\end{equation}
See the proof on page 3 of the   Supplementary Material.
 This is equivalent to testing whether
\begin{equation}\label{equvlent-weight}
\lim_{\Delta t\rightarrow 0^+}\int_0^\tau\int_{\mathbb{R}^p} \| \varphi_{\mathbf{X} |\Delta N_t,Y_t}(\mathbf{u},t)-\varphi_{\mathbf{X} | Y_t}(\mathbf{u},t)\|^2 d w_1(\mathbf{u}) dw_2(t)=0,
\end{equation}
where   $w_1(\mathbf{u})$ and $w_2(t)$ are   two positive  weight functions.

It is important to choose the weight functions $w_1(\mathbf{u})$ and $w_2(t)$ in~\eqref{equvlent-weight} properly. For $w_1(\mathbf{u})$, here we choose the weight function proposed by \citet{szekely2007measuring}
\begin{equation}\label{weight-szekely2007}
dw_1(\mathbf{u})= \frac{1}{ c(p, \beta)\|\textbf{u}\|^{\beta+p} } d\mathbf{u}
~ \text{ with }~ c(p, \beta)=\frac{2 \pi^{p/ 2} \Gamma(1-\beta / 2)}{\beta 2^\beta \Gamma((p+\beta) / 2)},
\end{equation}
for any $\beta\in(0,2)$. For $w_2(t)$, we choose $dw_2(t)=a(t)dP\{Y\leq t, \delta=1 \}$, where $a(t)$ is a given nonnegative function with the support $[0,\tau)$. The choice of $a(t)$ is discussed in Section~\ref{chapter-Estimation}.
 Using the two weight functions, we define the following divergence.

\begin{definition}\label{defin-SDCov}
Let $\mathbf{X}$ be a $p$-dimensional random vector.
The  SID based on the characteristic functions between $T$ and $\mathbf{X}$ is defined as
\begin{equation}\label{def-SDC}
\operatorname{SID}_\beta(T,\mathbf{X})= \lim_{\Delta t\rightarrow 0^+} \int_0^\tau \int_{\mathbb{R}^p}  \| \varphi_{\mathbf{X} |\Delta N_t,Y_t}(\mathbf{u},t)-\varphi_{\mathbf{X} | Y_t}(\mathbf{u},t)\|^2
 a(t) dw_1(\mathbf{u})  d \nu(t),
\end{equation}
 where  $a(t)$ is a given nonnegative function.
\end{definition}

An important difference between $\operatorname{SID}_\beta(T,\mathbf{X})$ and the distance covariance \citep{szekely2007measuring} is that the former is built on the conditional characteristic functions and the latter is based on the joint characteristic functions. Our approach using conditional characteristic functions,  has also been explored in uncensored data settings, as seen in   \citet{Wang2015Conditional} and \citet{ke2019expected}.

Note that the integrals in~\eqref{def-SDC} may suffer from computational issues
due to their intractability. Fortunately, we can apply the following
  properties of $\operatorname{SID}_\beta(T,\mathbf{X})$ to overcome such issues.

\begin{theorem} \label{Theorem2}
Suppose that $(Y_1, \delta_1, \mathbf{X}_1)$ and $(Y_2,\delta_2, \mathbf{X}_2)$ are independently and identically distributed (i.i.d.) copies of $(Y, \delta, \mathbf{X})$.  Under Assumptions \ref{assum1}-\ref{condition-3} and     $E\{\|\mathbf{X}\|^\beta\}<\infty$  for  $\beta\in(0,2)$,
   we have  that
\begin{enumerate}
 \item[(i)] $\operatorname{SID}_\beta(T,\mathbf{X})$ can be expressed as
\begin{multline}\label{SCIC-res1}
\begin{aligned}
\!\!\!\!\!\!\!\!\!\!\!\!\!\!\!\operatorname{SID}_\beta(T,\mathbf{X})
&= - \ \int_0^\tau E\{ \|\mathbf{X}_1-\mathbf{X}_2\|^\beta \mid Y_1=t, \delta_1=1, Y_2=t, \delta_2=1 \} a(t) d\nu(t) \\
 & \quad + \ 2\int_0^\tau E\{\Vert\mathbf{X}_1-\mathbf{X}_2\Vert^\beta \mid Y_1=t, \delta_1=1, Y_2(t)=1\} a(t) d\nu(t) \\
 & \quad - \ \int_0^\tau E\{\Vert\mathbf{X}_1-\mathbf{X}_2\Vert^\beta \mid Y_1(t)=1, Y_2(t)=1\}a(t) d\nu(t).
\end{aligned}
\end{multline}

  \item[(ii)] $\operatorname{SID}_\beta(T,\mathbf{X})\geq 0$ and   $\operatorname{SID}_\beta(T,\mathbf{X})=0$ if and only if $ T\indep \textbf{X}$.
 \end{enumerate}
\end{theorem}

Theorem~\ref{Theorem2}(i) suggests that $\operatorname{SID}_\beta(T,\mathbf{X})$ has a closed form and is, thus, easily estimated from the data.
Theorem~\ref{Theorem2}(ii) implies that $\operatorname{SID}_\beta(T,\mathbf{X})$ is an applicable measure for testing independence between $T$ and $\mathbf{X}$.
Note that  Theorem~\ref{Theorem2}(ii)  holds only  for
 $0<\beta<2$ but may be not true for $\beta=2$. In fact, for $\beta=2$,  by some algebra, we  can easily  show that~\eqref{SCIC-res1} is equal to 0 if and only if
 $\lim_{\Delta t\rightarrow 0^+}E\{\mathbf{X} \mid \Delta N(t)=1,Y(t)=1\}=E\{\mathbf{X} \mid Y(t)=1\}$, which does not imply $\lim_{\Delta t\rightarrow 0^+}P_{\mathbf{X}|\Delta N_t,Y_t}=Q_{\mathbf{X}|Y_t}$.

\subsection{Kernel-based approach}\label{Methodology-sub2}
In recent decades, a Hilbert space embedding of a distribution has emerged as a useful tool for many nonparametric hypothesis test problems. For example, it has been developed and successfully used for independence \citep{gretton2007kernel}, two-sample \citep{gretton2012kernel}, and goodness-of-fit \citep{chwialkowski2016kernel} testing frameworks. The effectiveness of such embeddings prompts us to generalize the embedding method to test problem~\eqref{hypthesis-condition-twosample}.

We first briefly review the reproducing kernel Hilbert space (RKHS). A RKHS  $\mathcal{H}_K$ on $\mathcal{X}$ with a kernel $K(\mathbf{x}, \mathbf{x}^{\prime})$ is a Hilbert space of functions $f(\cdot): \mathcal{X}  \rightarrow \mathbb{R}$ with inner product $\langle\cdot, \cdot\rangle_{\mathcal{H}_K}$. The kernel $K(\mathbf{x}, \mathbf{x}^{\prime})$ satisfies (1) $K(\cdot,\mathbf{x})\in \mathcal{H}_K$ for all $\mathbf{x}\in\mathcal{X}$ and (2) $\langle f, K(\cdot,\mathbf{x})\rangle_{\mathcal{H}_K}=f(\mathbf{x})$ for all $ \mathbf{x}\in\mathcal{X}$ and $ f\in\mathcal{H}_K$. We call the mapping $\phi: \mathbf{x} \mapsto\mathcal{H}_K$ given by $\phi(\mathbf{x}) =K(\cdot, \mathbf{x})$ the canonical feature map of $K(\mathbf{x}, \mathbf{x}^{\prime})$.

Let $\mathcal{M}(\mathcal{X})$ be the set of all finite signed Borel measures on $\mathcal{X}$ and $\mathcal{M}^1_+(\mathcal{X})$ the set of all Borel probability measures on $\mathcal{X}$.
The basic idea behind the embedding of probability measures is to map measures into a RKHS. Specifically, for $\nu\in\mathcal{M}(\mathcal{X})$, the kernel embedding of $v$ into the RKHS $\mathcal{H}_K$ is $\mu_K(\nu)\in \mathcal{H}_K$ such that $\int f(\mathbf{x}) d\nu(\mathbf{x})=\langle f,\mu_K(\nu)\rangle_{\mathcal{H}_K}$, for all $f\in \mathcal{H}_K$. For more details, refer to \cite{sejdinovic2013equivalence}.

For a measurable kernel $K(\cdot, \cdot)$ on $\mathcal{X}$ and $\theta>0$, we need to restrict the kernel into the following particular class of measures
$
\mathcal{M}^\theta_{K}(\mathcal{X}) =  \{v\in\mathcal{M}(\mathcal{X}):\int K^{\theta}(\mathbf{x},\mathbf{x})d|v|(\mathbf{x})<\infty  \}.
$
We can see that each element in $\mathcal{M}^\theta_{K}(\mathcal{X})$ is required to have a finite $\theta$-moment with respect to the kernel $K(\cdot, \cdot)$.
Using the Riesz representation theorem, it can be shown that $\mu_K(\nu)$ exists if $\nu \in\mathcal{M}_K^{1/2}(\mathcal{X})$; see Lemma~3 in \citet{gretton2012kernel}.
Thus, we can impose the assumption that $P_{\mathbf{X}|\Delta N_t,Y_t}$, $Q_{\mathbf{X}|Y_t}\in \mathcal{M}^1_{+}(\mathcal{X})\cap\mathcal{M}^{1/2}_{K}(\mathcal{X})$ to ensure that the kernel embeddings of $P_{\mathbf{X}|\Delta N_t,Y_t}$ and $Q_{\mathbf{X}|Y_t}$ exist. From this assumption, we introduce the following Hilbert space distance.

\begin{definition}\label{Survival_HSIC}
Let $K(\cdot, \cdot)$ be a measurable kernel on $\mathcal{X}$.
 Assuming $P_{\mathbf{X}|\Delta N_t,Y_t}$ and $Q_{\mathbf{X}|Y_t}\in \mathcal{M}^1_{+}(\mathcal{X})\cap\mathcal{M}^{1/2}_{K}(\mathcal{X})$, the SID based on $K(\cdot, \cdot)$ is defined as
\begin{align*}
\operatorname{SID}_K(T,\mathbf{X})=& \lim_{\Delta t\rightarrow 0^+} \int_0^\tau \Vert \mu_K(P_{\mathbf{X}|\Delta N_t,Y_t})-\mu_K(Q_{\mathbf{X}|Y_t})\Vert^2_{\mathcal{H}_K} a(t) d\nu(t),
\end{align*}
where $\nu(t)=P\{Y\leq t, \delta=1 \},$ and $a(t)$ is a given nonnegative function.
\end{definition}

To find the properties of $\operatorname{SID}_{K}(T,\mathbf{X})$, the kernel has to be a characteristic function. A kernel function is said to be characteristic if the map $\mu_K: \nu\rightarrow \mu_K(\nu) $ is injective. The assumption is commonly used in the literature on kernel methods, for example, \citet{gretton2007kernel,gretton2012kernel}.

\begin{theorem} \label{Theorem4}
Assuming $P_{\mathbf{X}|\Delta N_t,Y_t}$ and $Q_{\mathbf{X}|Y_t}\in \mathcal{M}^1_+(\mathcal{X}) \bigcap \mathcal{M}^{1/2}_{K}(\mathcal{X})$,  under   Assumptions \ref{assum1}-\ref{condition-3},  we have
\begin{enumerate}
\item[(i)] $\operatorname{SID}_K(T,\mathbf{X})$ can be expressed as
\begin{multline}\label{SKC-K}
\begin{aligned}
 \operatorname{SID}_K(T,\mathbf{X})
&= \int_0^\tau E\{  K(\mathbf{X}_1,\mathbf{X}_2) \mid Y_1=t, \delta_1=1, Y_2=t, \delta_2=1\} a(t) d\nu(t) \\
 & \quad - \ 2\int_0^\tau E\{K(\mathbf{X}_1,\mathbf{X}_2) \mid Y_1=t, \delta_1=1, Y_2(t)=1\} a(t) d\nu(t) \\
 & \quad + \ \int_0^\tau E\{K(\mathbf{X}_1,\mathbf{X}_2) \mid Y_1(t)=1, Y_2(t)=1\} a(t) d\nu(t).
\end{aligned}
\end{multline}

\item[(ii)] $\operatorname{SID}_{K}(T,\mathbf{X})\geq 0$.
     If $K(\cdot,\cdot)$ is characteristic,
    $\operatorname{SID}_{K}(T,\mathbf{X})=0$ if and only if $ T\indep \textbf{X}$.
 \end{enumerate}
\end{theorem}

\subsection{Connection between $\operatorname{SID}_\beta(T,\mathbf{X})$ and $\operatorname{SID}_{K}(T,\mathbf{X})$}

In this section, we first show that $\operatorname{SID}_\beta(T,\mathbf{X})$ is a special class of negative type semimetric-based measures \citep{sejdinovic2013equivalence}. Through the link between negative type semimetrics and kernels, we then build up the connection between $\operatorname{SID}_\beta(T,\mathbf{X})$ and $\operatorname{SID}_{K}(T,\mathbf{X})$.

First, we recap a negative-type semimetric and space. Let $\mathcal{X}$ be an arbitrary space endowed with a semimetric $\rho: \mathcal{X} \times \mathcal{X} \rightarrow [0,+\infty)$,   satisfying $\sum_{i=1}^n\sum_{j=1}^n\alpha_i\alpha_j\rho(\mathbf{x}_i,\mathbf{x}_j)\leq0,$ where $\mathbf{x}_i \in \mathcal{X}$ and $\alpha_i \in \mathbb{R}$ for all $i = 1,...,n$,  with $\sum_{i=1}^n \alpha_i=0$. Then, $\rho$ is called a semimetric of negative type on $\mathcal{X}$, and
$(\mathcal{X},\rho)$ is called a space of negative type. In addition, like the finite moment condition on kernels, we also need to restrict $\rho$ to a particular class of measures. Specifically, we define a new class of Borel measures
$
\mathcal{M}^\theta_{\rho}(\mathcal{X}) =  \{v\in\mathcal{M}(\mathcal{X}):\exists \mathbf{x}_0\in\mathcal{X} \text{ s.t.\ } \int\rho^{\theta}(\mathbf{x},\mathbf{x}_0) d|v|(\mathbf{x})<\infty  \},
$
for a given negative-type semimetric $\rho$ and $\theta>0$. Then, we introduce the following divergence.

\begin{definition}\label{Survival_distance}
Let $(\mathcal{X},\rho)$ be a semimetric space of negative type. Assuming $P_{\mathbf{X}|\Delta N_t,Y_t}, Q_{\mathbf{X}|Y_t}\in \mathcal{M}^1_+(\mathcal{X}) \cap\mathcal{M}^1_\rho(\mathcal{X}),$ the SID with the negative-type semimetric $\rho$ is defined as
\begin{equation*}
\operatorname{SID}_{\rho}(T,\mathbf{X})
\!\!=\!\!\! \lim_{\Delta t\rightarrow 0^+}\!\!\!\! -\int_0^\tau \!\!\!\! \int_{\mathcal{X}\times \mathcal{X}}\rho(\mathbf{X},\mathbf{X}') d([P_{\mathbf{X}|\Delta N_t,Y_t}-Q_{\mathbf{X}|Y_t}]\times[P_{\mathbf{X}'|\Delta N_t,Y_t}-Q_{\mathbf{X}'|Y_t}]) a(t) d\nu(t),
\end{equation*}
where $\nu(t)=P\{Y\leq t, \delta=1 \},$ and $a(t)$ is a given nonnegative function.
\end{definition}

Like Theorems~\ref{Theorem2} and~\ref{Theorem4}, the following theorem gives a closed-form expression for $\operatorname{SID}_{\rho}(T,\mathbf{X})$.

\begin{theorem} \label{Theorem3}
Assume that $(\mathcal{X},\rho)$ is a semimetric space of negative type. Then,   we have that
\begin{equation}\label{SDC-rho}
\begin{aligned}
\operatorname{SID}_{\rho}(T,\mathbf{X}) &= -\int_0^\tau E\{ \rho(\mathbf{X}_1,\mathbf{X}_2) \mid Y_1=t, \delta_1=1, Y_2=t, \delta_2=1\} a(t) d\nu(t) \\
 & \quad + \ 2\int_0^\tau E\{\rho(\mathbf{X}_1,\mathbf{X}_2) \mid Y_1=t, \delta_1=1, Y_2(t)=1\} a(t) d\nu(t) \\
 & \quad - \ \int_0^\tau E\{\rho(\mathbf{X}_1,\mathbf{X}_2) \mid Y_1(t)=1, Y_2(t)=1\}a(t) d\nu(t).
\end{aligned}
\end{equation}%
\end{theorem}

By Proposition~3 in \cite{sejdinovic2013equivalence}, we obtain that $\rho_\beta(\mathbf{x},\mathbf{x}')=\Vert \mathbf{x}-\mathbf{x}'\Vert^\beta$ is a semimetric of negative type for $\beta\in(0,2)$. Thus, we can see that $\operatorname{SID}_\beta(T,\mathbf{X})$ belongs to the family of negative-type semimetric-based measures.  We next develop the connection between $\operatorname{SID}_\beta(T,\mathbf{X})$ and $\operatorname{SID}_K(T,\mathbf{X})$ using the following link between negative-type semimetrics and kernels. Specifically, Proposition~3 in \citet{sejdinovic2013equivalence} suggests that, for any negative-type semimetric $\rho$, there exists a nondegenerate kernel $K(\cdot,\cdot)$ such that
\begin{align}\label{semimetric-kernel}
\rho(\mathbf{x},\mathbf{x}')=K(\mathbf{x},\mathbf{x})+K(\mathbf{x}',\mathbf{x}')-2K(\mathbf{x},\mathbf{x}')=\Vert K(\cdot,\mathbf{x})-K(\cdot,\mathbf{x}')\Vert^2_{\mathcal{H}_K}.
\end{align}
Conversely, if $K(\cdot,\cdot)$ is any nondegenerate kernel, then $\rho(\cdot,\cdot)$ defined by~\eqref{semimetric-kernel} is a valid semimetric of negative type. Whenever the kernel $K(\cdot,\cdot)$ and the semimetric $\rho(\cdot,\cdot)$ satisfy~\eqref{semimetric-kernel}, we say that $K(\cdot,\cdot)$ generates $\rho(\cdot,\cdot)$. From~\eqref{SKC-K},~\eqref{SDC-rho}, and~\eqref{semimetric-kernel}, we have the following result.

\begin{theorem}\label{Relationship}
Let $(\mathcal{X},\rho)$ be a semimetric space of negative type and $K(\cdot,\cdot)$
any kernel that generates $\rho$. Then,  we have that
$\operatorname{SID}_{\rho}(T,\mathbf{X})=2\operatorname{SID}_{K}(T,\mathbf{X})$.
\end{theorem}

Let $K_\beta(\mathbf{x}, \mathbf{x}^{\prime}) = (\|\mathbf{x}\|^\beta+\|\mathbf{x}^{\prime}\|^\beta
-\|\mathbf{x}-\mathbf{x}^{\prime}\|^\beta)/2,$  for $\beta\in(0,2)$. From~\eqref{semimetric-kernel},
  $\rho_\beta(\mathbf{x},\mathbf{x}')=\Vert \mathbf{x}-\mathbf{x}'\Vert^\beta$ is the semimetric generated by
$K_\beta (\mathbf{x}, \mathbf{x}^{\prime} )$. By Theorem~\ref{Relationship}, we obtain that
 $\operatorname{SID}_\beta(T,\mathbf{X})=\operatorname{SID}_{\rho_{_\beta}}(T,\mathbf{X})
 =2\operatorname{SID}_{K_\beta}(T,\mathbf{X})$. Thus,
 $\operatorname{SID}_\beta(T,\mathbf{X})$ is a special case of $\operatorname{SID}_{K}(T,\mathbf{X})$.

\section{Estimation approaches}\label{chapter-Estimation}
In this section, we introduce the empirical estimators of $\operatorname{SID}_\beta(T,\mathbf{X})$ and $\operatorname{SID}_{K}(T,\mathbf{X})$. We provide details of the derivation of only the estimator of $\operatorname{SID}_{K}(T,\mathbf{X})$, since the estimator of $\operatorname{SID}_\beta(T,\mathbf{X})$ can be similarly obtained. Two kinds of estimates are proposed. The first is obtained by plugging the empirical functions into~\eqref{SKC-K}, and the second is based on a $U$-statistic.

The closed form in~\eqref{SKC-K}   consists of the following three conditional expectations
\begin{align*}
S_{K,1}(t) &= E\{ K(\mathbf{X}_1,\mathbf{X}_2) \mid Y_1=t, \delta_1=1, Y_2=t, \delta_2=1\}, \\
S_{K,2}(t) &= E\{K(\mathbf{X}_1,\mathbf{X}_2) \mid Y_1=t, \delta_1=1, Y_2(t)=1\}, \\
 S_{K,3}(t) &= E\{K(\mathbf{X}_1,\mathbf{X}_2) \mid Y_1(t)=1, Y_2(t)=1\}.
\end{align*}
To estimate $S_{K,1}(t)$ and $S_{K,2}(t)$, we use the Nadaraya--Watson kernel-smoothing methods.
Note that the estimates of $S_{K,1}(t)$ and $S_{K,2}(t)$ do not suffer from the so-called curse of dimensionality, because the estimation depends on only the 1-dimensional variable $t$.

Suppose
$\left\{\left(Y_{i}, \delta_{i}, \mathbf{X}_{i}\right), i= 1,\dots, n \right\}$ are
 independent random samples drawn from $(Y, \delta, \mathbf{X})$. Let $W(y)$ be a symmetric density function and $W_{h}(y)=h^{-1}W(y/h)$, where $h>0$ is a bandwidth.
In our numerical studies, a Gaussian kernel is used for $W(y)$. The Nadaraya--Watson kernel estimators of $S_{K,1}(t)$ and $S_{K,2}(t)$ are given by
\begin{align*}
\widehat{S}_{K,1}(t) &= \frac{\sum_{i,j=1}^n K(\mathbf{X}_i,\mathbf{X}_j) W_{h}( Y_i-t) W_{h}( Y_j-t)\delta_i\delta_j}{\Big[\sum_{i=1}^n W_{h}( Y_i-t)\delta_i \Big]^2}, \\
\widehat{S}_{K,2}(t) &= \frac{\sum_{i,j=1}^n K(\mathbf{X}_i,\mathbf{X}_j) W_{h}( Y_i-t)\delta_i I(Y_j\geq t)}{\Big[ \sum_{i=1}^n W_{h}(Y_i-t)\delta_i  \sum_{j=1}^n I(Y_j\geq t)\Big]}.
\end{align*}
And, $S_{K,3}(t)$ can be estimated by
\begin{equation*}
\widehat{S}_{K,3}(t)= \frac{\sum_{i,j=1}^n K(\mathbf{X}_i,\mathbf{X}_j) I(Y_i\geq t)I(Y_j\geq t)}{\Big[\sum_{i=1}^n I(Y_i\geq t) \Big]^2}.
\end{equation*}

Note that the above estimators $\widehat{S}_{K,j}(t),$ $j=1,2,3$, suffer from random denominator issues, which may lead to a large bias near 0. As suggested by \citet{SU2007807}, we can choose a proper weight function $a(\cdot)$ in~\eqref{SKC-K} to overcome such issues.
By the definitions of $\widehat{S}_{K,j}(t),$ we choose $a(\cdot)$ as
\begin{equation}\label{weight-function}
 a^*(t)= [f_{Y,\delta}(t,1)E \{  I(Y>t)\}]^2,
\end{equation}
where  $f_{Y,\delta}(t,1)$ is  the  subdensity   of     $(Y,\delta=1)$,   defined as
 $f_{Y,\delta}(t,1)= {d P\{Y \leq t, \delta=1\}}/{d t}$.
  Throughout this paper, we use $a(t)=a^*(t)$ in the definitions of $\operatorname{SID}_\beta(T,\mathbf{X})$, $\operatorname{SID}_{K}(T,\mathbf{X})$, and $\operatorname{SID}_{\rho}(T,\mathbf{X})$.

Using the weight function $a^*(t)$, we provide a plug-in estimator
\begin{equation*}
\widetilde{\operatorname{SID}}_{K}(T,\mathbf{X})
  =  \int_0^\tau\Big[\widehat{S}_{K,1}(t)-2\widehat{S}_{K,2}(t) + \widehat{S}_{K,3}(t)\Big]\widehat{a}^*(t) d\overline{N}(t),
\end{equation*}
where $\overline{N}(t)=\sum_{i=1}^n I(Y_i\leq t, \delta_i=1)/n$ and
$\widehat{a}^*(t) = [n^{-2}\sum_{i=1}^n W_{h}(Y_i-t)\delta_i  \sum_{j=1}^n I(Y_j\geq t) ]^2.$
Furthermore, we can easily obtain that
\begin{align}\label{SDC-van-denom}
\widetilde{\operatorname{SID}}_{K}(T,\mathbf{X}) =\frac{1}{n}\sum_{k=1}^n\Big[\widetilde{S}_{K,1}( Y_k) -2\widetilde{S}_{K,2}(Y_k)
+\widetilde{S}_{K,3}(Y_k) \Big]\delta_k,
\end{align}
where
\begin{align*}
& \widetilde{S}_{K,1}(t) =\frac{1}{n^4}\sum_{i,j=1}^{n}  K(\mathbf{X}_i,\mathbf{X}_j) W_{h}( Y_i-t)\delta_i W_{h}( Y_j-t)\delta_j\sum_{i,j=1}^{n} I(Y_{i} \geq t)I(Y_{j}\geq t),
\\
&\widetilde{S}_{K,2}(t)=\frac{1}{n^4}\sum_{i,j=1}^{n}  K(\mathbf{X}_i,\mathbf{X}_j) W_{h}( Y_i-t) \delta_i I(Y_{j}\geq t)\sum_{i,j=1}^{n} W_{h}( Y_i-t)\delta_iI( Y_{j} \geq t),
\\
&\widetilde{S}_{K,3}(t)=\frac{1}{n^4}\sum_{i,j=1}^{n}  K(\mathbf{X}_i,\mathbf{X}_j) I( Y_{i}\geq t)I(Y_{j}\geq t)\sum_{i,j=1}^{n} W_{h}( Y_i-t)\delta_i W_{h}( Y_j-t)\delta_j.
\end{align*}

After some algebra,   the above plug-in estimator satisfies
\begin{equation}\label{SDC-second-form}
\widetilde{\operatorname{SID}}_{K}(T,\mathbf{X})= \frac{1}{n^5}\sum_{i,j, k,l,r=1}^{n}[b_{ikr}-b_{kir}]K_{ij} [b_{jlr}-b_{ljr}]\delta_r,
\end{equation}
where $K_{ij}= K(\mathbf{X}_i,\mathbf{X}_j)$ and $b_{ijk}= \delta_i W_{h}( Y_i-Y_k)I(Y_{j} \geq Y_k)$. From~\eqref{SDC-second-form}, we can see that the plug-in estimator
 $\widetilde{\operatorname{SID}}_{K}(T,\mathbf{X})$ is essentially a $V$-statistic.
Then, its corresponding $U$-statistic-based estimator can be obtained as
\begin{equation}\label{SK-U-est}
\operatorname{\widehat{SID}}_{K}(T,\mathbf{X})= \frac{1}{(n)_5}\sum_{i\neq j\neq k\neq l\neq r} [b_{ikr}-b_{kir}]K_{ij} [b_{jlr}-b_{ljr}]\delta_r,
\end{equation}
where $(n)_5=n(n-1)(n-2)(n-3)(n-4)$.

With the same arguments used for~\eqref{SDC-van-denom} and~\eqref{SK-U-est},
the plug-in and $U$-statistic estimators for $\operatorname{SID}_{\beta}(T,\mathbf{X})$ are given by
\begin{align*}
\widetilde{\operatorname{SID}}_{\beta}(T,\mathbf{X}) &= \frac{1}{n}\sum_{k=1}^n\Big[-\widetilde{S}_{\beta,1}( Y_k) +2\widetilde{S}_{\beta,2}(Y_k)
-\widetilde{S}_{\beta,3}(Y_k)\Big]\delta_k, \\
\operatorname{\widehat{SID}}_{\beta}(T,\mathbf{X}) &= \frac{1}{(n)_5}\sum_{i\neq j\neq k\neq l\neq r} \Vert \mathbf{X}_i-\mathbf{X}_j\Vert^\beta[b_{ikr}-b_{kir}] [b_{jlr}-b_{ljr}]\delta_r,
\end{align*}
where $\widetilde{S}_{\beta,1}(t)$, $\widetilde{S}_{\beta,2}(t)$, and $\widetilde{S}_{\beta,3}(t)$ are similarly defined as
$\widetilde{S}_{K,1}(t)$, $\widetilde{S}_{K,2}(t)$, and $\widetilde{S}_{K,3}(t)$ by replacing $K(\mathbf{X}_i,\mathbf{X}_j)$ with $\Vert \mathbf{X}_i-\mathbf{X}_j\Vert^\beta$ for $\beta\in(0,2)$.

\section{Asymptotic properties}\label{chapter-asymptotic}
In this section, we first show that the above empirical estimators are consistent with their population counterparts.
Thus, we need the following assumptions
\begin{assumption}\label{condition-4}
 The kernel function $W(u)$ is a bounded and symmetric density function with a bounded derivative and satisfies $\int_{-\infty}^{\infty}|u|^j W(u) dW(u)<\infty, j=1,2,\dots$
\end{assumption}
 \begin{assumption}\label{condition-5}
 $h \rightarrow 0$ and $ \sqrt{nh}/\ln (1 / h) \rightarrow \infty$ as $n \rightarrow \infty$.
 \end{assumption}

The assumptions are common   in   nonparametric literature.  Many kernel functions satisfy Assumption 4, for example, the standard Gaussian kernel $W(u)=1 / \sqrt{ 2 \pi} \exp(-u^2 / 2)$ and the Epanechnikov kernel $W(u)=3/4(1-u^2) I(|u| \leq 1)$.  Assumption 5 requires the bandwidth to be chosen appropriately according to $n$. It is well known that the choice of the bandwidth in kernel-smoothing methods is a challenging and unsolved problem. In our numerical studies, we choose the optimal bandwidth \citep{silverman2018density}, given by $h=\{4/3\}^{-1/5}\hat{\sigma}_y n^{-1/5}$, where $\hat{\sigma}_y$ is the sample standard deviation of $\{y_1,\ldots, y_n\}$.

\begin{theorem} \label{Theorem-consistency}
Under  Assumptions   \ref{condition-3}-\ref{condition-5},  we
have, as $n\rightarrow \infty$,
\begin{align*}
\widetilde{\operatorname{SID}}_{\beta}(T,\mathbf{X})
&\stackrel{P}{\longrightarrow} {\operatorname{SID}}_{\beta}(T,\mathbf{X}),
& \operatorname{\widehat{SID}}_{\beta}(T,\mathbf{X}) &\stackrel{P}{\longrightarrow} {\operatorname{SID}}_{\beta}(T,\mathbf{X}), \\
\widetilde{\operatorname{SID}}_{K}(T,\mathbf{X})
& \stackrel{P}{\longrightarrow}{\operatorname{SID}}_{K}(T,\mathbf{X}),
& \operatorname{\widehat{SID}}_{K}(T,\mathbf{X})
& \stackrel{P}{\longrightarrow}{\operatorname{SID}}_{K}(T,\mathbf{X}).
\end{align*}
\end{theorem}

We next develop the asymptotic distribution of $\operatorname{\widehat{SID}}_{K}(T,\mathbf{X})$ under the null hypothesis. In the proof of Theorem~\ref{Theorem-consistency}, we constructed a 5-order $U$-statistic by symmetrizing $\operatorname{\widehat{SID}}_{K}(T,\mathbf{X})$.
When establishing the asymptotic null distribution, we can simplify the theoretical analysis and approximate $\operatorname{\widehat{SID}}_{K}(T,\mathbf{X})$ with the following 4-order $U$-statistic
\begin{equation}\label{sid-k-nu}
\operatorname{\widehat{SID}}_{K}^\nu(T,\mathbf{X}) =
\begin{pmatrix} n \\ 4 \end{pmatrix}^{-1}
\sum_{i<j<k<l}\int^\tau_0 h_{n}(\mathbf{Z}_{i}, \mathbf{Z}_{j}, \mathbf{Z}_{k}, \mathbf{Z}_{l};t)d \nu(t),
\end{equation}
 where $\mathbf{Z}_i = (Y_i, \delta_i, \mathbf{X}_i)$ and
\begin{equation}\label{sid-k-kernel}
h_{n}(\mathbf{Z}_{i}, \mathbf{Z}_{j}, \mathbf{Z}_{k}, \mathbf{Z}_{l};t) = \frac{1}{4 !} \sum_{\pi}P_{n}(\mathbf{Z}_{i'}, \mathbf{Z}_{j'}, \mathbf{Z}_{k'}, \mathbf{Z}_{l'};t),
\end{equation}
with $
 P_{n}(\mathbf{Z}_{i}, \mathbf{Z}_{j}, \mathbf{Z}_{k}, \mathbf{Z}_{l};t)
  =[b_{i k}(t)- b_{ki}(t)] {K}_{ij}[b_{j l}(t)- b_{lj}(t)]$ and
 $b_{i k}(t)=\delta_{i} W_h(Y_{i}-t)I(Y_{k}\geq t)$.
Here, $\sum_{\pi}$ denotes summation over all the $4!$ permutations $(i',j',k',l')$ of $(i,j,k,l)$. Some calculations yield that
\begin{multline}\label{symmetric-kernel-detail}
\begin{aligned}
\!\!\!\!\!\!\!\! h_{n}(\mathbf{Z}_{i},\! \mathbf{Z}_{j}, \!\mathbf{Z}_{k}, \!\mathbf{Z}_{l};\!t)
&= \frac{1}{12}\Big\{[K_{ij}-K_{il}-K_{jk}+K_{kl}][b_{ik}(t)-b_{ki}(t)][b_{jl}(t)-b_{lj}(t)] \\
& \quad + \ [K_{ij}-K_{jl}-K_{ik}+K_{kl}][b_{il}(t)-b_{li}(t)][b_{jk}(t)-b_{kj}(t)] \\
& \quad + \ [K_{ik}-K_{il}-K_{jk}+K_{jl}][b_{ij}(t)-b_{ji}(t)][b_{kl}(t)-b_{lk}(t)]\Big\}.
\end{aligned}
\end{multline}

Let $h_{n c}(z_{1},\ldots, z_{c}) = E\{ h_{n}(\mathbf{Z}_{1},\mathbf{Z}_{2},\mathbf{Z}_{3},\mathbf{Z}_{4}) | \mathbf{Z}_{1}=z_{1}, \ldots, \mathbf{Z}_{c}=z_{c}\}$ be the $c$-order projection of $h_{n}(\mathbf{Z}_{1},\mathbf{Z}_{2},\mathbf{Z}_{3},\mathbf{Z}_{4})$, for $c=1,2,3,4.$ The following lemma provides closed-form expressions of the first- and second-order projections under $H_{0}$.

\begin{lemma}\label{Hn2-degenation}
Let $\mathbf{z}=(y, \tilde{\delta}, \mathbf{x})$ and $\mathbf{z}'=(y', \tilde{\delta}', \mathbf{x}')$, for any $y,y'\in\mathbb{R}^+$, $\tilde{\delta},\tilde{\delta}'\in\{0,1\}$, and $\mathbf{x},\mathbf{x}'\in \mathcal{X}$. Under $H_{0},$ we have
 \begin{equation*}
 {h}_{n 1}(\mathbf{z};t)=0; \quad {h}_{n 2}(\mathbf{z},\mathbf{z}';t) = \frac{1}{6}   U(\mathbf{x}, \mathbf{x}^{\prime};t) V(y, \tilde{\delta}; y^{\prime}, \tilde{\delta}^{\prime};t),
\end{equation*}
 where
\begin{align*}
\!\!\!\!\!\!U(\mathbf{x}, \mathbf{x}^{\prime};t) &= K(\mathbf{x}, \mathbf{x}^{\prime}) - E\{K (\mathbf{x}, \mathbf{X}_1) \mid Y_1(t)=1\}    - \ E\{K (\mathbf{x}', \mathbf{X}_1) \mid Y_1(t)=1\} \\
& \quad + \ E\{K ( \mathbf{X}_1, \mathbf{X}_2) \mid Y_1(t)=1, Y_2(t)=1\}, \\
V(y, \tilde{\delta}; y^{\prime}, \tilde{\delta}^{\prime};t) &= [\tilde{\delta} W_h(y-t)S(t)-I(y\geq t)F_h(t)]   [\tilde{\delta}^{\prime} W_h(y^{\prime}-t)S(t)-I(y^{\prime}\geq t)F_h(t)],
\end{align*}
 with $S(t)=P\{Y\geq t\}$ and $F_h(t)=E\{ \delta W_h(Y-t)\}$.
\end{lemma}

Lemma~\ref{Hn2-degenation} suggests that $\operatorname{\widehat{SID}}^\nu_{K}(T,\mathbf{X})$ is a degenerate $U$-statistic under $H_0$. Using the Hoeffding decomposition \citep{lee1990u}, we can approximate $\operatorname{\widehat{SID}}^\nu_{K}(T,\mathbf{X})$ with the $U$-statistic based on the second-order projection ${h}_{n 2}(\mathbf{z},\mathbf{z}';t)$. Thus, the closed-form expression of ${h}_{n 2}(\mathbf{z},\mathbf{z}';t)$ is key to establishing the asymptotic null distribution. Additionally, note that $h_{n}(\mathbf{Z}_{i}, \mathbf{Z}_{j}, \mathbf{Z}_{k}, \mathbf{Z}_{l};t)$ depends on the bandwidth $h$ and thus, is a variable kernel. We can apply the asymptotic theory for a degenerate variable kernel $U$-statistic established by \citet{Li96Fan} to study the asymptotic null distribution of $\operatorname{\widehat{SID}}_{K}(T,\mathbf{X})$.

\begin{theorem} \label{Theorem-asymptotic-null}
Assume that Assumptions    \ref{condition-3}-\ref{condition-5}     hold.
Under $H_{0},$ we have
$$
n h^{1/ 2} \operatorname{\widehat{SID}}_{K}(T,\mathbf{X})\stackrel{d}{\longrightarrow} N(0,2\sigma^{2}),
$$
where $\sigma^2\! =\!E\{E\{U^2(\mathbf{X}_{1}, \mathbf{X}_{2};Y) \!\mid\! Y,\delta=1\}S^4(Y)f^3_{Y,\delta}(Y,1)\}\int_0^\tau  [\int_0^\tau W(u)W(u+v) du ]^2 dv.
$
\end{theorem}

Theorem~\ref{Theorem-asymptotic-null} suggests that $\operatorname{\widehat{SID}}_{K}(T,\mathbf{X})$ converges in distribution to $N(0,2\sigma^{2})$ at the rate of $n h^{1/ 2}$, under $H_0$.
 This is consistent with the existing literature on nonparametric tests, such as \citet{SU2007807}, and \citet{ke2019expected}.
We next study the asymptotic behavior of $\operatorname{\widehat{SID}}_{K}(T,\mathbf{X})$ under $H_{1}$.

\begin{theorem} \label{Theorem-asymptotic-alternatives}
Assume that Assumptions    \ref{condition-3}-\ref{condition-5}
 hold.
Under $H_{1},$ we have that $nh^{1/ 2}\operatorname{\widehat{SID}}_{K}(T,\mathbf{X})\stackrel{P}{\rightarrow} \infty$, and
 $  \lim _{n \rightarrow \infty} P\{ nh^{1/ 2}\operatorname{\widehat{SID}}_{K}(T,\mathbf{X})>\gamma|H_{1}\}= 1
 $, for any $\gamma>0$.
\end{theorem}

It can be seen from Theorems~\ref{Theorem-asymptotic-null} and~\ref{Theorem-asymptotic-alternatives} that $n h^{1/ 2} \operatorname{\widehat{SID}}_{K}(T,\mathbf{X})$ is stochastically bounded under the null hypothesis, whereas it diverges to infinity under fixed alternative hypotheses. Thus, $n h^{1/ 2} \operatorname{\widehat{SID}}_{K}(T,\mathbf{X})$ is able to detect any type of dependency between $T$ and $\mathbf{X}$.
Additionally, under the assumption $E\{\Vert \mathbf{X}\Vert^\beta\}<\infty$, $\beta\in(0,2)$, we can derive similar asymptotic properties for $\operatorname{\widehat{SID}}_{\beta}(T,\mathbf{X})$ as those in Theorems~\ref{Theorem-asymptotic-null} and~\ref{Theorem-asymptotic-alternatives}.

\section{Wild bootstrap}\label{chapter-bootstrap}
Note that the limiting null distribution of $\operatorname{\widehat{SID}}_{K}(T,\mathbf{X})$ is unknown. In this section, we propose a wild bootstrap approach to approximate the limiting null distribution.
To develop our wild bootstrap approach,  a strategy commonly used  is to perturb the symmetrized version of $\operatorname{\widehat{SID}}_{K}(T,\mathbf{X})$ through a zero-mean and unit-variance random variable.
However, note that the symmetrical kernel of $\operatorname{\widehat{SID}}_{K}(T,\mathbf{X})$ in~\eqref{sid-k-kernel} or~\eqref{symmetric-kernel-detail} is computationally intensive and thus, the above strategy is infeasible. To overcome this problem,
 we use Lemma~\ref{Hn2-degenation} and perturb the second-order projection
 $U$-statistic to approximate the null distribution. Specifically, with some abuse of notation,  define the wild bootstrap test as
\begin{equation*}
\operatorname{\widehat{SID}}^*_{K}(T,\mathbf{X})= \frac{2}{n(n-1)}  \sum_{i< j}e_i e_j \int^\tau_0  \widehat{ U}(\mathbf{X}_i, \mathbf{X}_j;t) \widehat{V}(Y_i, \delta_i; Y_j, \delta_j;t) d\overline{N}(t),
\end{equation*}
where $\{e_i\}_{i=1}^n$ are i.i.d.\ samples drawn from a zero-mean and unit-variance variable. Here, $\widehat{ U}(\mathbf{x}, \mathbf{x}^{\prime};t)$ and $ \widehat{V}(y, \tilde{\delta}; y^{\prime}, \tilde{\delta}^{\prime};t)$ are plug-in estimators of
$ { U}(\mathbf{x}, \!\mathbf{x}^{\prime};\!t)$ and $  {V}(y, \!\tilde{\delta}; \!y^{\prime}, \!\tilde{\delta}^{\prime};t)$,
where $E\{K (\mathbf{x}, \mathbf{X}_1) \!\mid\! Y_1(t)\!=\!1\}$, $E\{K ( \mathbf{X}_1, \mathbf{X}_2) \!\mid\! Y_1(t)\!=\!1, Y_2(t)\!=\!1\}$, ${S}(t)$  and ${F}_h(t)$ are estimated by
\begin{eqnarray*}
&&\widehat{E}\{K (\mathbf{x}, \mathbf{X}_1)|Y_1(t)=1\}=  {\sum_{i=1}^nK (\mathbf{x}, \mathbf{X}_i)Y_i(t)}/{\sum_{i=1}^nY_i(t)},\\
&&\widehat{E}\{K ( \mathbf{X}_1,  \mathbf{X}_2) |Y_1(t)=1,  Y_2(t)=1\}= {\sum_{i,j=1}^n K(\mathbf{X}_i, \mathbf{X}_j)Y_i(t)Y_j(t)}/{\{\sum_{i=1}^nY_i(t)\}^2},\\
&&\widehat{S}(t)=\frac{1}{n} \sum_{i=1}^n I(Y_i\geq t),
~~\widehat{F}_h(t)=\frac{1}{n} \sum_{i=1}^n  \delta_i W_h(Y_i-t).
\end{eqnarray*}

The wild bootstrap test procedure is given by
\begin{enumerate}
  \item Generate the bootstrap samples of $\operatorname{\widehat{SID}}^*_{K}(T,\mathbf{X})$
  \begin{equation}\label{SK-U-est-boot-al}
\operatorname{\widehat{SID}}^{*(b)}_{K}(T,\mathbf{X})=
 \frac{1}{n^2}  \sum_{i< j }e^{(b)}_i e^{(b)}_j \int^\tau_0  \widehat{ U}(\mathbf{X}_i, \mathbf{X}_j;t) \widehat{V}(Y_i, \delta_i; Y_j, \delta_j;t) d\overline{N}(t),
\end{equation}
where $\{e_i^{(b)}\}_{i=1}^n$ are i.i.d.\ samples
from a distribution with zero mean and unit variance.

\item Repeat step 1 for $B$ times and obtain $\{ n h^{1/ 2} \operatorname{\widehat{SID}}^{*(b)}_{K}(T,\mathbf{X}) \}_{b=1}^B$.

\item Calculate the $(1-\alpha)$-th quantile of $\{ n h^{1/ 2} \operatorname{\widehat{SID}}^{*(b)}_{K}(T,\mathbf{X}) \}_{b=1}^B$, denoted as $\widehat{Q}_{n,(1-\alpha)}^*$, for a given significance level $\alpha$.

\item Reject the null hypothesis if $n h^{1/ 2}\operatorname{\widehat{SID}}_{K}(T,\mathbf{X})\geq \widehat{Q}_{n,(1-\alpha)}^*$.
\end{enumerate}

We next state the validity of the above wild bootstrap procedure.
From the proof of Theorem~\ref{Theorem-asymptotic-null}, we show that $\operatorname{\widehat{SID}}^\nu_{K}(T,\mathbf{X})$ is a degenerate $U$-statistic of second order under $H_0$, and its limiting distribution can be approximated by that of the   second-order projection-based $U$-statistic
\begin{equation}\label{second-order-projection}
\frac{12}{n(n-1)} \sum_{i< j} \int^\tau_0  {h}_{n 2}(\mathbf{Z}_i,\mathbf{Z}_j;t) d\nu(t).
\end{equation}
From Lemma~\ref{Hn2-degenation} we can see that $\operatorname{\widehat{SID}}^*_{K}(T,\mathbf{X})$ is essentially a bootstrap approximation to~\eqref{second-order-projection}.
This is confirmed by the following theorem.
\begin{theorem} \label{Theorem-asymptotic-boostrap}
Assume that Assumptions    \ref{condition-3}-\ref{condition-5}  hold. Under $H_{0},$  we have
$$
n h^{1/ 2}\operatorname{\widehat{SID}}^*_{K}(T,\mathbf{X}) \stackrel{d^*}{\longrightarrow} N\left(0,2\sigma^{2}\right),
$$
where $\stackrel{d^*}{\longrightarrow}$ denotes convergence in distribution with respect to  $\{e_i\}_{i=1}^n$ conditional on  $\{(Y_{i}, \delta_{i}, \mathbf{X}_{i}) \}_{i=1}^n$.
\end{theorem}

Theorem \ref{Theorem-asymptotic-boostrap} implies that $\operatorname{\widehat{SID}}^*_{K}(T,\mathbf{X})$ is consistent with the limiting
 distribution of $\operatorname{\widehat{SID}}_{K}(T,\mathbf{X})$ under $H_{0}$.
 The following theorem establishes the asymptotic behavior of  $\operatorname{\widehat{SID}}^*_{K}(T,\mathbf{X})$  under $H_{1}$.

\begin{theorem} \label{Theorem-power-boostrap}
Assume that Assumptions    \ref{condition-3}-\ref{condition-5} hold. Under $H_1$, we have that
$$ \lim _{n \rightarrow \infty}P \{n h^{1/ 2} \operatorname{\widehat{SID}}_{K}(T,\mathbf{X})\geq \widehat{ Q}_{n,(1-\alpha)}^*|H_1 \} = 1.$$
\end{theorem}

Theorem \ref{Theorem-power-boostrap} suggests that the $U$-wild bootstrap statistics, $\widehat{\operatorname{SID}}^*_{K}(T,\mathbf{X})$, can consistently detect a dependency between survival times and covariates.  Similarly, we can construct a $V$-wild bootstrap statistic based on the $V$-statistic $\widetilde{\operatorname{SID}}_{K}(T,\mathbf{X})$, given by
\begin{equation*}
\widetilde{\operatorname{SID}}^*_{K}(T,\mathbf{X})= \frac{1}{n^2}  \sum_{i, j=1}^ne_i e_j \int^\tau_0  \widehat{ U}(\mathbf{X}_i, \mathbf{X}_j;t) \widehat{V}(Y_i, \delta_i; Y_j, \delta_j;t) d\overline{N}(t).
\end{equation*}
Let  $\widetilde{\operatorname{SID}}^{*(b)}_{K}(T,\mathbf{X}),$ $b=1,\ldots,B$, be the bootstrap samples of $\widetilde{\operatorname{SID}}^*_{K}(T,\mathbf{X})$.
Then, we reject the null hypothesis if $n h^{1/ 2}\widetilde{\operatorname{SID}}_{K}(T,\mathbf{X})\geq \widetilde{Q}_{n,(1-\alpha)}^*$, where
$\widetilde{Q}_{n,(1-\alpha)}^*$ is the $(1-\alpha)$-th quantile of $\{ n h^{1/ 2} \widetilde{\operatorname{SID}}^{*(b)}_{K}(T,\mathbf{X}) \}_{b=1}^B$.

Note that the computational complexity for both $\widehat{\operatorname{SID}}_{K}(T,\mathbf{X})$ and $\widetilde{\operatorname{SID}}_{K}(T,\mathbf{X})$ is \(O(n^5)\). In contrast, the $U$-wild bootstrap statistic, $\widehat{\operatorname{SID}}^*_{K}(T,\mathbf{X})$, offer an improvement with a  computational complexity of \(O(n^2)\). The $V$-wild bootstrap statistic, $\widetilde{\operatorname{SID}}^*_{K}(T,\mathbf{X})$, also maintain a computational complexity of \(O(n^2)\). However, it is more convenient   in practice compared to the $U$-wild bootstrap statistic. In the following numerical studies, we   carry on our tests based on $\widetilde{\operatorname{SID}}^*_{K}(T,\mathbf{X})$.

\section{Simulation studies}\label{chapter-simulation}

In this section, We conduct simulation studies to evaluate the finite-sample performance of the proposed SID-based test methods.
We consider the two classes of SID test statistics, which are characteristic function-based SID tests with the weight function in~\eqref{weight-szekely2007}
($\operatorname{SID}_\beta$) and kernel-based SID tests ($\operatorname{SID}_K$). For $\operatorname{SID}_\beta$, we consider the two special choices of $\beta$, $\beta=1$ ($\operatorname{SID}_1$) and $\beta=1/2$ ($\operatorname{SID}_{0.5}$). For $\operatorname{SID}_K$, we consider the two special kernel functions, namely the Gaussian kernel ($\operatorname{SID}_{\mathrm{gauss}}$) and the Laplacian kernel ($\operatorname{SID}_{\mathrm{lap}}$), defined by
 $K( \mathbf{x}_1, \mathbf{x}_2)=\exp(-\|\mathbf{x}_1- \mathbf{x}_2\|^2/\gamma^2_1) \text{ and } K(\mathbf{x}_1,\mathbf{x}_2)=\exp \left(- {\|\mathbf{x}_1-\mathbf{x}_2\|}/{\gamma_2}\right),$
where $\gamma_1, \gamma_2>0$ are tuning parameters.
In the numerical studies, we use the median-distance heuristic \citep{gretton2012kernel}, $\gamma_\mathrm{med}\!=\! (\text {median }\{\! \|\mathbf{X}_{i}\!-\!\mathbf{X}_{j}\|^2: i \!\neq\! j \}/ 2)^{1/2}$, to select $\gamma_1$ and $\gamma_2$.

We compare the results from our methods with those from a Cox's proportional hazards model-based test (CPH), the IPCW-based distance covariance \citep[IPCW;][]{edelmann2021consistent}, and the kernel log-rank test \citep[KLR;][]{Tamara}. According to \citet{edelmann2021consistent} and \citet{Tamara}, we compute the critical values of the IPCW and KLR test statistics with the permutation and wild bootstrap methods.
For KLR, we choose the Gaussian kernels as the kernel functions with respect to time and covariates to conserve space.
Here, we repeat each experiment 1000 times and set the permuted and bootstrap sample sizes $B=2000$. For the wild bootstrap method used by
the KLR and SIDs, $\{e_i\}_{i=1}^n$ are generated from the Rademacher distribution, i.e., $P\{e_i=\pm 1 \}=1/2.$

\begin{example}\label{exampl_1}
This example examines the type-I error rates of our SID-based tests.
We consider the following four cases, which were investigated by \citet{Tamara}
\begin{description}
 \item [\textbf{Case 1:}]   $T\sim\operatorname{Exp}(\lambda)$ and  $C\sim\operatorname{Exp}(1)$ with  $X\sim \mathrm{Unif}[-1,1];$ \item [\textbf{Case 2:}]  $ T  \sim\operatorname{Exp}(\lambda)$ and  $C\sim\operatorname{Exp}( e^{X / 3})$ with  $X\sim \mathrm{Unif}[-1,1];$
\item [\textbf{Case 3:}]    $T\sim\operatorname{Exp}(\lambda)$ and  $C\sim\operatorname{Weib}(3.35+1.75 X,1)$ with  $X \sim \mathrm{Unif}[-1,1];$
\item [\textbf{Case 4:}]  $T \sim\operatorname{Exp}(\lambda)$ and  $C\sim\operatorname{Exp}( e^{\textbf{1} _{10}^{T} \textbf{X}/20})$ with  $\textbf{X} \sim{N}_{10} (\mathbf{0}, \boldsymbol{\Sigma}_{10} ).$
\end{description}
Here, $\operatorname{Weib}(a,b)$ is the Weibull distribution with the parameters $a$ and $b$, and $\operatorname{Exp}(\lambda)$ is the exponential distribution with mean $\lambda$, where $\lambda$ is used to control the censoring rate approximately. Here, $\textbf{1}_{p}=(1, \ldots, 1)^T \in \mathbb{R}^{p}$ and $\boldsymbol{\Sigma}_p=(0.5^{|j-k|})$.
\end{example}

Tables~\ref{tab1-1} and~\ref{tab1-2} summarize the empirical type-I error rates for the seven methods, namely CPH, KLR, IPCW, $\operatorname{SID}_1$, $\operatorname{SID}_{0.5}$, $\operatorname{SID}_{\mathrm{gauss}}$, and $\operatorname{SID}_{\mathrm{lap}}$ at the significance levels $\alpha=0.01$, 0.05, or 0.1 with 30\% or 60\% censoring and $n=50$ or $150$.
 It can be seen that the empirical type-I error rates for $\operatorname{SID}_1$, $\operatorname{SID}_{0.5}$, $\operatorname{SID}_{\mathrm{gauss}}$, and $\operatorname{SID}_{\mathrm{lap}}$ are very close to the true significance levels, which confirms the asymptotical properties in Theorems~\ref{Theorem-asymptotic-null} and~\ref{Theorem-asymptotic-boostrap}. Additionally, the KLR has similar performance.

As seen from Tables \ref{tab1-1} and \ref{tab1-2}, the empirical type-I error rates
 for the IPCW method are inflated in Cases 2--4, especially when the censoring rate is 60\%. This is, perhaps, because IPCW assumes that $C$ is completely independent of $\mathbf{X}$.
Specifically, in Case 1, for which the completely independent assumption is satisfied, IPCW achieves approximately the true significance levels, whereas it has inflated the empirical type-I error rates in Cases 2--4, for which the assumption is invalid.
Additionally, the empirical type-I error rates of the CPH method are under reasonable control, except Case 4, where  $p$ is relatively high.

\begin{table}[http]
\caption{Empirical type-I error rate with 30\% censoring for Example~\ref{exampl_1}.}
\label{tab1-1}\scriptsize
\resizebox{\linewidth}{!}{%
\begin{tabular}{cccccccccccc}
 \hline
     &&  &  &   &  & &\multicolumn{2}{c}{$\operatorname{SID}_\beta$ } & &\multicolumn{2}{c}{$\operatorname{SID}_K$}    \\
 \cline{8-9}                   \cline{11-12}
 $\alpha$   &Case   &$n$ &CPH  &KLR  &IPCW  & &$\operatorname{SID}_1$ &$\operatorname{SID}_{0.5}$   &   & $\operatorname{SID}_{\mathrm{gauss}}$ &$\operatorname{SID}_{\mathrm{lap}}$   \\ \hline
 0.01 &Case 1        &50 &0.012&0.009 &0.007 &&0.008  &0.007 &&0.007 &0.007      \\
      &              &150&0.007&0.010 &0.011 &&0.007  &0.009 &&0.008 &0.009      \\
      &Case 2        &50 &0.012&0.005 &0.029 &&0.006  &0.007 &&0.006 &0.007      \\
      &              &150&0.010&0.010 &0.014 &&0.013  &0.010 &&0.011 &0.011      \\
      &Case 3        &50 &0.021&0.009 &0.017 &&0.012  &0.011 &&0.011 &0.015      \\
      &              &150&0.016&0.010 &0.014 &&0.013  &0.014 &&0.013 &0.012      \\
      &Case 4        &50 &0.011&0.010 &0.021 &&0.013  &0.013 &&0.013 &0.015      \\
      &              &150&0.033&0.013 &0.012 &&0.011  &0.012 &&0.013 &0.012      \\  \hline
 0.05 &Case 1        &50 &0.040&0.038 &0.056 &&0.041  &0.050 &&0.049 &0.054      \\
      &              &150&0.048&0.051 &0.052 &&0.048  &0.054 &&0.047 &0.045      \\
      &Case 2        &50 &0.042&0.042 &0.105 &&0.051  &0.050 &&0.048 &0.050      \\
      &              &150&0.045&0.056 &0.071 &&0.051  &0.049 &&0.054 &0.042      \\
      &Case 3        &50 &0.042&0.043 &0.059 &&0.053  &0.049 &&0.050 &0.048      \\
      &              &150&0.056&0.050 &0.053 &&0.061  &0.057 &&0.056 &0.058      \\
      &Case 4        &50 &0.121&0.083 &0.080 &&0.066  &0.065 &&0.055 &0.056      \\
      &              &150&0.076&0.058 &0.064 &&0.051  &0.052 &&0.051 &0.052      \\ \hline
 0.10 &Case 1        &50 &0.088&0.088 &0.103 &&0.101  &0.101 &&0.097 &0.108      \\
      &              &150&0.096&0.110 &0.092 &&0.092  &0.091 &&0.098 &0.096      \\
      &Case 2        &50 &0.089&0.086 &0.173 &&0.110  &0.107 &&0.091 &0.105      \\
      &              &150&0.099&0.112 &0.124 &&0.101  &0.102 &&0.107 &0.106      \\
      &Case 3        &50 &0.091&0.079 &0.124 &&0.109  &0.103 &&0.104 &0.109      \\
      &              &150&0.101&0.112 &0.106 &&0.100  &0.107 &&0.105 &0.096      \\
      &Case 4        &50 &0.213&0.115 &0.148 &&0.115  &0.112 &&0.105 &0.109      \\
      &              &150&0.140&0.146 &0.114 &&0.099  &0.101 &&0.100 &0.095      \\
\hline
\end{tabular}}
\end{table}

\begin{table}[http]
\caption{Empirical type-I error rate with 60\% censoring for Example~\ref{exampl_1}.}
\label{tab1-2} \scriptsize
\resizebox{\linewidth}{!}{%
\begin{tabular}{cccccccccccc}
\hline
     &&  &  &   &  & &\multicolumn{2}{c}{$\operatorname{SID}_\beta$ } & &\multicolumn{2}{c}{$\operatorname{SID}_K$}    \\
 \cline{8-9}                   \cline{11-12}
 $\alpha$   & Case   &$n$ &CPH  &KLR  &IPCW  & &$\operatorname{SID}_1$ &$\operatorname{SID}_{0.5}$   &   & $\operatorname{SID}_{\mathrm{gauss}}$ &$\operatorname{SID}_{\mathrm{lap}}$   \\ \hline
 0.01 &Case  1           &50 &0.012&0.003 &0.013 &&0.011  &0.009 &&0.007 &0.008    \\
      &             &150&0.007&0.008 &0.008 &&0.009  &0.010 &&0.008 &0.009    \\
      &Case  2           &50 &0.004&0.007 &0.027 &&0.016  &0.015 &&0.015 &0.011    \\
      &             &150&0.014&0.006 &0.022 &&0.009  &0.007 &&0.008 &0.009    \\
      &Case  3           &50 &0.014&0.008 &0.021 &&0.015  &0.013 &&0.012 &0.015    \\
      &             &150&0.008&0.004 &0.011 &&0.009  &0.009 &&0.011 &0.012    \\
      &Case  4           &50 &0.061&0.004 &0.006 &&0.010  &0.011 &&0.011 &0.011    \\
      &             &150&0.022&0.010 &0.012 &&0.013  &0.013 &&0.013 &0.012    \\ \hline
0.05  &Case  1           &50 &0.054&0.042 &0.054 &&0.044  &0.050 &&0.051 &0.046    \\
      &             &150&0.051&0.052 &0.046 &&0.051  &0.047 &&0.045 &0.050    \\
      &Case  2           &50 &0.053&0.038 &0.062 &&0.054  &0.051 &&0.048 &0.051    \\
      &             &150&0.048&0.050 &0.041 &&0.051  &0.056 &&0.053 &0.056    \\
      &Case  3           &50 &0.052&0.041 &0.035 &&0.046  &0.045 &&0.048 &0.042    \\
      &             &150&0.057&0.050 &0.051 &&0.058  &0.052 &&0.047 &0.049    \\
      &Case  4           &50 &0.151&0.047 &0.070 &&0.045  &0.044 &&0.043 &0.040    \\
      &             &150&0.084&0.055 &0.068 &&0.063  &0.058 &&0.055 &0.057    \\ \hline
0.10  &Case  1           &50 &0.099&0.091 &0.083 &&0.099  &0.100 &&0.095 &0.097    \\
      &             &150&0.098&0.110 &0.107 &&0.110  &0.127 &&0.112 &0.113    \\
      &Case  2           &50 &0.110&0.096 &0.152 &&0.113  &0.112 &&0.111 &0.090    \\
      &             &150&0.120&0.121 &0.132 &&0.103  &0.101 &&0.101 &0.102    \\
      &Case  3           &50 &0.093&0.121 &0.118 &&0.102  &0.106 &&0.115 &0.095    \\
      &             &150&0.118&0.095 &0.104 &&0.120  &0.125 &&0.115 &0.102    \\
      &Case  4           &50 &0.256&0.097 &0.121 &&0.110  &0.107 &&0.103 &0.108    \\
      &             &150&0.149&0.101 &0.113 &&0.100  &0.099 &&0.100 &0.104    \\
\hline
\end{tabular}}
\end{table}

\begin{example}\label{exampl_2}
This example examines the power of our methods in various dependence relations and compares them with IPCW under the completely independent assumption. We generate the data in a similar way as \citet{edelmann2021consistent}.
\begin{description}
\item[(1)] Log-linear:   $\log(T) = 0.5\eta(X)+  \varepsilon$ with  $\eta(X)= {X}$;
\item[(2)] Log-quadratic: $\log(T) = 1.2\eta(X)+   \varepsilon$ with $\eta(X)={X}^2$;
\item[(3)] Log-cubic:   $\log(T) =  \eta(X)+  \varepsilon$ with $\eta(X)= {X}^3$;
\item[(4)] Log-cosine:  $\log(T) = 0.5\eta(X)+  \varepsilon$ with $\eta(X)=\cos(3 {X})$;
\item[(5)] Log-twolines:   $\log(T) = 4 \eta(X)+  \varepsilon$ with $\eta(X)=I(A=1) {X}-I(A=0) {X}$, ${A}\sim B(1,0.5),$   $ {A} \indep  {X}$;
\item[(6)]  Log-circle:   $\log(T) = \eta({X}_1,{X}_2)+   \varepsilon$ with $\eta({X}_1,{X}_2)=1$ if ${X}_1^2+{X}_2^2\leq 0.5$ and  $\eta({X}_1,{X}_2)=0$ otherwise.
\end{description}
Here, $ {X},{X}_1, {X}_2\sim \operatorname{Unif}[-1,1]$, ${X}_1\indep{X}_2,$ $\varepsilon \sim {N}(0,1),$ and
 $C\sim  \operatorname{Exp}(\lambda)$.
\end{example}

\begin{figure}[http]
 \centering
\begin{minipage}[b]{0.35\textwidth}
\centering
\includegraphics[width=\textwidth]{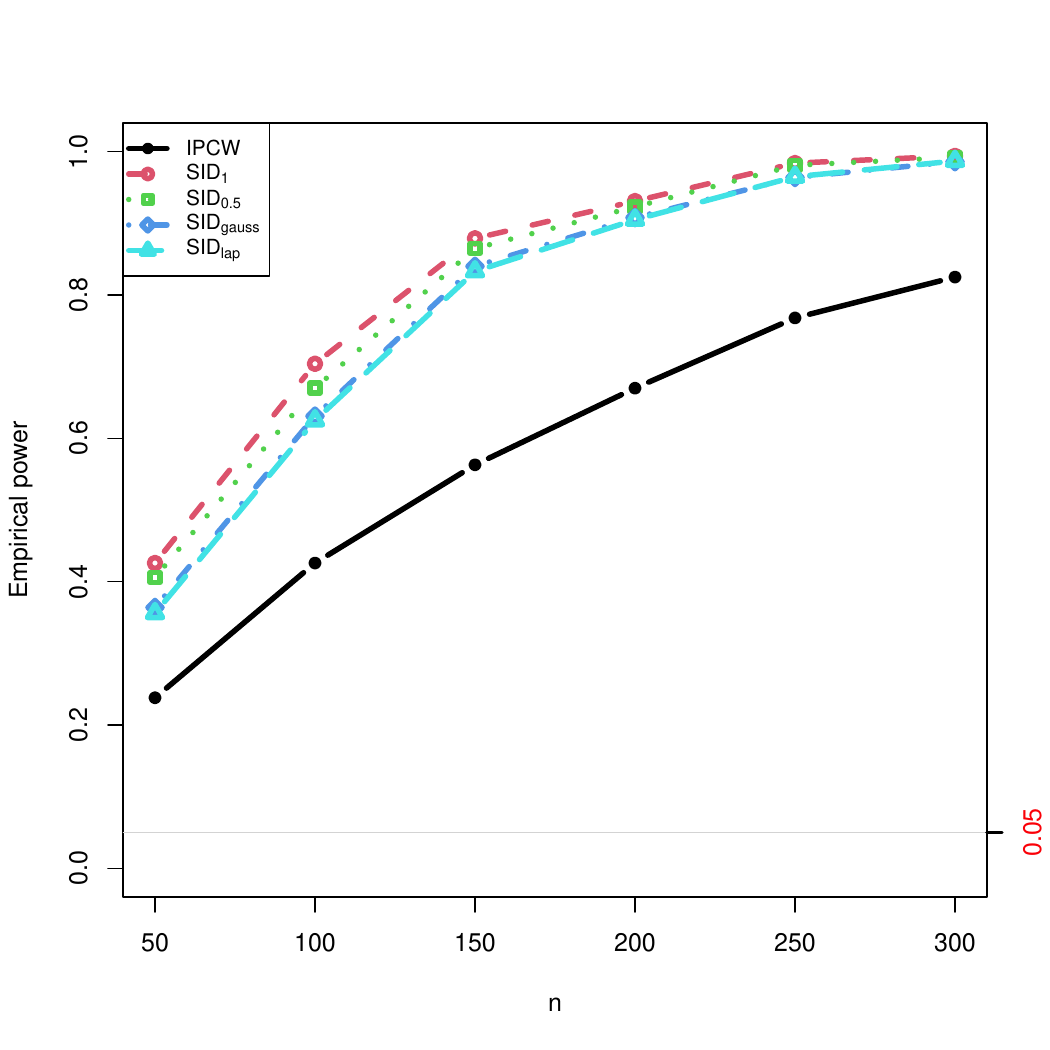}\vspace{-3mm}
{\footnotesize Log-linear }
\end{minipage}  \hspace{9mm}
 \begin{minipage}[b]{0.35\textwidth}
\centering
\includegraphics[width=\textwidth]{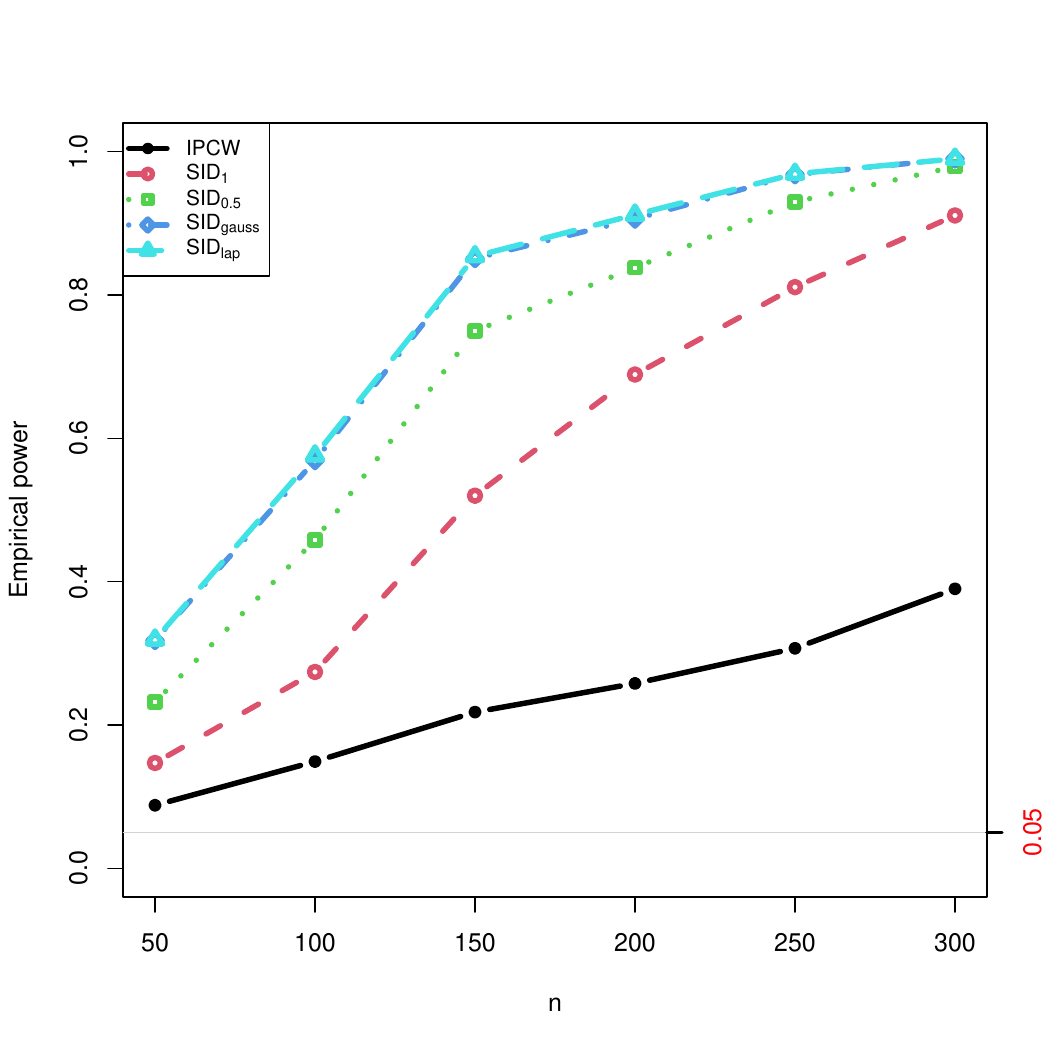}\vspace{-3mm}
{\footnotesize Log-quadratic }
\end{minipage}
 \begin{minipage}[b]{0.35\textwidth}
\centering
\includegraphics[width=\textwidth]{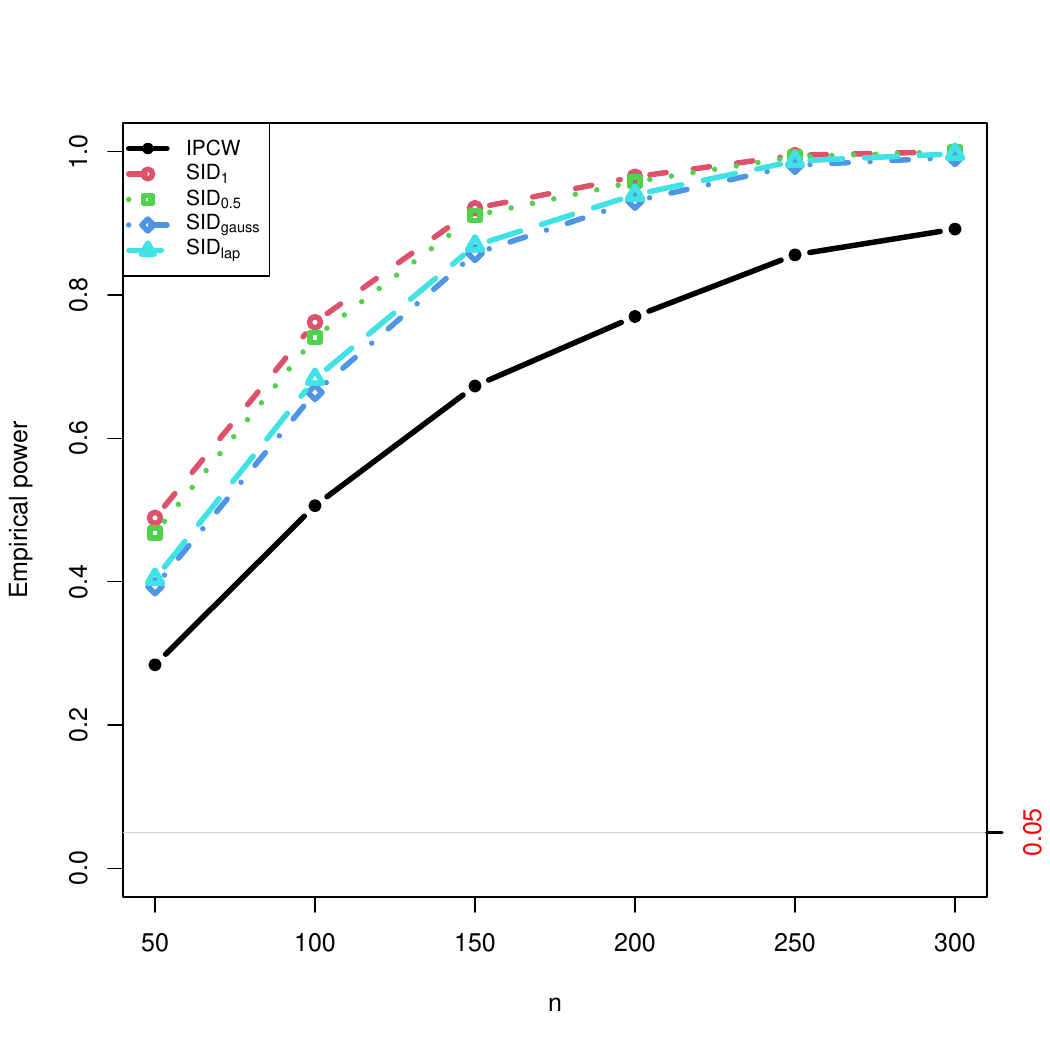}\vspace{-3mm}
{\footnotesize Log-cubic}
\end{minipage}  \hspace{9mm}
 \begin{minipage}[b]{0.35\textwidth}
\centering
\includegraphics[width=\textwidth]{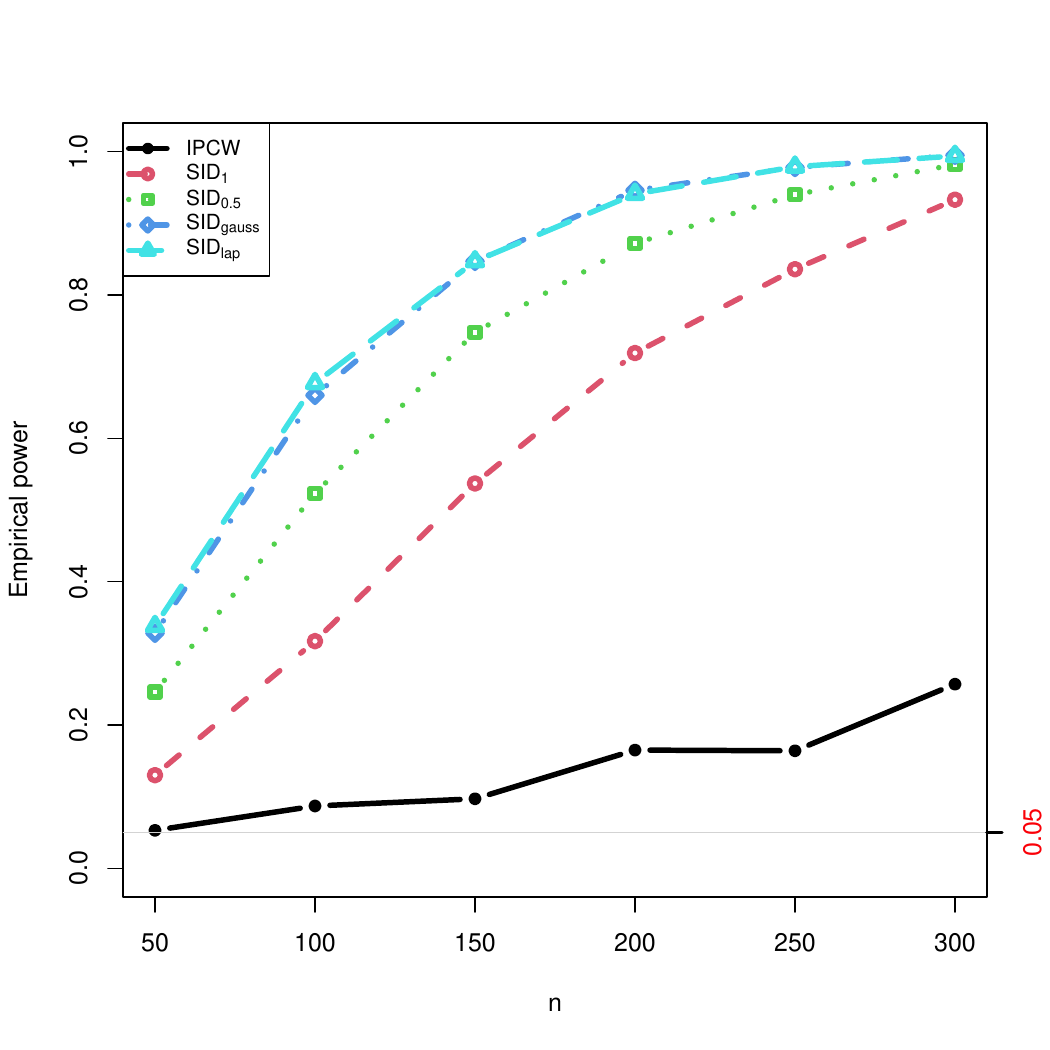}\vspace{-3mm}
{\footnotesize Log-cosine}
\end{minipage}
 \begin{minipage}[b]{0.35\textwidth}
\centering
\includegraphics[width=\textwidth]{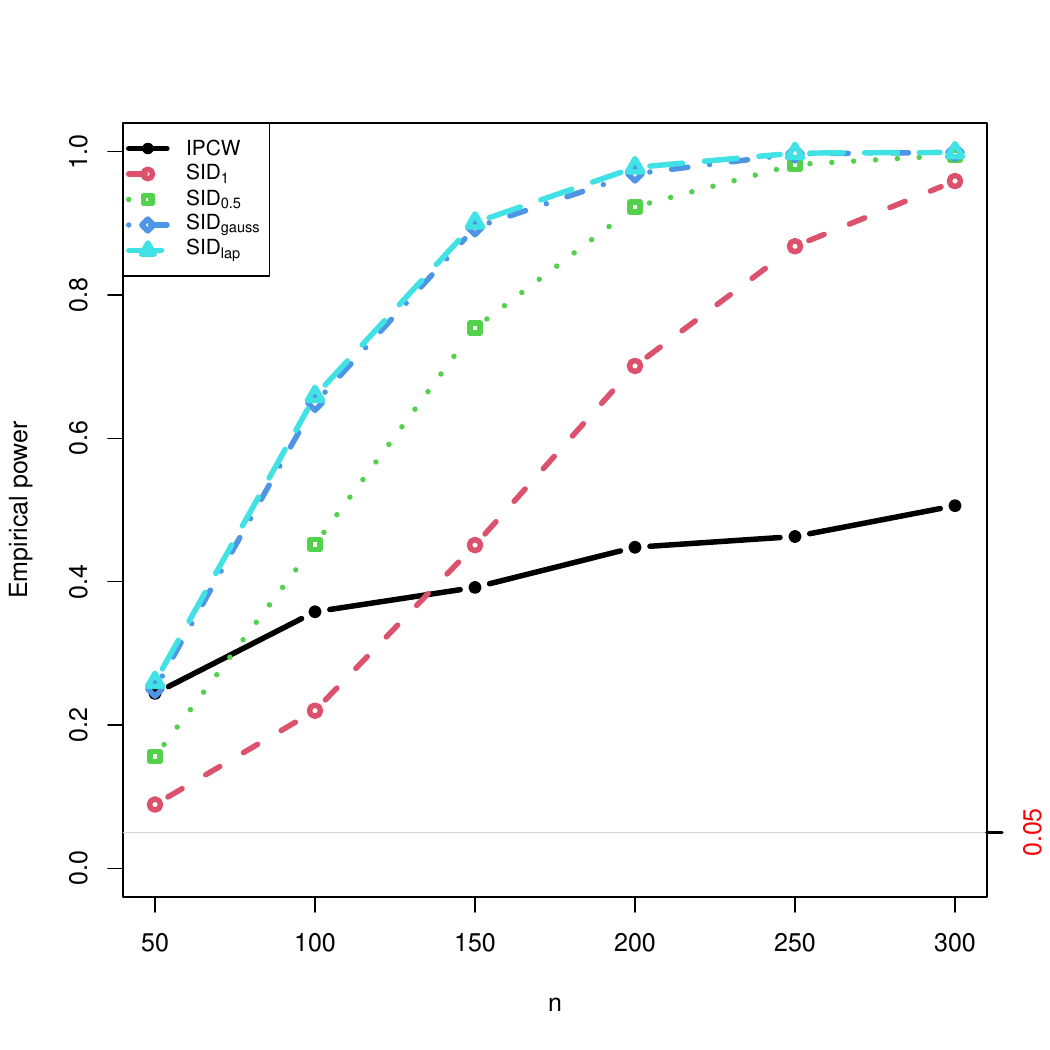}\vspace{-3mm}
{\footnotesize Log-twolines}
\end{minipage}  \hspace{9mm}
\begin{minipage}[b]{0.35\textwidth}
\centering
\includegraphics[width=\textwidth]{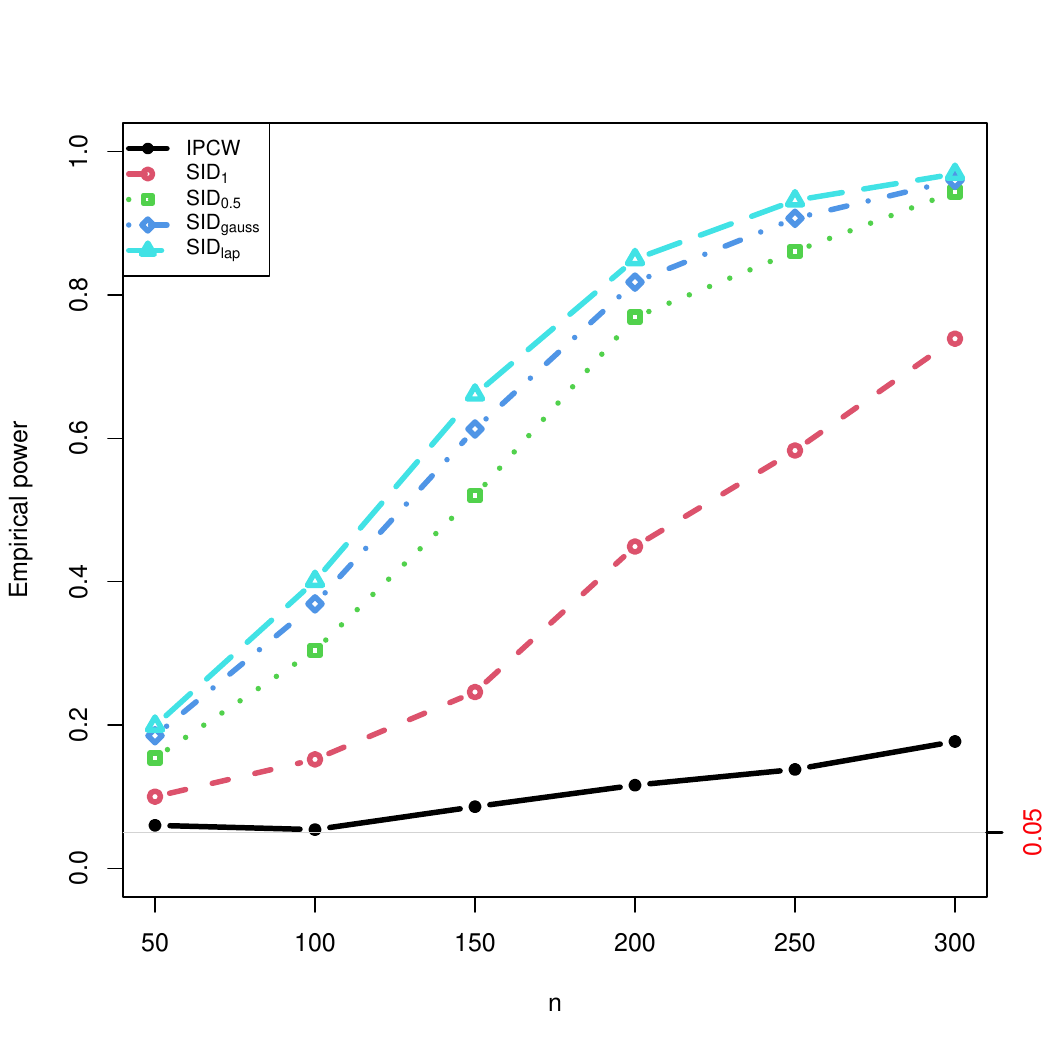}\vspace{-3mm}
{\footnotesize Log-circle }
\end{minipage}
\caption{Comparisons of empirical power at $\alpha=0.05$ with 30\% censoring for Example~\ref{exampl_2}.}
\label{fig:examp2}
\end{figure}

Figure~\ref{fig:examp2} shows plots of the power against sample size at $\alpha=0.05$ with 30\% censoring for Example~\ref{exampl_2}.
All four of our proposed test methods, $\operatorname{SID}_1$, $\operatorname{SID}_{0.5}$, $\operatorname{SID}_{\mathrm{gauss}}$, and $\operatorname{SID}_{\mathrm{lap}}$, perform  better than IPCW for the six dependence relations. The results suggest that our methods are all capable of detecting linear and nonlinear dependencies.

Figure~\ref{fig:examp2} also reveals that $\operatorname{SID}_1$ slightly outperform  $\operatorname{SID}_{0.5}$, $\operatorname{SID}_{\mathrm{gauss}}$, and $\operatorname{SID}_{\mathrm{lap}}$ for the log-linear and log-cosine cases, while it is inferior to these three methods for the other cases.
This is related to the norm $\|\mathbf{x}_1-\mathbf{x}_2\|$ used by $\operatorname{SID}_1$. In fact, $\operatorname{SID}_1$ works in a similar way to the well-known distance covariance \citep{szekely2007measuring}, which also depends on the norm $\|\mathbf{x}_1-\mathbf{x}_2\|$.
We can alleviate the poor performance of $\operatorname{SID}_\beta$ by properly choosing $\beta$.

\begin{example}\label{exampl_3}
This example compares the empirical power of our methods with that of KLR. The following cases are studied\vspace{-0.3cm}
\begin{align*}
\text{Case~1:}&~ T\sim\operatorname{Exp}(e^{\textbf{1}_{p}^{T}\textbf{X} / 10});
 ~~&&\text{Case~2:}~  T \sim \operatorname{Exp}(e^{ (\beta^T\textbf{X})^2/2}); \\
\text{Case~3:}&~\log(T)=0.2\beta^T\textbf{X}+ 2\varepsilon;
~~ &&\text{Case~4:}~ \log(T)=-0.5(\beta^T\textbf{X})^2+4\varepsilon;\\
\text{Case~5:}&~\log(T)=0.25\beta_1^T\textbf{X}+1.5(\beta_2^T\textbf{X})\varepsilon;
 ~~&&\text{Case~6:}~ \log(T) = 2(\beta_1^T\textbf{X})^2+ 0.15(\beta_2^T\textbf{X})\varepsilon.
\end{align*}
Here, $\textbf{X}=({X}_{1},\dots,{X}_{p})^T$ is generated from ${N}_{p}\left(0, \boldsymbol{\Sigma}_{p}\right)$ with $\boldsymbol{\Sigma}_{p}=(0.5^{|j-k|})$.   Consider that $p=6$, $\beta= (1,1,1,-1,-1,-1)^T$,
$\beta_1=(0, 0, 1, -1, 0, 0)^T$, $\beta_2=(1, 1, 0, 0, 0, 0)^T$, $\varepsilon \sim N(0,1)$, and
 $C\sim  \operatorname{Exp}(\lambda)$.
\end{example}

\begin{figure}
 \centering
\begin{minipage}[t]{0.35\textwidth}
\centering
\includegraphics[width=\textwidth]{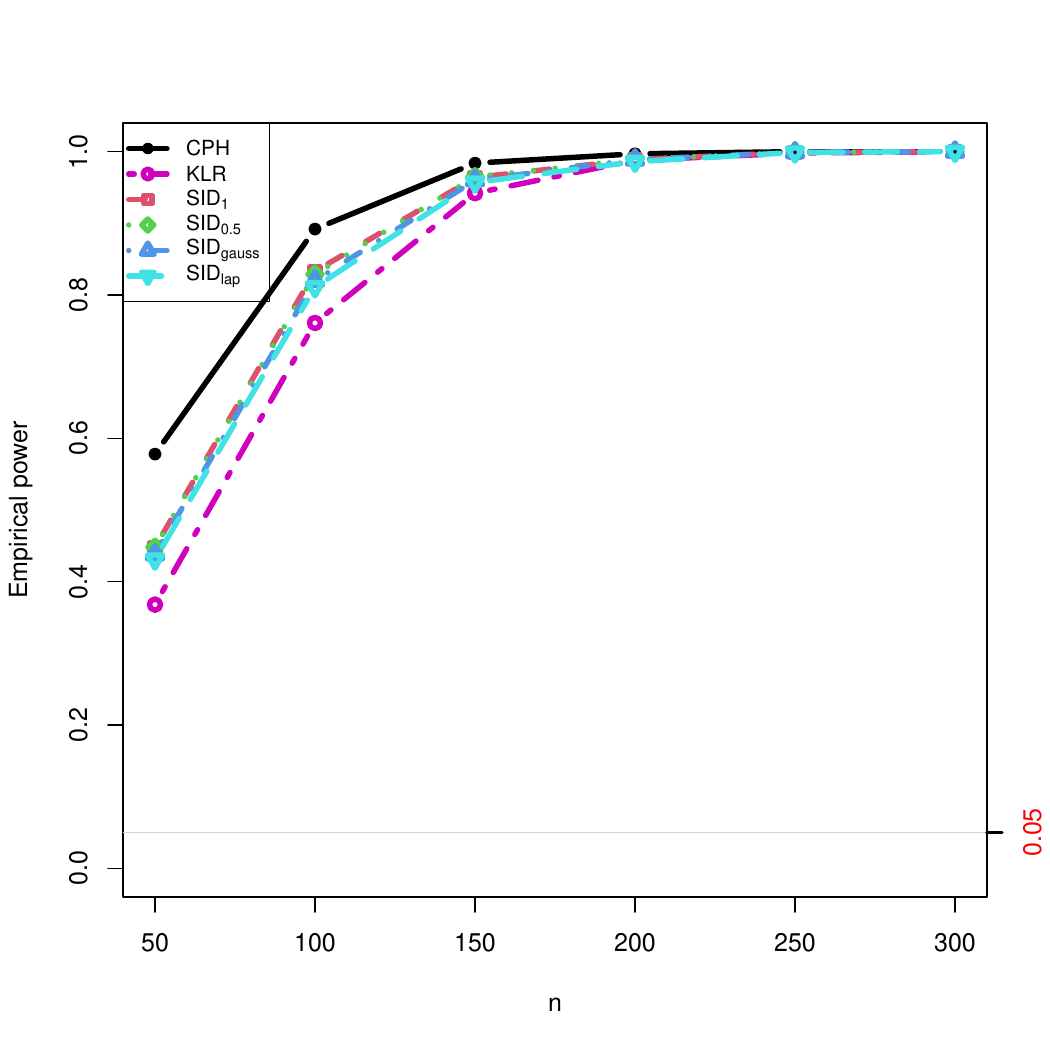}\vspace{-3mm}
{\footnotesize Case 1   }
\end{minipage} \hspace{9mm}
\begin{minipage}[t]{0.35\textwidth}
\centering
\includegraphics[width=\textwidth]{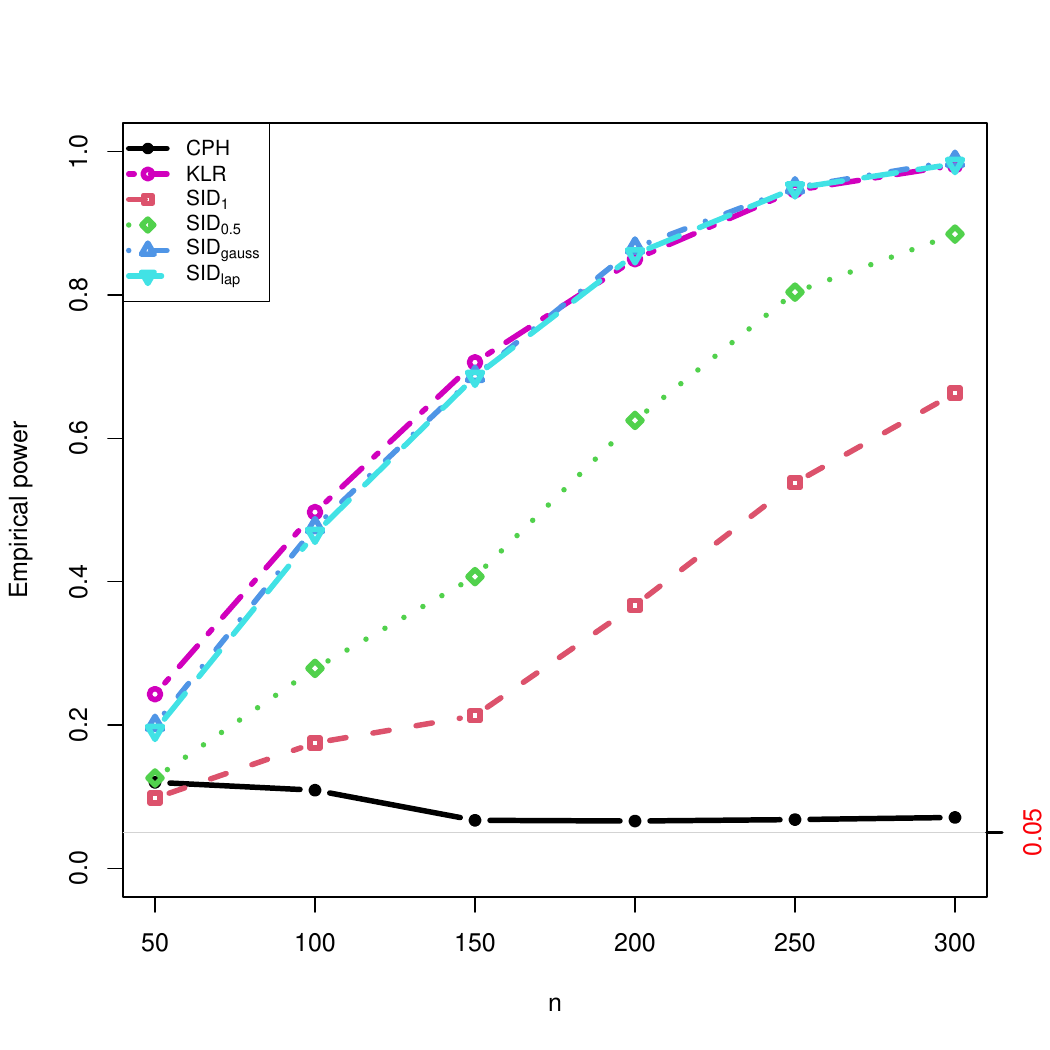}\vspace{-3mm}
{\footnotesize Case 2 }
\end{minipage}
\begin{minipage}[t]{0.35\textwidth}
\centering
\includegraphics[width=\textwidth]{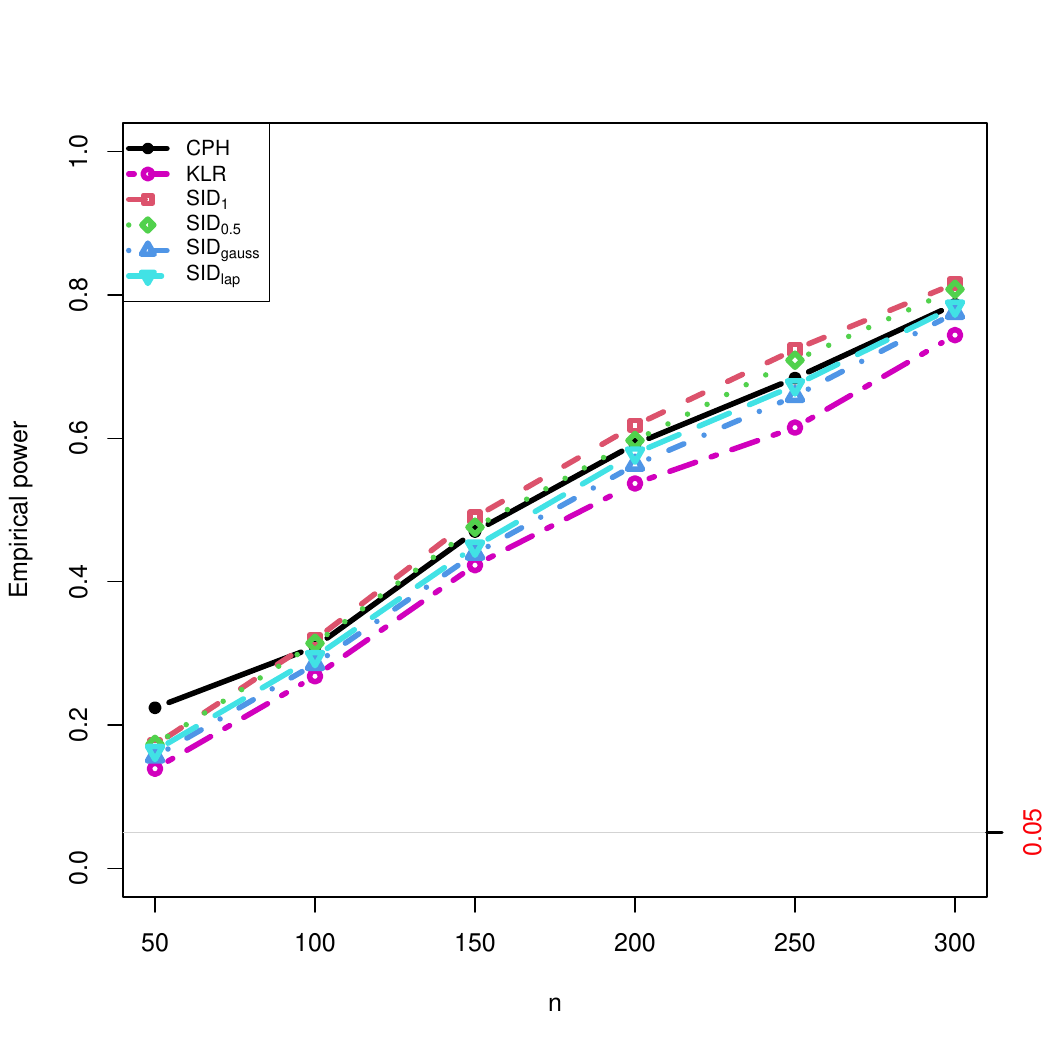}\vspace{-3mm}
{\footnotesize Case 3 }
\end{minipage}\hspace{9mm}
\begin{minipage}[t]{0.35\textwidth}
\centering
\includegraphics[width=\textwidth]{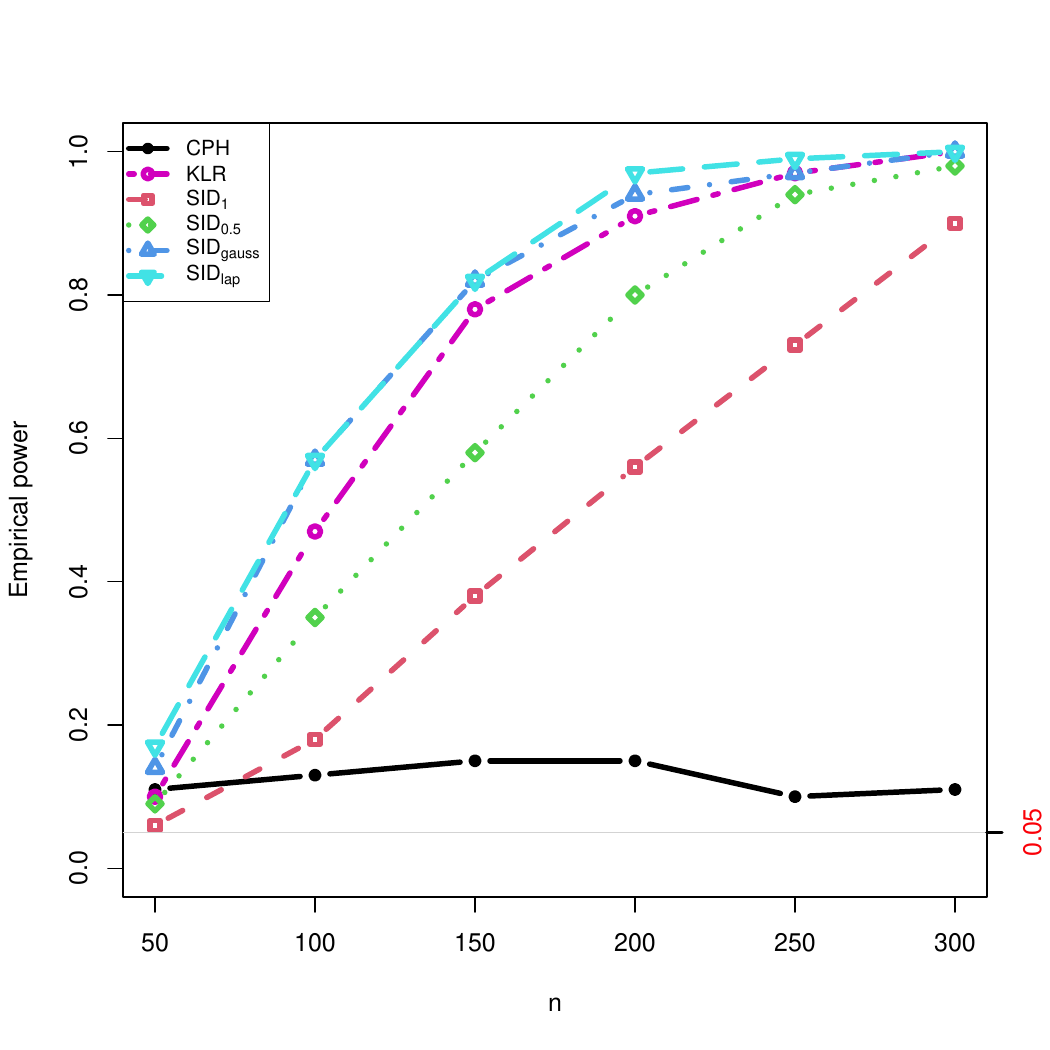}\vspace{-3mm}
{\footnotesize Case 4 }
\end{minipage}
\begin{minipage}[t]{0.35\textwidth}
\centering
\includegraphics[width=\textwidth]{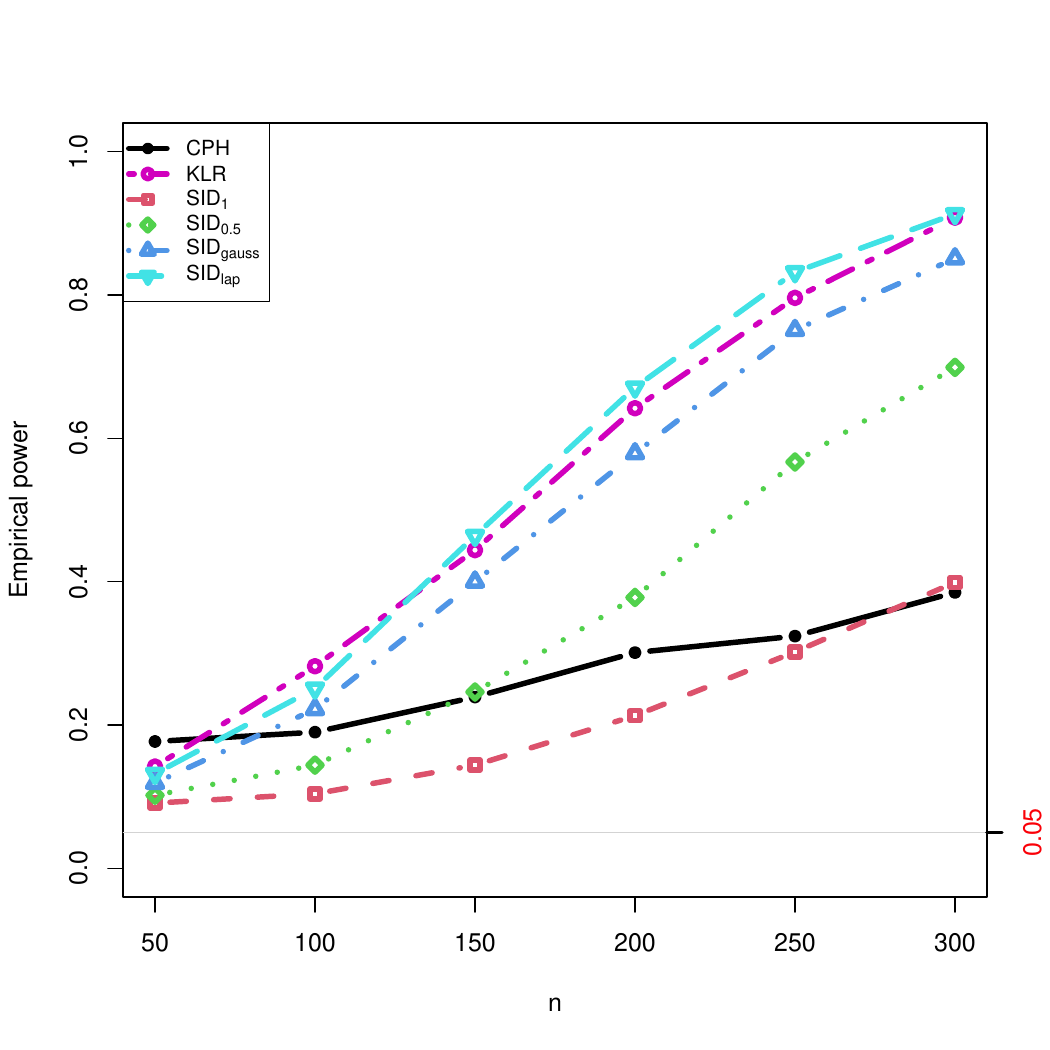}\vspace{-3mm}
{\footnotesize Case 5     }
\end{minipage} \hspace{9mm}
\begin{minipage}[t]{0.35\textwidth}
\centering
\includegraphics[width=\textwidth]{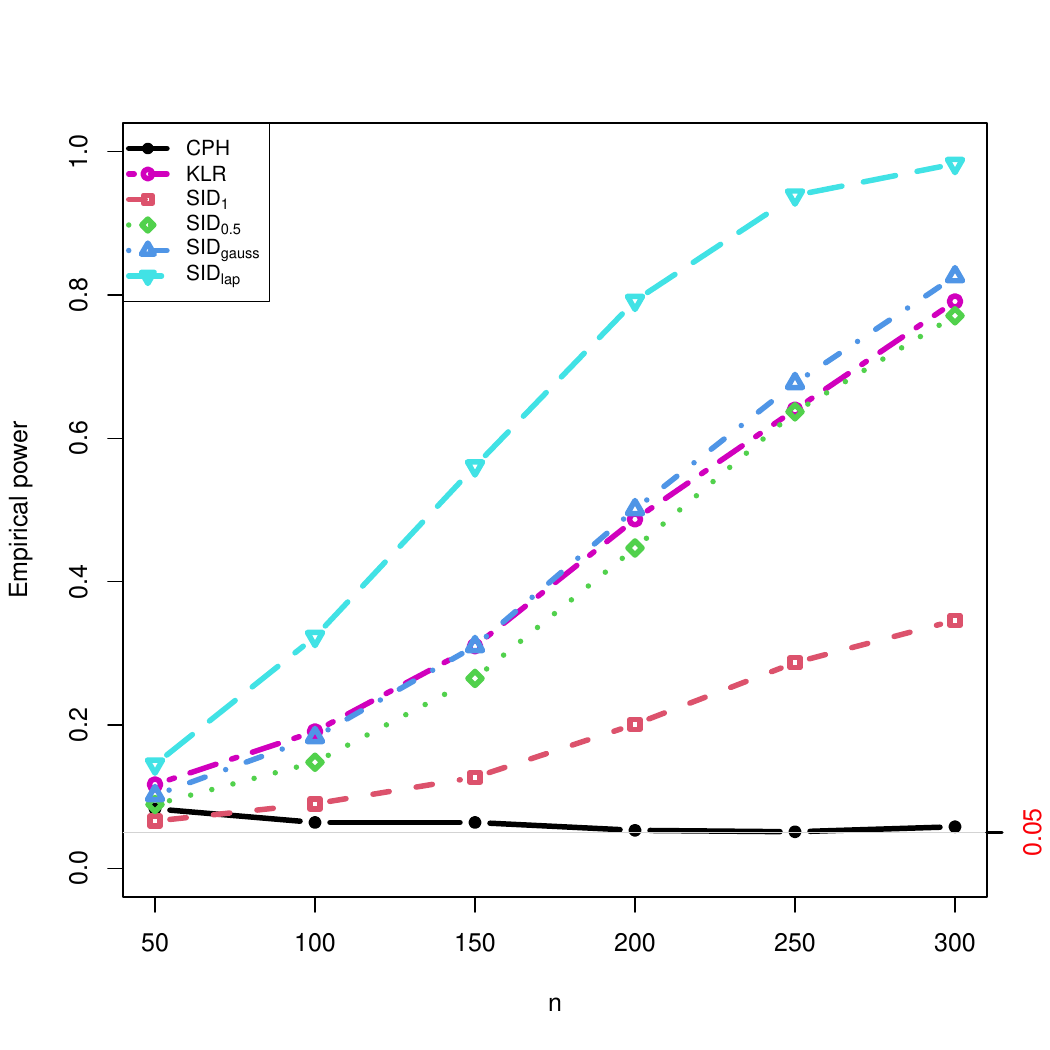}\vspace{-3mm}
{\footnotesize Case 6     }
\end{minipage}
\caption{Comparisons of empirical power at $\alpha=0.05$ with 30\% censoring for Example~\ref{exampl_3}.}
\label{fig:exampl_3}
\end{figure}

The  empirical comparisons of the power for Example~\ref{exampl_3} are summarized in Figure~\ref{fig:exampl_3} at $\alpha=0.05$ with 30\% censoring. Note that Cases~1 and~2   are Cox models, and Cases~3--6 are accelerated failure time models with homogeneous and heterogeneous errors. As expected,  CPH has the highest power in Case~1 but loses power in the other cases as the CPH assumption is invalid. Our methods, especially $\operatorname{SID}_{\mathrm{gauss}}$ and $\operatorname{SID}_{\mathrm{lap}}$, are comparable and even mostly superior to KLR. Additionally, $\operatorname{SID}_1$ has similar performance as in Example~\ref{exampl_2}, as it is powerful in log-linear dependence relations but behaves worse in log-nonlinear or more complex dependence relations.

\begin{example}\label{exampl_5}
In the example, we study the empirical power of our methods when the censoring time $C$ depends on the covariates. The data are generated from the models defined in Cases 1--6 of Example~\ref{exampl_3}, except that $C$ is generated from $\operatorname{Exp}(e^{\lambda+{X}_1})$, where $\lambda$ is used to control the censoring rate.
\end{example}

\begin{figure}
 \centering
\begin{minipage}[t]{0.35\textwidth}
\centering
\includegraphics[width=\textwidth]{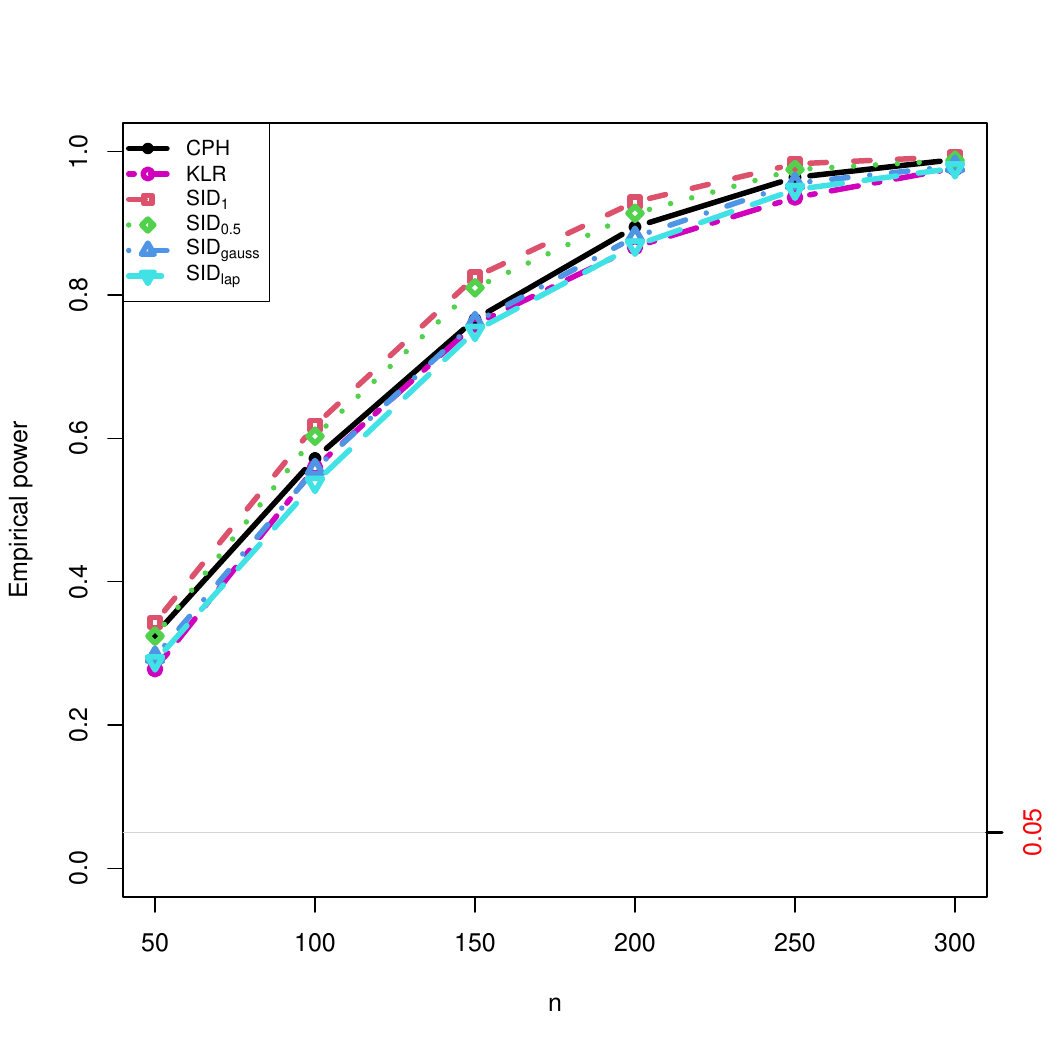}\vspace{-3mm}
{\footnotesize Case 1 }
\end{minipage} \hspace{9mm}
\begin{minipage}[t]{0.35\textwidth}
\centering
\includegraphics[width=\textwidth]{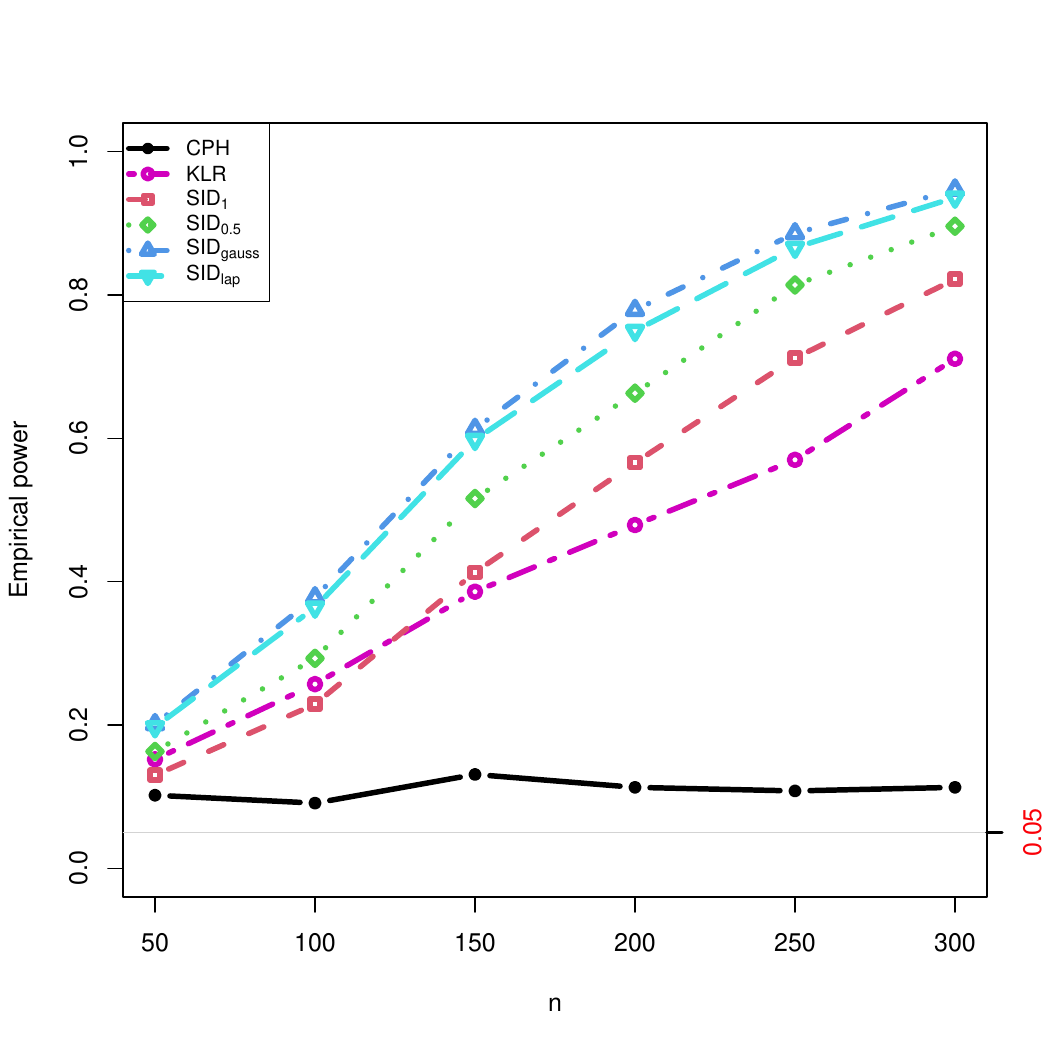}\vspace{-3mm}
{\footnotesize Case 2 }
\end{minipage}
\begin{minipage}[t]{0.35\textwidth}
\centering
\includegraphics[width=\textwidth]{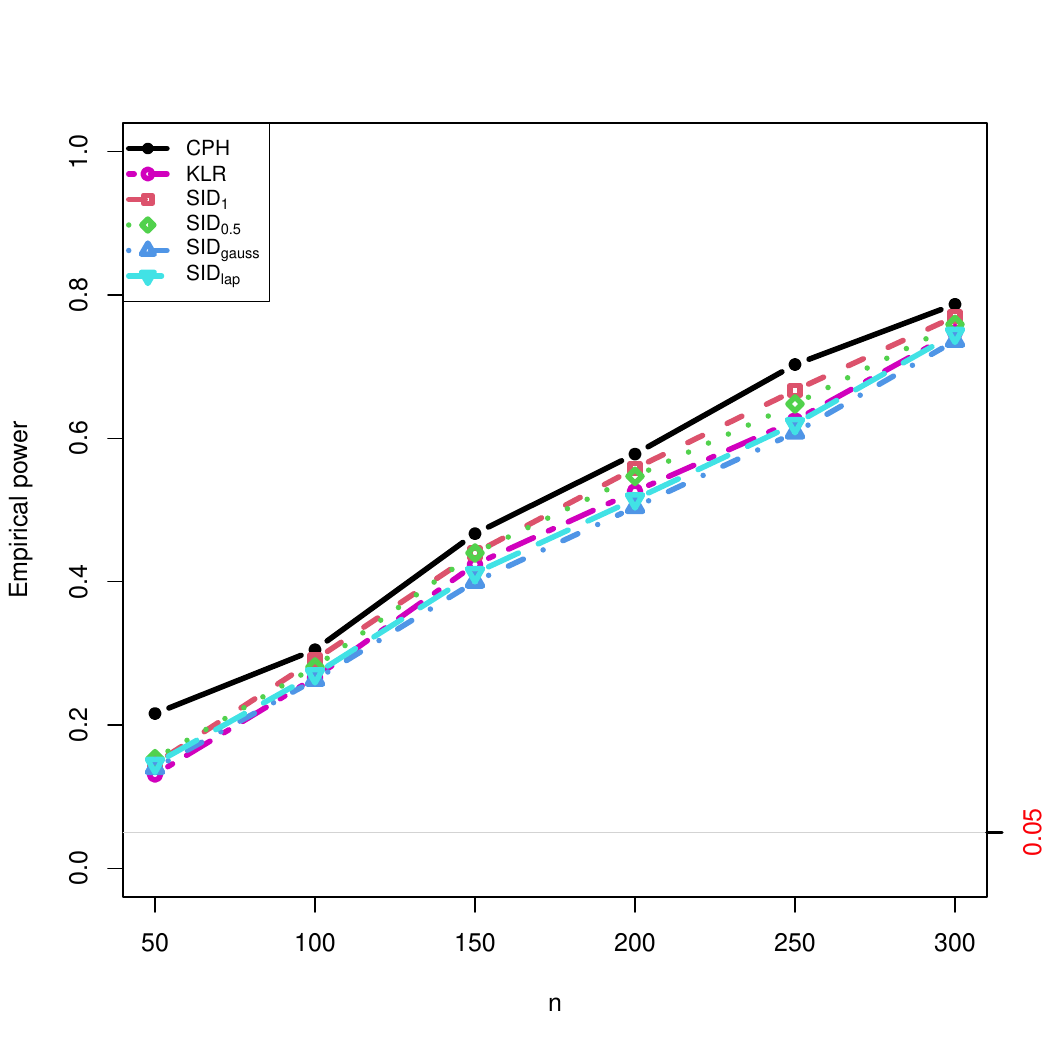}\vspace{-3mm}
{\footnotesize Case 3 }
\end{minipage} \hspace{9mm}
\begin{minipage}[t]{0.35\textwidth}
\centering
\includegraphics[width=\textwidth]{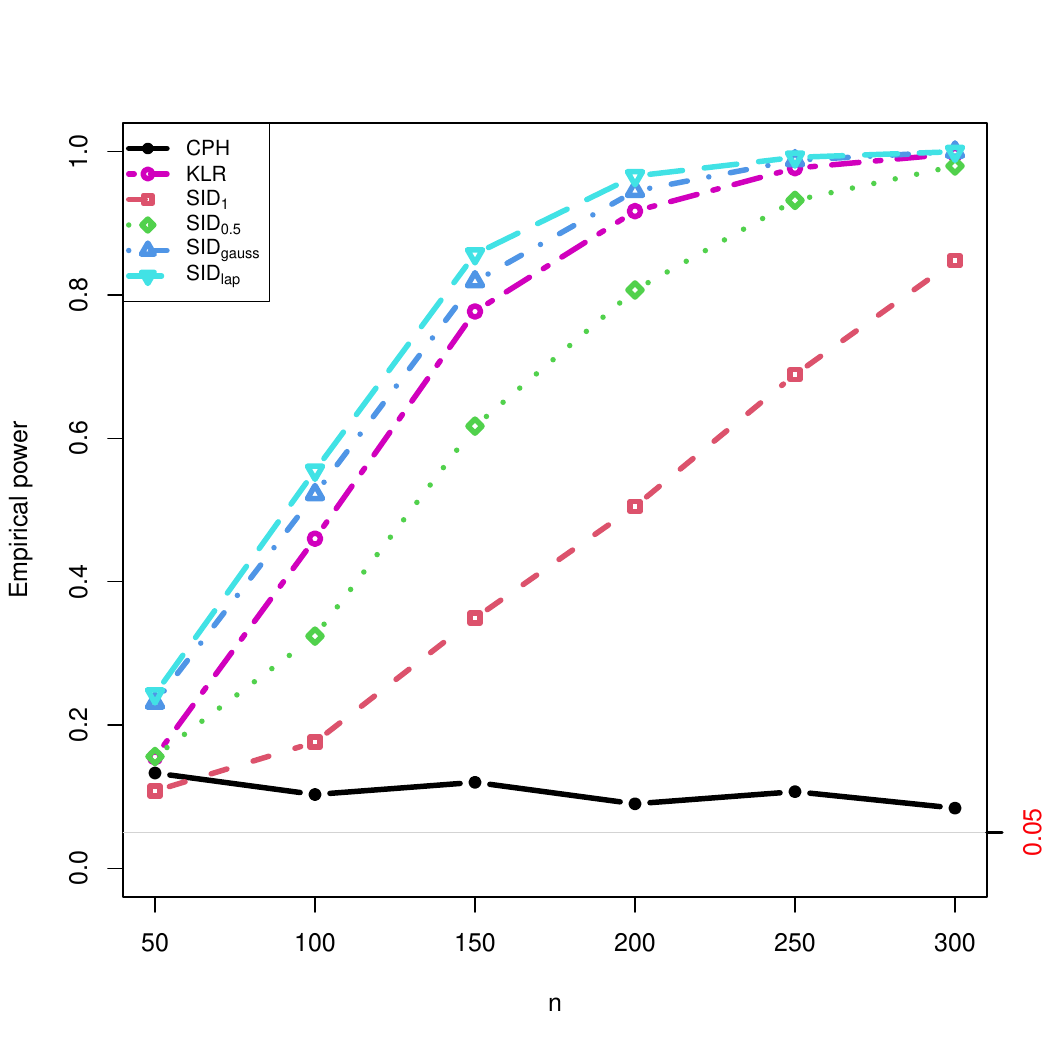}\vspace{-3mm}
{\footnotesize Case 4  }
\end{minipage}
\begin{minipage}[t]{0.35\textwidth}
\centering
\includegraphics[width=\textwidth]{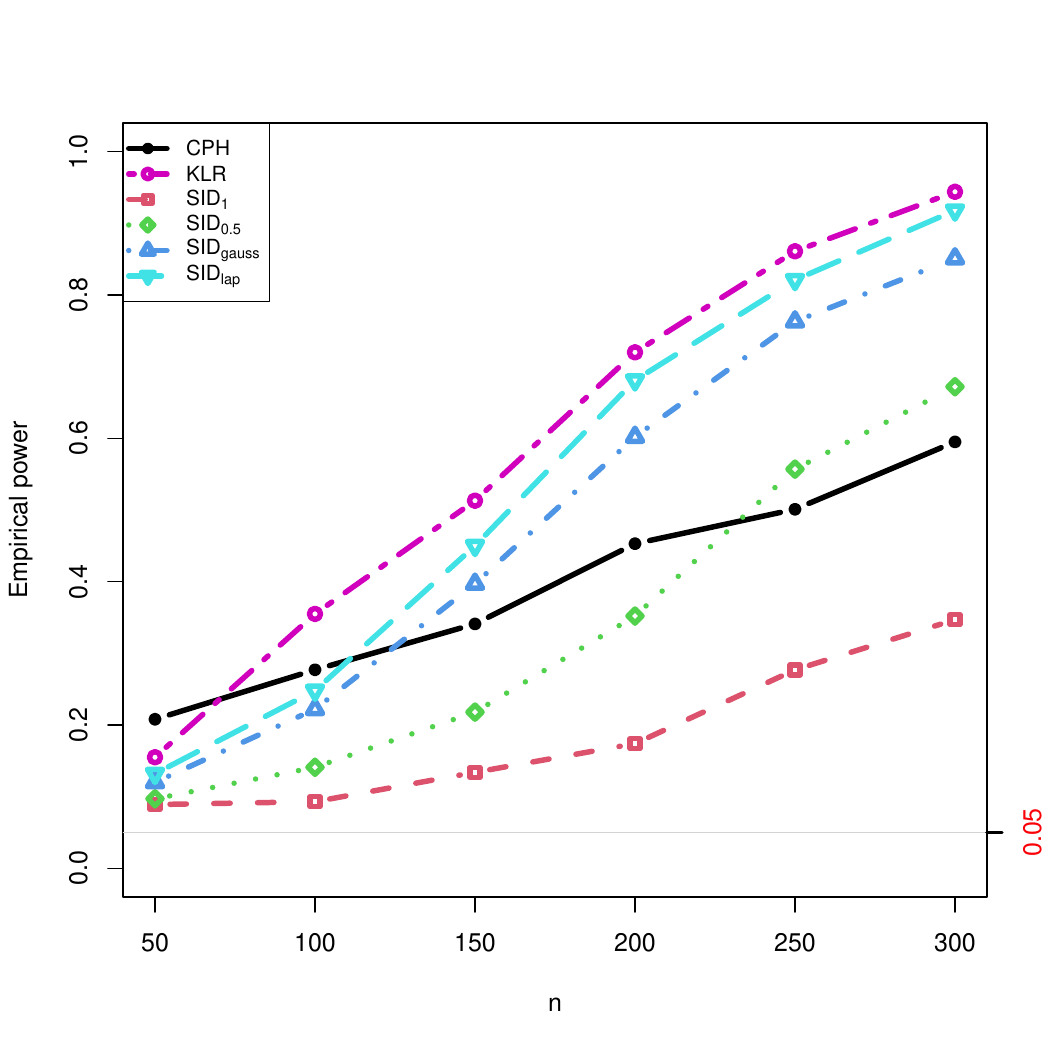}\vspace{-3mm}
{\footnotesize Case 5  }
\end{minipage} \hspace{9mm}
\begin{minipage}[t]{0.35\textwidth}
\centering
\includegraphics[width=\textwidth]{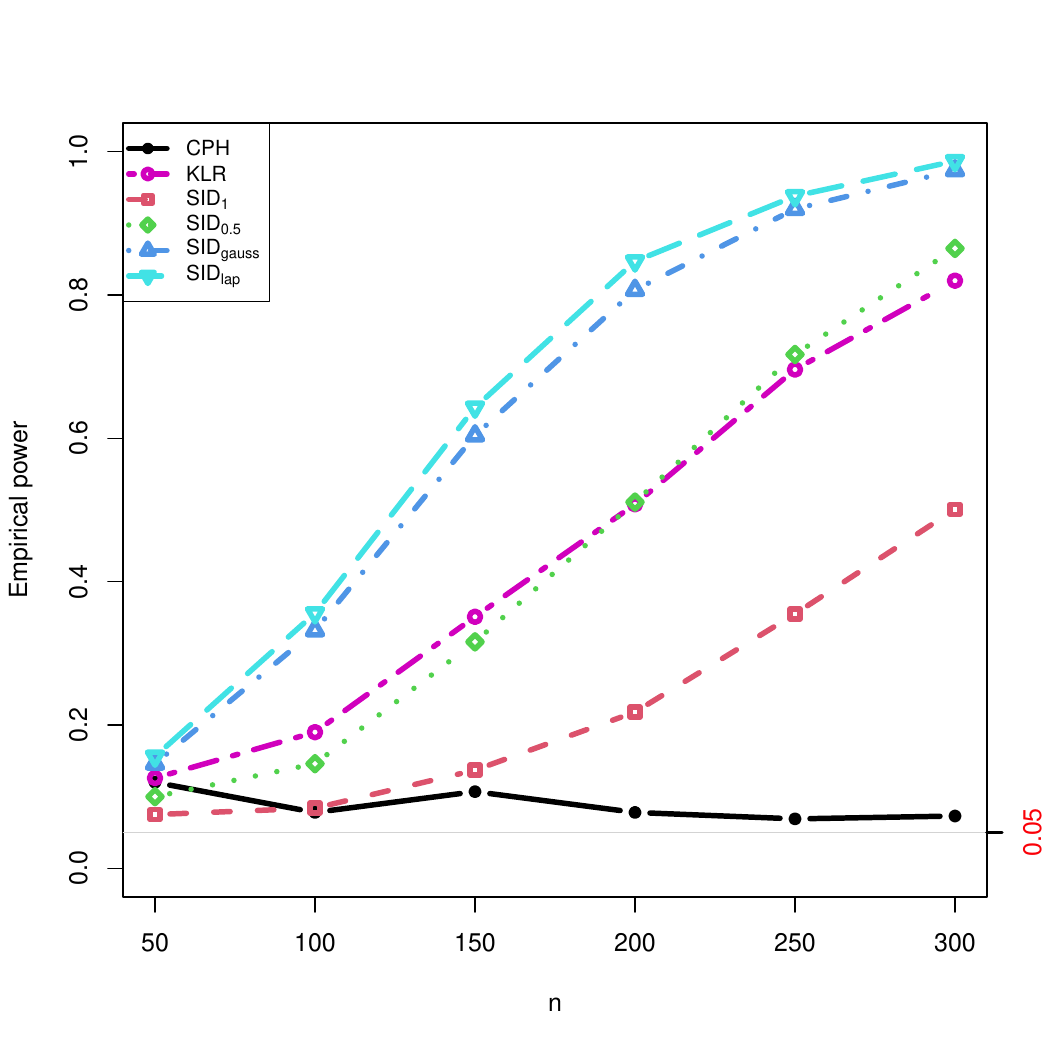}\vspace{-3mm}
{\footnotesize Case 6 }
\end{minipage}
\caption{Comparisons of empirical power at $\alpha=0.05$ with 30\% censoring for Example~\ref{exampl_5}.}
\label{fig:exampl_5}
\end{figure}

As seen from Figure~\ref{fig:exampl_5}, our four methods, as well as KLR, are all capable of detecting the different dependence relations between $T$ and $\mathbf{X}$ when the censoring time $C$ depends on the covariates. Their performances are basically similar to those presented in Figure~\ref{fig:exampl_3}. These results in the two figures indicate that our methods work well, regardless of the association between $C$ and $\mathbf{X}$.

\begin{example}\label{exampl_4}
 In this example, we further compare the performance of our methods with those of KLR in the following mixture cure rate models
\begin{description}
\item [Case 1:] $T=\eta T^*+(1-\eta)\infty,$ with
$T^*\mid X\sim\operatorname{Exp}(e^{0.5X})$,
  $X\sim N(0,1)$;

\item [Case 2:] $T=\eta T^*+(1-\eta)\infty,$ with $T^*\mid X\sim\operatorname{Exp}(e^{0.5X^2})$,
  $X\sim N(0,1)$;

\item [Case 3:] $T=\eta T^*+(1-\eta)\infty,$ with $\log(T^*)=0.5\beta^T\textbf{X}+ 3 \varepsilon$;

\item [Case 4:] $T=\eta T^*+(1-\eta)\infty,$ with $\log(T^*)=0.2(\beta^T\textbf{X})^3+  \varepsilon$.
\end{description}
Here,  $\eta\sim B(1,0.6)$ and
 $C\sim  \operatorname{Exp}(\lambda)$.
In Cases~3-4,  $\beta=(1, 1, 1, -1, 1, -1)^T$  and    $\textbf{X}\sim N_6(0,\Sigma_6)$ with $\boldsymbol{\Sigma}_{6}=(0.5^{|j-k|})$.
\end{example}

\begin{figure}
 \centering
\begin{minipage}[b]{0.35\textwidth}
\centering
\includegraphics[width=\textwidth]{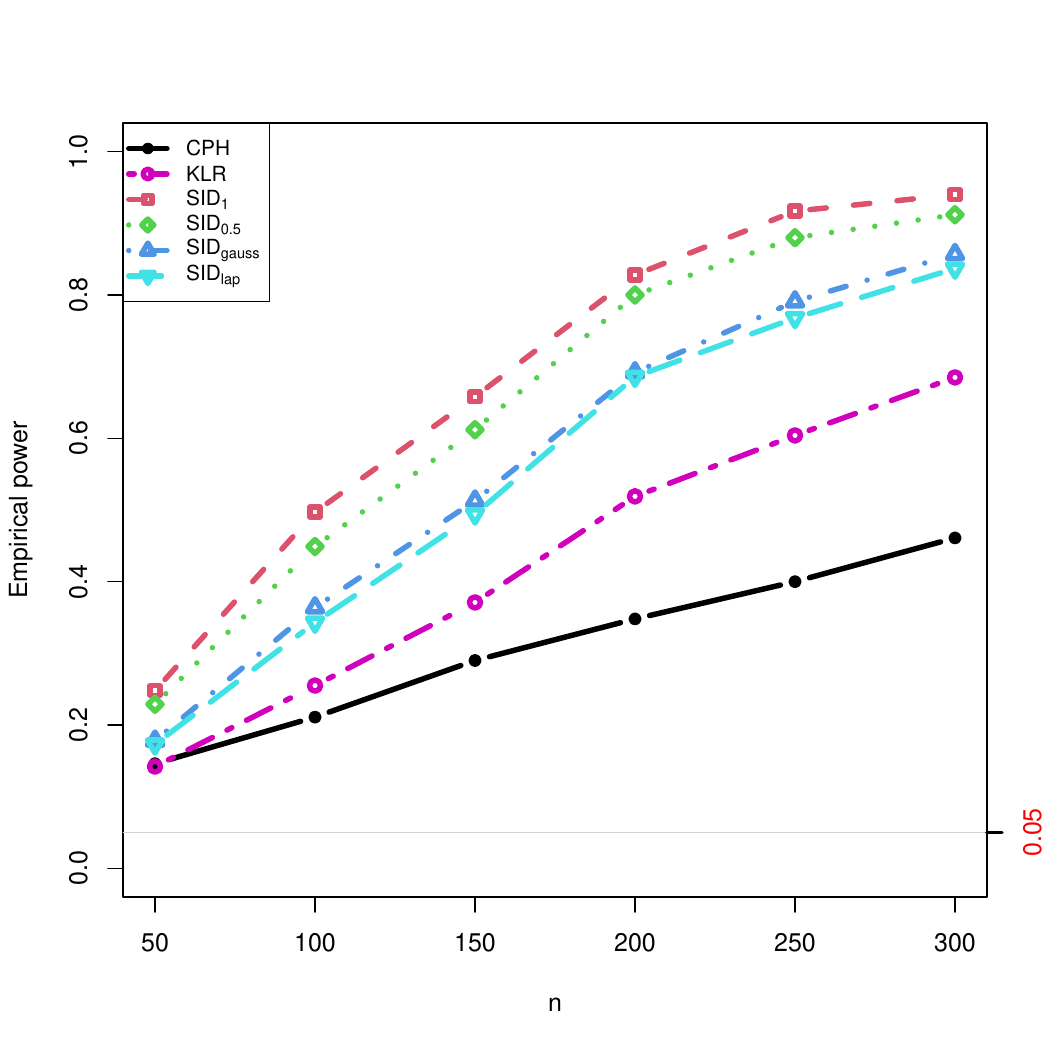}\vspace{-3mm}
{\footnotesize Case 1: $T=\eta T^*+(1-\eta)\infty,$ with
$T^*\mid X\sim\operatorname{Exp}(e^{0.5X})$}
\end{minipage} \hspace{9mm}
\begin{minipage}[b]{0.35\textwidth}
\centering
\includegraphics[width=\textwidth]{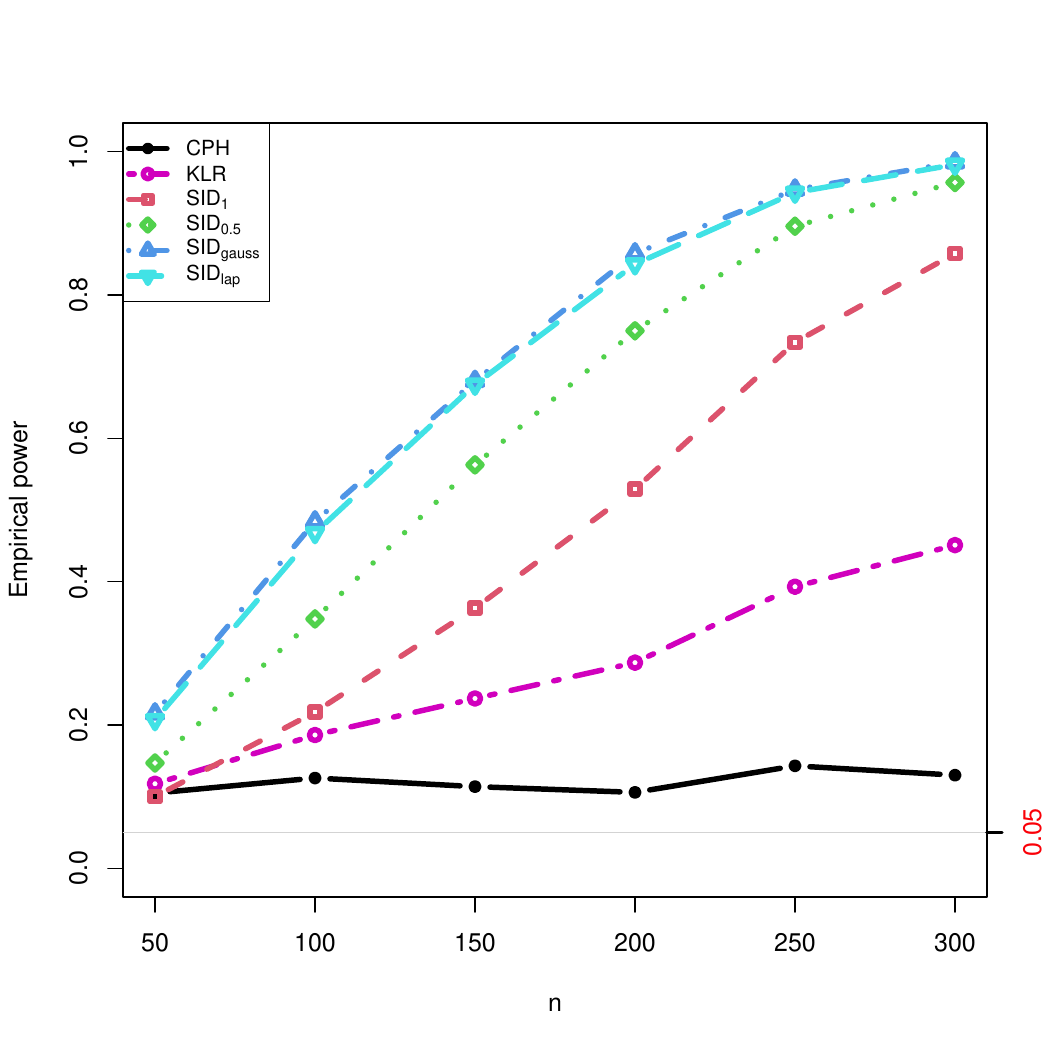}\vspace{-3mm}
{\footnotesize Case 2:  $T=\eta T^*+(1-\eta)\infty,$ with $T^*\mid X\sim\operatorname{Exp}(e^{0.5X^2})$}
\end{minipage}\hspace{9mm}
\begin{minipage}[b]{0.35\textwidth}
\centering
\includegraphics[width=\textwidth]{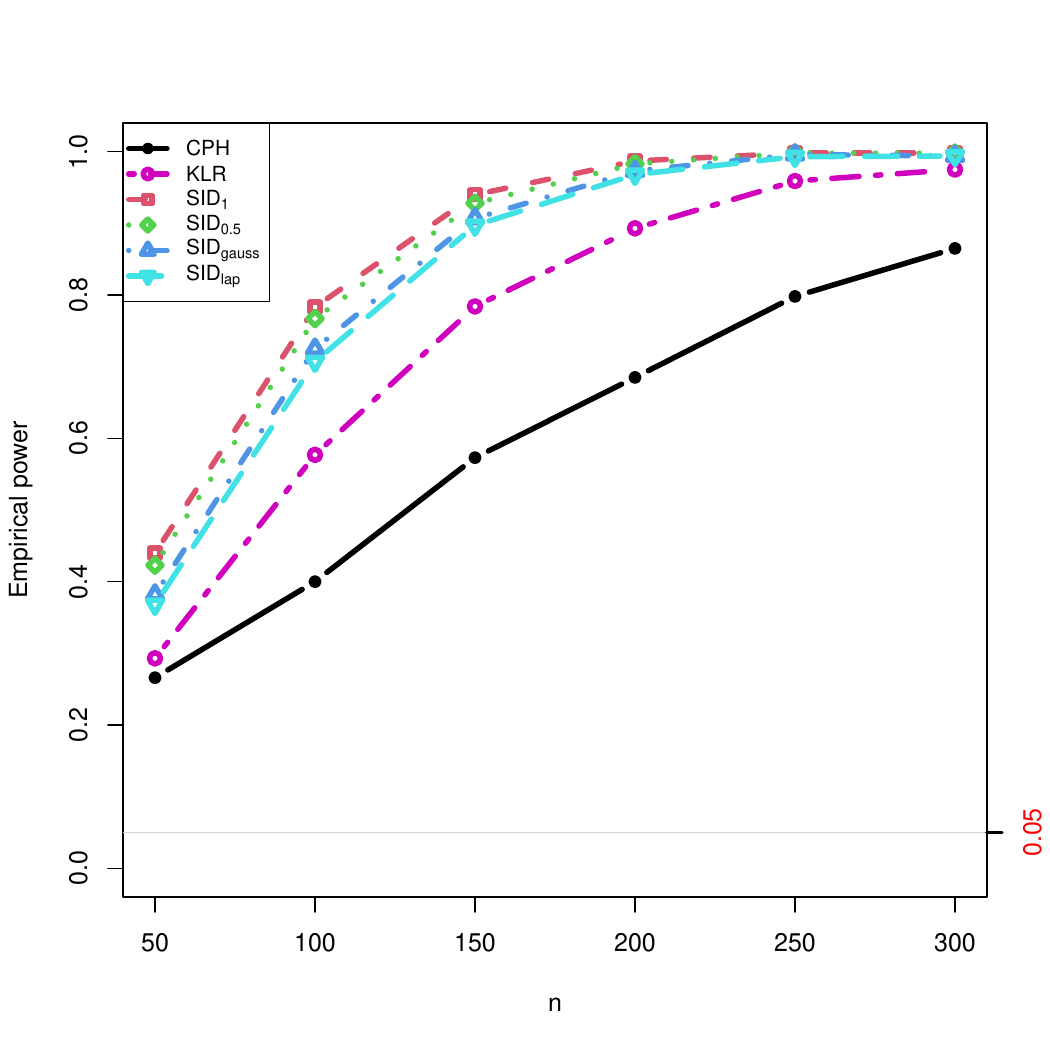}\vspace{-3mm}
{\footnotesize Case 3:  $T=\eta T^*+(1-\eta)\infty,$ with $\log(T^*)=0.5\beta^T\textbf{X}+ 3
\varepsilon$}
\end{minipage}\hspace{9mm}
\begin{minipage}[b]{0.35\textwidth}
\centering
\includegraphics[width=\textwidth]{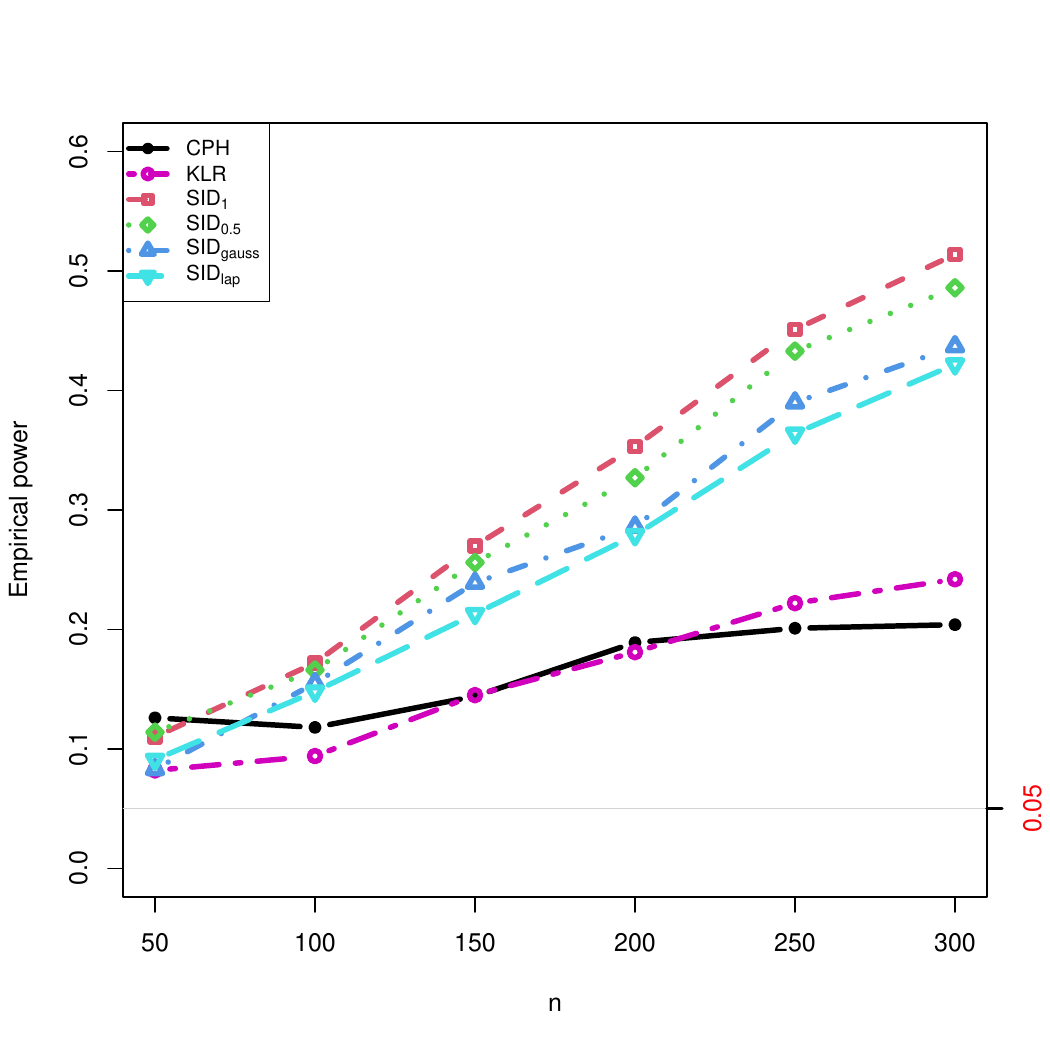}\vspace{-3mm}
{\footnotesize Case 4: $T=\eta T^*+(1-\eta)\infty,$ with $\log(T^*)=0.2(\beta^T\textbf{X})^3+ \varepsilon$}
\end{minipage}
\caption{Empirical power comparisons at $\alpha=0.05$ with 50\% censoring for Example~\ref{exampl_4}.}
\label{fig:exampl_4}
\end{figure}

 Figure \ref{fig:exampl_4} shows that the $\operatorname{SID}_1$, $\operatorname{SID}_{0.5}$, $\operatorname{SID}_{\mathrm{gauss}}$, and $\operatorname{SID}_{\mathrm{lap}}$ tests outperform  KLR in the four mixture cure rate models of Example \ref{exampl_4}.
Compared with the results for Examples \ref{exampl_3}-\ref{exampl_5}, KLR is significantly inferior to our methods. This is, presumably, because  the SIDs are local and  need not to calculate   pairwise distances  between observations, which have high volatility in the settings of Example \ref{exampl_4}.

\begin{example}\label{exampl_6}
The example investigates the empirical power of our methods for small deviations from the null hypothesis. We generate  data for two cases
\begin{align*}
\text{Case~1:}&~ \log(T)= \theta {X} + \varepsilon~   \text{ and } ~   C\sim \operatorname{Exp}(3);
 ~~&&\text{Case~2:}~\log(T)= \theta {X}^2 + \varepsilon ~    \text{ and } ~   C\sim \operatorname{Exp}(3).
\end{align*}
Here, $X$ and $\varepsilon$ are independently generated from $N(0,1)$.
\end{example}

\begin{figure}
 \centering
\begin{minipage}[b]{0.35\textwidth}
\centering
\includegraphics[width=\textwidth]{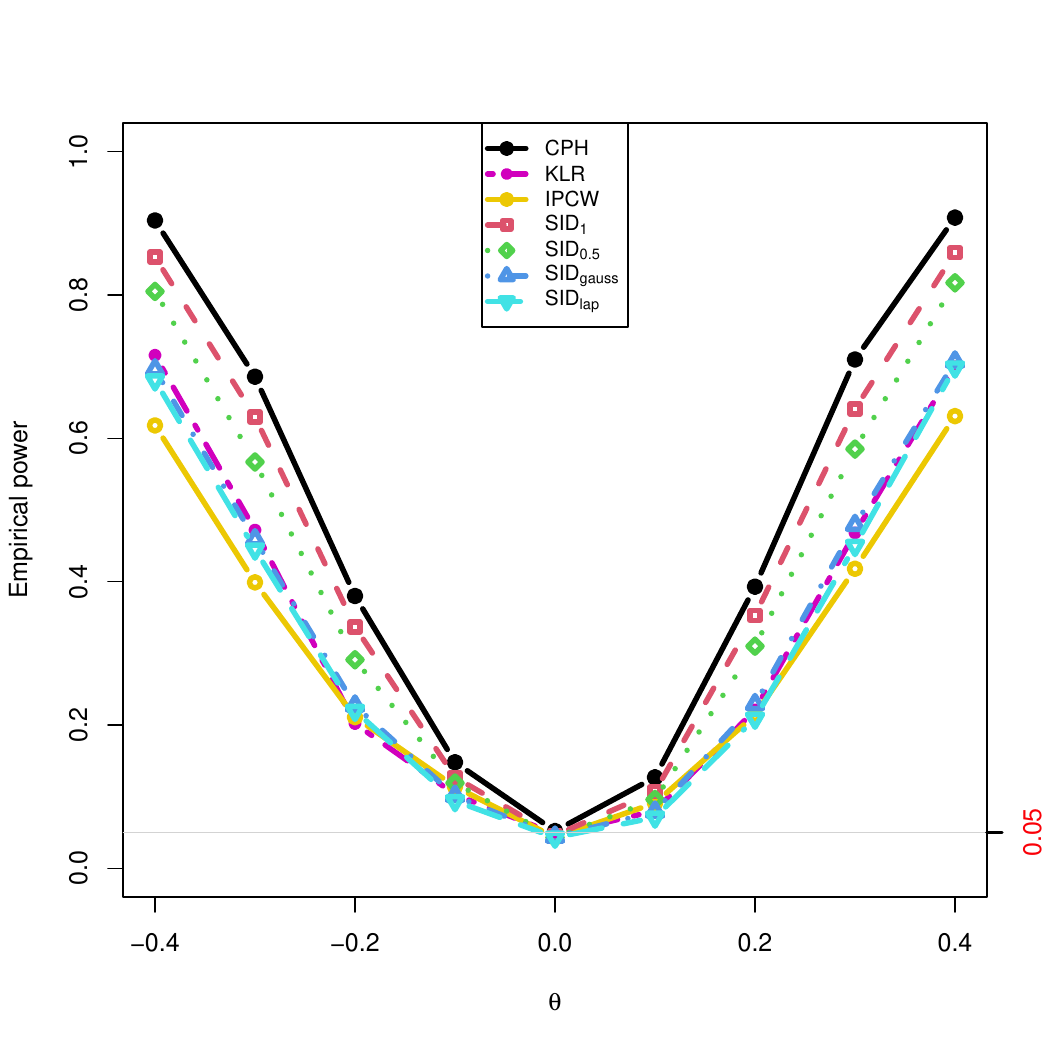}\vspace{-3mm}
{\footnotesize Case 1: \\  $\log(T)= \theta {X} + \varepsilon$ }
\end{minipage}     \hspace{9mm}
\begin{minipage}[b]{0.35\textwidth}
\centering
\includegraphics[width=\textwidth]{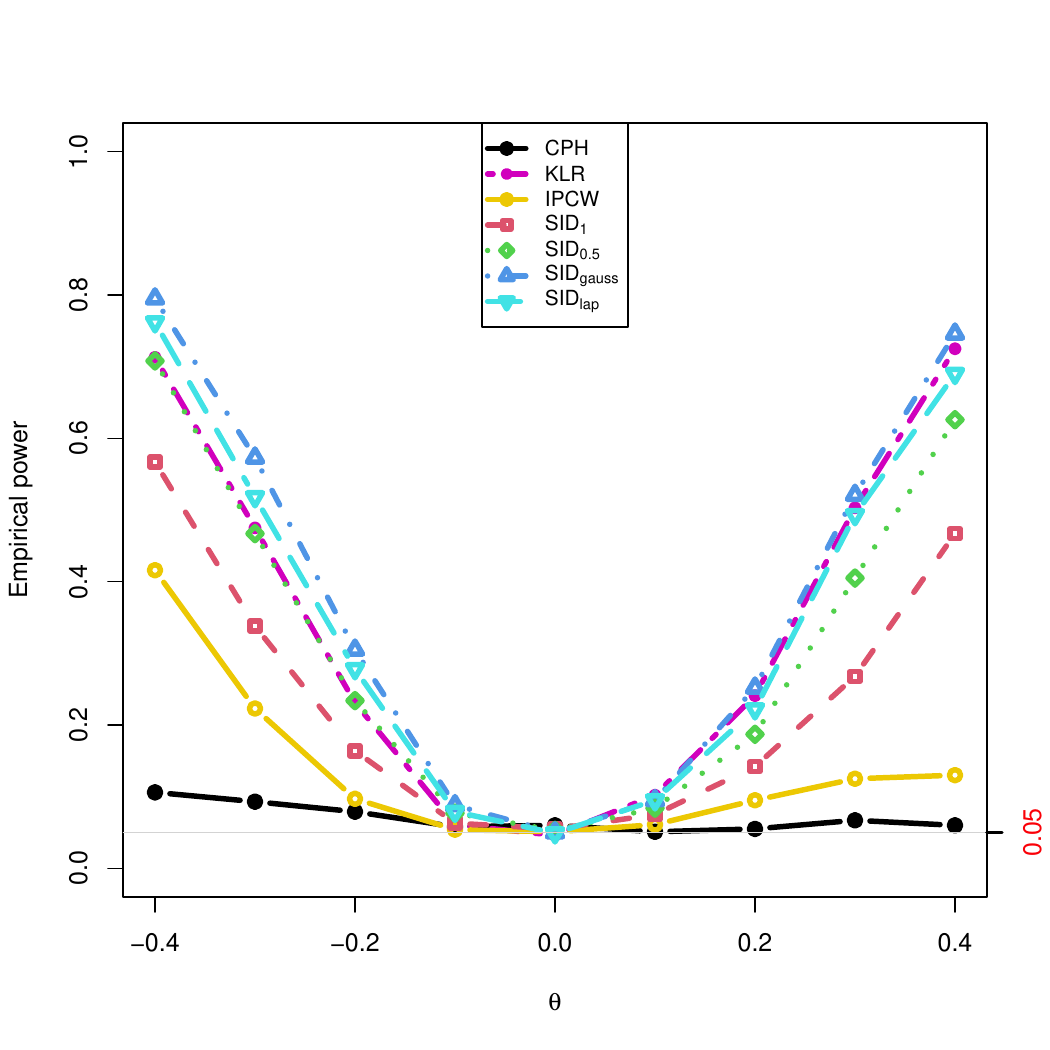}\vspace{-3mm}
{\footnotesize Case 2: \\  $\log(T)= \theta {X}^2 + \varepsilon$ }
\end{minipage}
 \caption{Comparisons of empirical power at $\alpha=0.05$ and $n=100$ for Example~\ref{exampl_6}.}
\label{fig:exampl_6}
\end{figure}

Figure~\ref{fig:exampl_6} displays the power with respect to $\theta$ at the significance level $\alpha=0.05$ and $n=100$. The power of our methods increases rapidly as $\theta$ moves away from 0, so that they are highly competitive with the KLR test. The results indicate that our methods can effectively detect the difference between the null and alternative hypotheses.

Additional simulations, which include various types of covariates such as discrete, heavy-tailed, and high-dimensional covariates, are detailed in Appendix C of the Supplementary Material; see Examples C.1--C.3. These simulation results further illustrate the effectiveness of the proposed SID method in managing diverse and high-dimensional covariates. Additionally, we also conduct a study to assess the impact of the parameter \(\beta\) on \(\operatorname{SID}_\beta\), as demonstrated in Example C.4.

From  all the   numerical results, we can draw the following conclusions
(1) Our proposed SID-based metrics are capable of detecting and testing various dependence relations between the censored outcome and covariates;
(2) $\operatorname{SID}_1$ has high power with linear dependence relations but behaves worse with nonlinear dependence relations. With a better choice of $\beta$, $\operatorname{SID}_\beta$ could be as powerful as $\operatorname{SID}_K$;
(3) In most settings, our test methods are highly comparable to the KLR test and more powerful than IPCW. Overall, our methods do not require strong assumptions on the censoring mechanism.

\section{Real data analysis}\label{chapter-Realdata}

In this section, we illustrate the SID-based tests with empirical analyses of two real datasets. We use 2000 wild bootstrap samples for the KLR and SID-based tests.

\begin{example}[BMT data]\label{example_bmt} This example considers the  bone marrow transplant (BMT) data in \citet{klein2003survival}, available as the \texttt{bmt} dataset of the R package \texttt{KMsurv}. The dataset is for the recovery process of 137 patients after BMT. At the time of transplantation, several risk factors were measured. Risk factors, denoted by $z_1$ to $z_{10}$, include the  age, gender,   cytomegalovirus immune status for both recipient and donor,   the waiting time from diagnosis to transplantation, the  French-American-British disease grade, hospital stay, and methotrexate dose.
In our analysis, we consider the time of death, denoted by $t_1$, as the event time.
 About 59.1\% of the failure times are censored in the study.
\end{example}

In the analysis, we are interested in detecting the dependence between $t_1$ and two of the covariates, namely recipient age in years ($z_1$) and the waiting time to transplantation in days  ($z_{7}$). Table~\ref{table-BMT} lists the $p$-values for the seven test methods for the BMT data.
Our four proposed methods, except $\operatorname{SID}_1$, have smaller $p$-values than the KLR test. In particular, $\operatorname{SID}_{\mathrm{gauss}}$ has the smallest $p$-values (0.037, 0.018, and 0.035) among the seven methods, which implies that $t_1$ and $z_1/z_{7}$ or both are significantly dependent.

\begin{table}
\caption{  $p$-values for the various tests for the BMT data.}
\label{table-BMT}\scriptsize
\resizebox{\linewidth}{!}{%
\begin{tabular}{cccccccccc}
\hline
  &  &   &  & &\multicolumn{2}{c}{$\operatorname{SID}_\beta$ } & &\multicolumn{2}{c}{$\operatorname{SID}_K$}    \\
 \cline{6-7}                   \cline{9-10}
  Null hypothesis & CPH  &KLR  &IPCW  & &$\operatorname{SID}_1$ &$\operatorname{SID}_{0.5}$   &   & $\operatorname{SID}_{\mathrm{gauss}}$ &$\operatorname{SID}_{\mathrm{lap}}$   \\ \hline
  $H_0:t_1\indep z_{1}$          &0.304&0.101 &0.090 &&0.061  &0.068 &&0.037 &0.059           \\
  $H_0:t_1\indep z_{7}$          &0.888&0.074 &0.231 &&0.180  &0.043 &&0.018 &0.018       \\
  $H_0:t_1\indep (z_{1}, z_{7})$ &0.589&0.115 &0.238 &&0.195  &0.075 &&0.035 &0.036               \\
\hline
\end{tabular}}
\end{table}

\begin{table}
\caption{  $p$-values of the various tests for  Colon data.}
\label{table-Colon}\scriptsize
\resizebox{\linewidth}{!}{%
\begin{tabular}{cccccccccc}
\hline
  &  &   &  & &\multicolumn{2}{c}{$\mathrm{SID}_\beta$ } & &\multicolumn{2}{c}{$\mathrm{SID}_K$}    \\
 \cline{6-7}                   \cline{9-10}
  Null hypothesis   &CPH  &KLR  &IPCW  & &$\mathrm{SID}_1$ &$\mathrm{SID}_{0.5}$   &   & $\mathrm{SID}_{\mathrm{gaus}}$ &$\mathrm{SID}_{\mathrm{lap}}$   \\\hline
  $H_0:T\indep \text{Age}$                  &0.626 &0.108 &0.552 &&0.027  &0.042 &&0.047 &0.070  \\
  $H_0:T\indep (\text{Age,  Perfor, Adhere})$ &0.102 &0.016 &0.543 &&0.015  &0.020 &&0.025 &0.012\\
  \hline
  $H_0:C\indep \text{Age}$                    &0.222 &0.566 &/  &&0.799 &0.758 &&0.832 &0.842    \\
  $H_0:C\indep (\text{Age,  Perfor, Adhere})$ &0.555 &0.652 &/  &&0.793 &0.745 &&0.752 &0.952    \\
  \hline
\end{tabular}}
\end{table}

\begin{figure}
 \centering
\begin{minipage}[b]{0.30\textwidth}
\centering
\includegraphics[width=\textwidth]{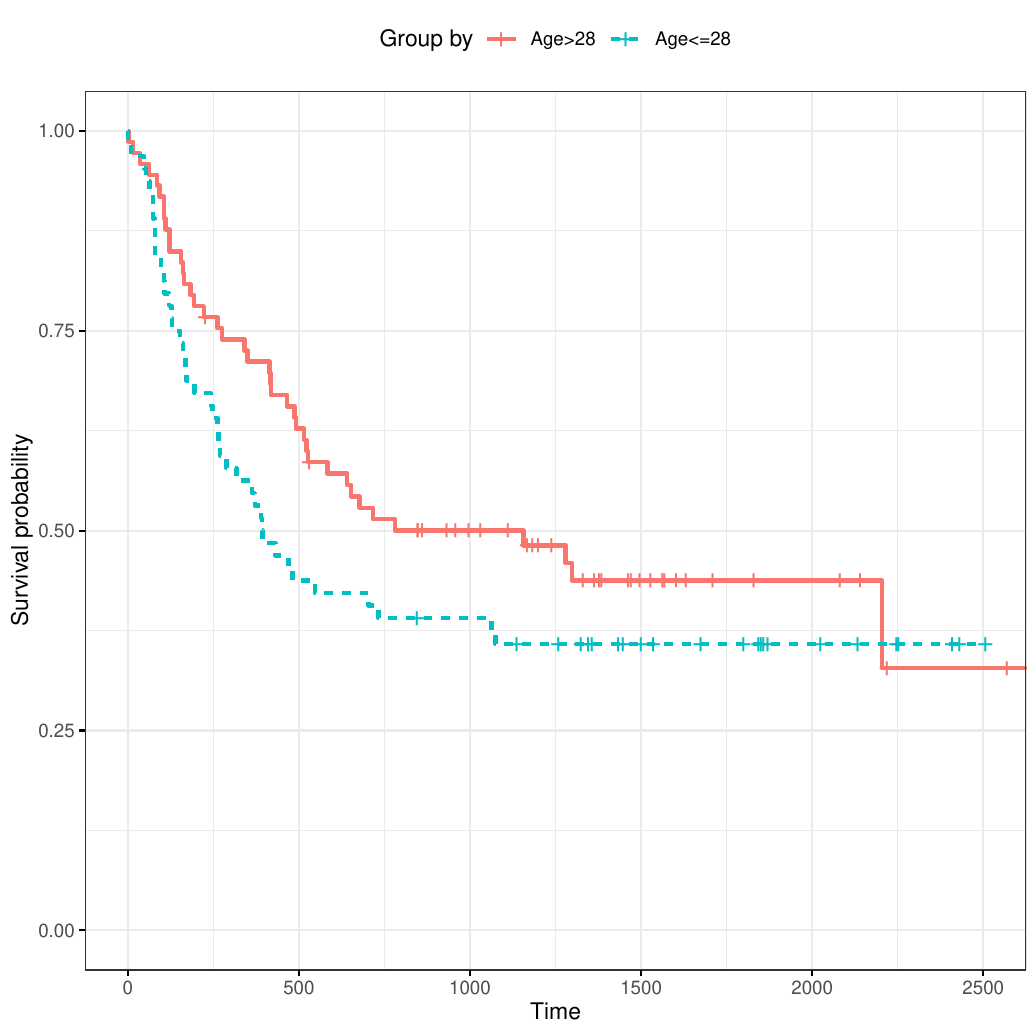}\vspace{-2mm}
 \end{minipage}    \hspace{9mm}
\begin{minipage}[b]{0.30\textwidth}
\centering
\includegraphics[width=\textwidth]{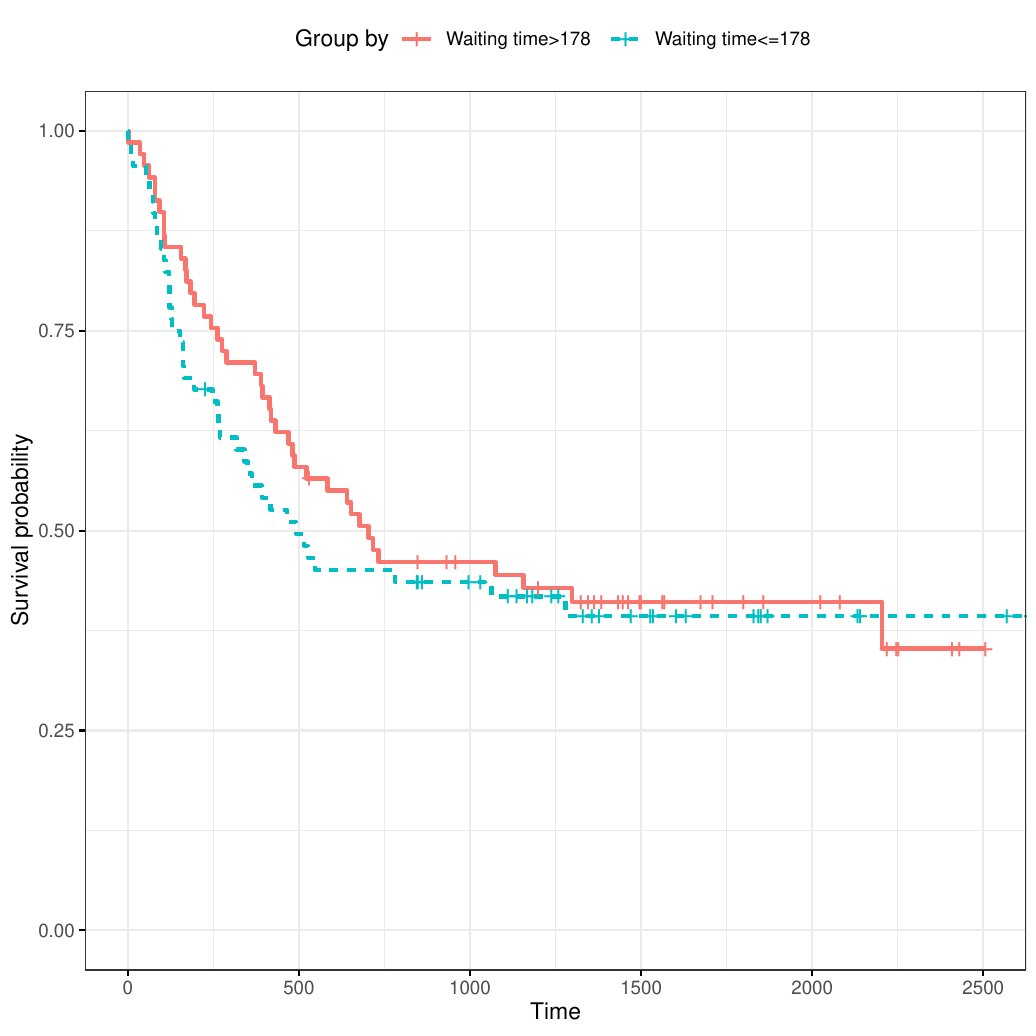}\vspace{-2mm}
 \end{minipage}
\caption{Kaplan--Meier estimates of the survival times grouped by the medians of the recipient age  (left) and the waiting time to transplantation   (right) for the BMT data. 
}
\label{fig:BTM}
\end{figure}

To illustrate the alleged dependence relations, we classify the dataset into four subgroups by the medians of $z_1$ and $z_{7}$. Figure~\ref{fig:BTM} displays the Kaplan--Meier estimates of the survival times for each subgroup, where $Q_{0.5}(z_1)=28$ (years) and $Q_{0.5}(z_7)=178$ (days). The survival curves can be divided by the medians of $z_1$ and $z_{7}$,  implying  that $t_1$ depends on these two covariates. These results are consistent with the findings for the SID tests in Table~\ref{table-BMT}. 

\begin{example}[Colon data]\label{example_Colon}
This example considers a colon cancer dataset originally described in
\citet{laurie1989surgical}, available as the \texttt{colonCS} dataset of the R package \texttt{condSURV}. The dataset consists of information about the recurrence of tumors and survival in 929 patients undergoing treatment for stage B/C colon cancer.
Each patient record contains the time to cancer recurrence, the survival time, and 11 covariates. In the example, the survival time is the event of interest.
\end{example}

The dataset has been analyzed by \citet{Tamara}.
Like \citet{Tamara}, we aim to test independence between the survival time $T$ and the three covariates Age, Perfor, and Adhere.
 The  $p$-values from seven test methods in our simulation studies are summarized in Table~\ref{table-Colon}.
 These results indicate that  under  the hypothesis \(H_0: T \perp\!\!\!\perp \text{Age}\), the \(\mathrm{SID}_1\), \(\mathrm{SID}_{0.5}\), and \(\mathrm{SID}_{\mathrm{gaus}}\) tests show   $p$-values below the 0.05 significance level, providing evidence that    \( T \)  depends on  $\text{Age}$. In contrast, the KLR, CPH, and IPCW tests fail to identify a significant association  at this level. For the hypothesis   \(H_0: T \perp\!\!\!\perp (\text{Age, Perfor, Adhere})\), all four SID methods, along with the KLR test, demonstrate   $p$-values below 0.05. This suggests a significant dependence between   \( T \) and  $(\text{Age, Perfor, Adhere})$.

It is a little surprising that IPCW has $p$-values of 0.552 and 0.543 and seems to fail to detect the above dependence. A direct doubt is whether the condition that the censoring time $C$ is independent of $\mathbf{X}$ is invalid. To answer this problem, we consider the other two test problems,
$H_0:C\indep \text{Age}$ and $H_0:C\indep (\text{Age, Perfor, Adhere})$. The $p$-values of all tests except IPCW are summarized in the third and fourth rows of Table~\ref{table-Colon}. All the test methods have high $p$-values. These indicate that there is no evidence of any correlation between $T$ and $\text{Age}$ or $(\text{Age, Perfor, Adhere})$ and thus, the doubt does not hold. Why IPCW does not work with colon data needs more research, which is not covered here.

\section{Discussion}\label{Discussion}
The proposed   SID measures focus on   continuous covariates.  In
survival data analysis, we often encounter datasets with  categorical and   continuous covariates.
To extend the SID metrics into such cases, a simple approach is  to
use kernel functions for the categorical variables.  In fact, such kernel functions have been proposed  in   literatures of support vector machine for  classification problems, see, \citet{Muoz2013KernelFF}.   Additionally,
for $\operatorname{SID}_\beta(T,\mathbf{X})$ and $\operatorname{SID}_{K}(T,\mathbf{X})$,
how to choose the optimal $\beta$ and  $K(\cdot,\cdot)$, including $a(t)$, is still an open question. These will be  interesting topics  for future research.

Although this paper primarily focuses on right-censored data, our method can be extended to handle left-truncated data, and doubly censored data. The key to these    extensions is using counting process techniques to derive a representation similar to that in Theorem \ref{theorem1.1}. Further detailed studies are needed to fully explore these extensions.

The  test statistic proposed  in this paper   is  a 5-order $U$-statistic, which is computationally intensive. In Section 6, we employ the second-order projection of the $U$-statistic to derive the critical values for our test. An alternative approach could involve the application of incomplete $U$-statistics, as suggested by \cite{janson1984asymptotic}. For this method, we can draw insights from recent work by \cite{schrab2022efficient}. Exploring this alternative method  is worth considering for future research.

\section*{Acknowledgments}
The authors would like to thank an associate editor and two referees for constructive comments.

\section*{Funding}
Jicai Liu's research was  supported by the Humanities and Social Sciences Youth Foundation of Ministry of Education of China (23YJC910003).   Jinhong You's research was supported by the National Natural Science Foundation of China (NSFC) (No.11971291) and
Innovative Research Team of Shanghai University of Finance and Economics.  Riquan Zhang's research was supported by the National Natural Science Foundation of China  (12371272)  and the Basic Research Project of Shanghai Science and Technology Commission (22JC1400800). Jicai Liu is the corresponding author.

\section*{Supplementary Material}

The supplement \citep{Jinhong2025SID} provides the proofs of all results of this paper and additional simulations.

\end{document}


\begin{frontmatter}
\title{Supplement to ``SID: a novel class of nonparametric   tests of independence for censored outcomes''}
\runtitle{Survival Independence Divergence}

\begin{aug}
\author[A]{\fnms{Jinhong}~\snm{Li}\ead[label=e1]{jinhongli0106@gmail.com}}
\author[B]{\fnms{Jicai}~\snm{Liu}\ead[label=e2]{liujicai1234@126.com}}
\author[C]{\fnms{Jinhong}~\snm{You}\ead[label=e3]{johnyou07@163.com}}
\author[D]{\fnms{Riquan}~\snm{Zhang}\ead[label=e4]{zhangriquan@163.com}}
\address[A]{School of Statistics and Mathematics, Zhejiang Gongshang University\printead[presep={,\ }]{e1}}
\address[B]{School of Statistics and Mathematics,
Shanghai Lixin University of Accounting and Finance\printead[presep={,\ }]{e2}}
\address[C]{School of Statistics and Management,
Shanghai University of Finance and Economics\printead[presep={,\ }]{e3}}
\address[D]{School of Statistics and Information,
Shanghai University of International Business and Economics\printead[presep={,\ }]{e4}}
\runauthor{Jinhong Li et al.}
\end{aug}

\begin{abstract}
Appendix \ref{Appendix-A} provides detailed technical proofs of the main  theoretical results of the paper. Appendix \ref{Appendix-B} contains several lemmas for the proof of Theorem 5.3. Appendix \ref{Appendix-C} includes some additional simulation studies.
\end{abstract}

\end{frontmatter}

\begin{appendix}

\section{Proofs of the theoretical results}\label{Appendix-A}
\renewcommand{\theequation}{A.\arabic{equation}}
\renewcommand{\thelemma}{A.\arabic{lemma}}

In the section, we prove the theoretical results presented  in Sections \ref{Methodology-Identifiability}-\ref{chapter-bootstrap}.

\subsection{Proofs of    results in Sections~\ref{Methodology-Identifiability}-\ref{Methodology}}

\begin{proof}[\textbf{Proof of Theorem~\ref{identifiability}.}]
Let \(\tau_{_{T,\mathbf{x}}}=\sup\{t: S_{T|\mathbf{X}=\mathbf{x}}(t)>0\}\), \(\tau_{_{C,\mathbf{x}}}=\sup\{t: S_{C|\mathbf{X}=\mathbf{x}}(t)>0\}\), and \(\tau_{_{Y,\mathbf{x}}}=\sup\{t: S_{Y|\mathbf{X}=\mathbf{x}}(t)>0\}\). We prove the theorem by dividing the proof into two parts: (i) \(\tau_{_{T,\mathbf{x}}}\leq\tau_{_{C,\mathbf{x}}}\) holds for almost all \(\mathbf{x} \in \mathcal{X}\); (ii) there exists a set \(\mathcal{A}_X\subset\mathcal{X}\) with \(P\{\mathbf{X}\in\mathcal{A}_X\}>0\), such that \(\tau_{_{T,\mathbf{x}}}>\tau_{_{C,\mathbf{x}}}\) for \(\mathbf{x}\in \mathcal{A}_X\).

(i) If \(\tau_{_{T,\mathbf{x}}}\leq\tau_{_{C,\mathbf{x}}}\) holds for almost all \(\mathbf{x} \in \mathcal{X}\), we obtain that \(S_{C|\mathbf{X}=\mathbf{x}}(t)>0\) for \(t\leq \tau_{_{T,\mathbf{x}}}\), and \(\lambda(t|\mathbf{x})=0\) for \(t>\tau_{_{T,\mathbf{x}}}\). For any two conditional hazard functions \(\lambda_1(t|\mathbf{x})\) and \(\lambda_2(t|\mathbf{x})\), satisfying \(\lambda_1(t|\mathbf{x})\neq\lambda_2(t|\mathbf{x})\), the above result  implies that there exists a set \(\mathcal{T}\subseteq [0,\tau_{_{T,\mathbf{x}}})\), such that \(\lambda_1(t|\mathbf{x})\neq\lambda_2(t|\mathbf{x})\) for \(t\in\mathcal{T}\). Then, by the definition of \(p_{_{Y,\delta,\mathbf{X}}}(t,i,\mathbf{x};\lambda)\), we have \(p_{_{Y,\delta,\mathbf{X}}}(t,i,\mathbf{x};\lambda_1)\neq p_{_{Y,\delta,\mathbf{X}}}(t,i,\mathbf{x};\lambda_2)\) for \(t\leq \tau_{_{T,\mathbf{x}}}\).  Using Definition \ref{defin-identifiable}, we conclude that problem (1) is identifiable.

(ii) When \(\tau_{_{T,\mathbf{x}}}>\tau_{_{C,\mathbf{x}}}\) for any \(\mathbf{x}\in \mathcal{A}_X\), we have \(\tau_{_{Y,\mathbf{x}}}=\min\{\tau_{_{T,\mathbf{x}}},\tau_{_{C,\mathbf{x}}}\}=\tau_{_{C,\mathbf{x}}}\), where the first equality is obtained from Assumption \ref{assum1}. Additionally,  Proposition C.1 in \citet{Tamara} implies that \(\tau= \text{ess}\sup_{\mathbf{x}}\tau_{_{Y,\mathbf{x}}}\) and thus \(\tau_{_{C,\mathbf{x}}}\leq\tau\). If \(T\cdot I(T>\tau)\) is independent of \(\mathbf{X}\), we have \(\lambda(t|\mathbf{x})=\lambda(t)\) for \(t>\tau\). For any two conditional hazard functions \(\lambda_1(t|\mathbf{x})\) and \(\lambda_2(t|\mathbf{x})\), satisfying \(\lambda_1(t|\mathbf{x})\neq\lambda_2(t|\mathbf{x})\), the above result implies that \(\lambda_1(t|\mathbf{x})=\lambda_1(t)\) and \(\lambda_2(t|\mathbf{x})=\lambda_2(t)\) for \(t>\tau\). Then, the following cases may hold:
\begin{itemize}
    \item Case 1: \(\lambda_1(t|\mathbf{x})\neq\lambda_2(t|\mathbf{x})\) for \(t\leq\tau\), while \(\lambda_1(t)=\lambda_2(t)\) for \(t>\tau\);
    \item Case 2: \(\lambda_1(t|\mathbf{x})=\lambda_2(t|\mathbf{x})\) for \(t\leq\tau\), while \(\lambda_1(t)\neq\lambda_2(t)\) for \(t>\tau\);
    \item Case 3: \(\lambda_1(t|\mathbf{x})\neq\lambda_2(t|\mathbf{x})\) for \(t\leq\tau\), and \(\lambda_1(t)\neq\lambda_2(t)\) for \(t>\tau\).
\end{itemize}

For Cases 1 and 3, by using a similar argument of proof as in (i) and replacing \(\tau_{_{T,\mathbf{x}}}\) with \(\tau_{_{C,\mathbf{x}}}\), we can prove that problem (1) is identifiable. For Case 2, we find that \(f_{C|\mathbf{X}=\mathbf{x}}(t)=0\) for \(t>\tau\) and \(p_{_{Y,\delta,\mathbf{X}}}(t,i,\mathbf{x};\lambda_1)= p_{_{Y,\delta,\mathbf{X}}}(t,i,\mathbf{x};\lambda_2)\) for \(t>0\).
Consequently, in this scenario,   we cannot determine whether \(\lambda_1(t|\mathbf{x})\) and \(\lambda_2(t|\mathbf{x})\) are equal for \(t>\tau\)
(here, their difference for \(t\leq\tau\) can still be discerned). However, the condition that \(T\cdot I(T>\tau)\) is independent of \(\mathbf{X}\) provides additional information to overcome this issue. Thus, we can still identify problem (1) in Case 2.\qedhere
\end{proof}

\begin{proof}[\textbf{Proof of Theorem~\ref{theorem1.1}.}]
By  properties of conditional probability, we have
  \begin{eqnarray}\label{lemm1-1-1}
\!\!\! && P\{\mathbf{X}\leq \mathbf{x}| \Delta N(t)=1,   Y(t)=1\}P\{\Delta N(t)=1,  Y(t)=1\}\nonumber\\
&=& P\{\Delta N(t)=1|\mathbf{X}\leq \mathbf{x},  Y(t)=1\} P\{\mathbf{X}\leq \mathbf{x},  Y(t)=1\} \nonumber\\
&=&  P\{\Delta N(t)=1|\mathbf{X}\leq \mathbf{x},  Y(t)=1\}P\{  Y(t)=1\} P\{\mathbf{X}\leq \mathbf{x} |  Y(t)=1\}.
\end{eqnarray}

If $P\{\mathbf{X}\leq \mathbf{x} \mid \Delta N(t)=1,Y(t)=1\} = P\{\mathbf{X}\leq \mathbf{x} \mid Y(t)=1\}$ holds for any $\mathbf{x}\in\mathcal{X}$, then~\eqref{lemm1-1-1} implies that
\begin{equation*}
P\{\Delta N(t)=1 \mid \mathbf{X}\leq \mathbf{x}, Y(t)=1\}
 =\frac{P\{\Delta N(t)=1, Y(t)=1\}}{P\{  Y(t)=1\}},
\end{equation*}
for any   $t<\tau$.
Thus, we obtain that
\begin{equation*}
\begin{aligned}
 &P\{\Delta N(t)=1 \mid \mathbf{X}\leq \mathbf{x}, Y(t)=1\}= P\{\Delta N(t)=1 \mid Y(t)=1\}, \text{ for any } \mathbf{x}\in\mathcal{X},
\end{aligned}
\end{equation*}
which is equivalent to
\begin{equation}\label{lemm1-1}
\begin{aligned}
 &P\{\Delta N(t)=1 \mid \textbf{X}, Y(t)=1\} = P\{\Delta N(t)=1 \mid Y(t)=1\}.
\end{aligned}
\end{equation}

With   Assumption 1 and the definition of the conditional hazard function, we have that
\begin{equation}\label{lemm1-1-2}
\begin{aligned}
\lim_{\Delta t\rightarrow 0^+} \frac{1}{\Delta t} P\{\Delta N(t)=1 \mid \mathbf{X}, Y(t)=1\} &  {=}\lim_{\Delta t\rightarrow 0^+} \frac{1}{\Delta t}\frac{P\{\Delta N(t)=1, Y(t)=1 \mid \mathbf{X}\}}{P\{  Y(t)=1 \mid \mathbf{X}\}} \\
&=\lim_{\Delta t\rightarrow 0^+} \frac{1}{\Delta t}\frac{P\{t+\Delta t > T\geq t, T\leq C \mid \mathbf{X}\}}{P\{ T\wedge C\geq t \mid \mathbf{X}\}} \\
& {=}\lim_{\Delta t\rightarrow 0^+} \frac{1}{\Delta t}\frac{\int_t^{t+\Delta t}P\{  C\geq u|\mathbf{X}\} \dx P\{ T\leq u \mid \mathbf{X}\}}{P\{ T>t \mid \mathbf{X}\}P\{ C>t \mid \mathbf{X}\}} \\
&= \frac{ P\{ C\geq t \mid \mathbf{X}\} f_{T \mid \mathbf{X}}(t) }{P\{ T\geq t \mid \mathbf{X}\}P\{ C\geq t \mid \mathbf{X}\}}\\
&= \lambda(t \mid \mathbf{X}).
\end{aligned}
 \end{equation}
  Then,~\eqref{lemm1-1} and~\eqref{lemm1-1-2} suggest that we have
 \begin{equation}\label{lemm1-1-3}
 \lambda(t \mid \mathbf{X}) =\lim_{\Delta t\rightarrow 0^+} \frac{1}{\Delta t}P\{\Delta N(t)=1 \mid Y(t)=1\},
 \end{equation}
 for any $t \in [0,\tau)$. Note that the right side of \eqref{lemm1-1-3} is independent of $\mathbf{X}$. Thus,
 this shows that  $\mathbf{X}$ and $T$ are independent due to the one-to-one relation between the distribution function and the hazard function.

When  $ T\indep \textbf{X}$, we have that $\lambda(t \mid \mathbf{X})= \lambda(t).$
By the same argument of~\eqref{lemm1-1-2}, we can prove that $\lambda(t)  =\lim_{\Delta t\rightarrow 0^+} P\{\Delta N(t)=1 \mid Y(t)=1\}/{\Delta t} ,$  and thus~\eqref{lemm1-1-3} holds. Then, by~\eqref{lemm1-1-2},
we obtain that~\eqref{lemm1-1} holds. This, together with~\eqref{lemm1-1-1}, yields $\lim_{\Delta t\rightarrow 0^+}P\{\mathbf{X}\leq \mathbf{x} \mid \Delta N(t)=1,Y(t)=1\}=P\{\mathbf{X}\leq \mathbf{x} \mid Y(t)=1\}$.\qedhere

\end{proof}

\begin{proof}[\textbf{Proof of
 the results in \eqref{eq:remark2}  and \eqref{eq:char-function-equal}.}]
(1)
Note that \( \{\Delta N(t) = 1, Y(t) = 1\} = \{t \leq Y < t + \Delta t, \delta = 1\}\). Then,
 we have that
\begin{align}
P\{\mathbf{X} \leq \mathbf{x} \mid \Delta N(t) = 1, Y(t) = 1\} &= P\{\mathbf{X} \leq \mathbf{x} \mid t \leq Y \leq t + \Delta t, \delta = 1\} \nonumber \\
&= \frac{P\{\mathbf{X} \leq \mathbf{x}, t \leq Y - t + \Delta t, \delta = 1\}}{P\{Y < t + \Delta t, \delta = 1\}} \nonumber \\
&= \frac{\int_{\mathcal{X}} \int_{t}^{t + \Delta t} \mathbf{1}(\mathbf{s} \leq \mathbf{x}) f(\mathbf{s}, y, \delta = 1)  d\mathbf{s}\, dy}{\int_{\mathbb{R}} \int_{t}^{t + \Delta t} f(\mathbf{s}, y, \delta = 1)   d\mathbf{s} \, dy} \nonumber\\
&\to \frac{\int_{\mathcal{X}} \mathbf{1}(\mathbf{s} \leq \mathbf{x}) f(\mathbf{s}, t, \delta = 1)   d\mathbf{s}}{\int_{\mathcal{X}} f(\mathbf{s}, t, \delta = 1)  d\mathbf{s}}  \quad \text{(as } \Delta t \to 0^+\text{)} \nonumber\\
&= P\{\mathbf{X} \leq \mathbf{x} \mid Y = t, \delta = 1\}. \label{eq5:prob-proof}
\end{align}
 The limit holds due to the continuity of \( f_{\mathbf{X},Y,\delta}(\mathbf{x},t,1) \) with respect to \( t \) and the mean value theorem for integrals. This proves \eqref{eq:remark2}.

(2) Similarly, we can obtain that
\begin{align}
\varphi_{\mathbf{X}| \Delta N_t,Y_t} (\mathbf{u}, t) &=E\{\exp\{i\mathbf{u}^T\mathbf{X}\} \mid \Delta N(t) = 1, Y(t) =1\} \nonumber\\
&= E\{\exp\{i\mathbf{u}^T\mathbf{X}\} \mid t \leq Y \leq t + \Delta t, \delta = 1\}\nonumber\\
&= \frac{\int_{\mathcal{X}} \int_{t}^{t + \Delta t} \exp\{i\mathbf{u}^T\mathbf{s}\} f(\mathbf{s}, y, \delta = 1) \, d\mathbf{s} \, dy}{\int_{\mathbb{R}} \int_{t}^{t + \Delta t} f(\mathbf{s}, y, \delta = 1)   d\mathbf{s} \, dy} \nonumber\\
&\to \frac{\int_{\mathcal{X}} \exp\{i\mathbf{u}^T\mathbf{s}\} f(\mathbf{s}, t, \delta = 1)   d\mathbf{s}}{\int_{\mathcal{X}} f(\mathbf{s}, t, \delta = 1)  d\mathbf{s}}
~~~~( \text{ as }   \Delta t \to 0^+) \nonumber \\
&=E\{\exp\{i\mathbf{u}^T\mathbf{X}\} \mid Y = t, \delta = 1\}. \label{eq6:char-proof}
\end{align}
By Theorem \ref{theorem1.1} and \eqref{eq5:prob-proof}, we have that
$\lim\limits_{ \Delta t\rightarrow 0^+}P_{\mathbf{X}|\Delta N_t,Y_t}=Q_{\mathbf{X}|Y_t}$ if and only if
$$ \lim\limits_{ \Delta t\rightarrow 0^+}\varphi_{\mathbf{X} |\Delta N_t,Y_t}(\mathbf{u},t)=\varphi_{\mathbf{X} |Y_t}(\mathbf{u},t),$$
 for all    $\mathbf{u}\in \mathbb{R}^p$    and $t\in[0,\tau)$. This completes the proof of   \eqref{eq:char-function-equal}.\qedhere

\end{proof}

\begin{proof}[\textbf{Proof of Theorem~\ref{Theorem2}.}]
(i)  Note that
\begin{eqnarray*}
   &&  \| \varphi_{\mathbf{X} |\Delta N_t,Y_t}(\mathbf{u},t)-\varphi_{\mathbf{X} | Y_t}(\mathbf{u},t)\|^2  \\
   &=& \Big[ E\{\exp\{i\mathbf{u}^T\mathbf{X}\}|\Delta N(t)=1,Y(t)=1\} -E\{\exp\{i\mathbf{u}^T\mathbf{X}\}| Y(t)=1\} \Big]\\
   &&\times \Big[E\{\exp\{-i\mathbf{u}^T\mathbf{X}\}|\Delta N(t)=1,Y(t)=1\} -E\{\exp\{-i\mathbf{u}^T\mathbf{X}\}| Y(t)=1\}\Big]\\
  &=&  E \{\exp\{i\mathbf{u}^T(\mathbf{X}_1-\mathbf{X}_2)\} |\Delta N_1(t)=1,Y_1(t)=1,\Delta N_2(t)=1,Y_2(t)=1 \}\\
   &&-  2E \{ \exp\{i\mathbf{u}^T(\mathbf{X}_1-\mathbf{X}_2)\} |\Delta N_1(t)=1,Y_1(t)=1, Y_2(t)=1  \}    \\
  && + E \{ \exp\{i\mathbf{u}^T(\mathbf{X}_1-\mathbf{X}_2)\} | Y_1(t)=1, Y_2(t)=1  \}.
\end{eqnarray*}
If $E\{\|\mathbf{X}\|^\beta\}<\infty$, this, together with Lemma~1 in \citet{szekely2007measuring}, yields that
\begin{eqnarray*}\label{integal-cons}
  &&\frac{1}{c(p, \beta)} \int_{\mathbb{R}^p}  \| \varphi_{\mathbf{X} |\Delta N_t,Y_t}(\mathbf{u},t)-\varphi_{\mathbf{X} | Y_t}(\mathbf{u},t)\|^2  \frac{1}{ \|\textbf{u}\|^{\beta+p} }d\mathbf{u}\nonumber\\
  &=& -E \{\|\mathbf{X}_1-\mathbf{X}_2\|^\beta\mid \Delta N_1(t)=1, \Delta N_2(t)=1\}\nonumber\\
  &&   +2  E\{\|\mathbf{X}_1-\mathbf{X}_2\|^\beta |\Delta N_1(t)=1, Y_2(t)=1\}  -  E\{\|\mathbf{X}_1-\mathbf{X}_2\|^\beta   | Y_1(t)=1, Y_2(t)=1\}.
\end{eqnarray*}
By similar arguments of \eqref{eq5:prob-proof} and \eqref{eq6:char-proof}, we obtain that
\begin{eqnarray*}
&&\lim_{\Delta t\rightarrow 0} \frac{1}{c(p, \beta)} \int_{\mathbb{R}^p}  \| \varphi_{\mathbf{X} |\Delta N_t,Y_t}(\mathbf{u},t)-\varphi_{\mathbf{X} | Y_t}(\mathbf{u},t)\|^2  \frac{1}{ \|\textbf{u}\|^{\beta+p} }d\mathbf{u}\\
 &=&  -E \{ \|\mathbf{X}_1-\mathbf{X}_2\|^\beta  \mid  Y_1=t, \delta_1=1,   Y_2=t, \delta_2=1\} \\&&  +2  E\{\|\mathbf{X}_1-\mathbf{X}_2\|^\beta   | Y_1=t, \delta_1=1, Y_2(t)=1\}    -  E\{\|\mathbf{X}_1-\mathbf{X}_2\|^\beta   | Y_1(t)=1, Y_2(t)=1\},
\end{eqnarray*}
for any $t\in[0,\tau)$. This immediately shows part (i).

(ii) By the proof of part (i), we can see that when $E\{\|\mathbf{X}\|^\beta\}<\infty$ for $\beta\in(0,2)$, the integral
\begin{equation*}
  \frac{1}{c(p, \beta)} \int_{\mathbb{R}^p}  \| \varphi_{\mathbf{X} |\Delta N_t,Y_t}(\mathbf{u},t)-\varphi_{\mathbf{X} | Y_t}(\mathbf{u},t)\|^2  \frac{1}{ \|\textbf{u}\|^{\beta+p} } \dx\mathbf{u}
\end{equation*}
 exists. Thus, $\operatorname{SID}_\beta(T,\mathbf{X})$ is well defined. Then, by the definition of $\operatorname{SID}_\beta(T,\mathbf{X})$, we have that $\operatorname{SID}_\beta(T,\mathbf{X})\geq 0$ and
\begin{equation*}
\operatorname{SID}_\beta(T,\mathbf{X})=0 \Longleftrightarrow \lim_{ \Delta t\rightarrow 0^+}\varphi_{\mathbf{X} |\Delta N_t,Y_t}(\mathbf{u},t)=\varphi_{\mathbf{X} |Y_t}(\mathbf{u},t),
 \text{ for all }  \mathbf{u}\in \mathbb{R}^p  \text{ and } t\in[0,\tau).
\end{equation*}
Theorem~\ref{theorem1.1} indicates that $\operatorname{SID}_\beta(T,\mathbf{X})=0$ if and only if $T$ and $\mathbf{X}$ are  independent.\qedhere
\end{proof}

\begin{proof}[\textbf{Proof of Theorem~\ref{Theorem4}.}]
(i) By the definition of $\mu_K(\nu)$ and the property that $\mu_K(\nu)=\int K(\cdot,z) \dx{}\nu(z)$,
 we have that
\begin{eqnarray*}
&&\Vert \mu_K(P_{\mathbf{X}|\Delta N_t,Y_t})-\mu_K(Q_{\mathbf{X}|Y_t})\Vert^2_{\mathcal{H}_K} \nonumber \\
&=&\left\langle\mu_K(P_{\mathbf{X}|\Delta N_t,Y_t}),  \mu_K(P_{\mathbf{X}|\Delta N_t,Y_t})\right\rangle_{\mathbb{H}_K}+\left\langle\mu_K(Q_{\mathbf{X}|Y_t}),  \mu_K(Q_{\mathbf{X}|Y_t})\right\rangle_{\mathbb{H}_K}\nonumber\\ &&-2\left\langle\mu_K(P_{\mathbf{X}|\Delta N_t,Y_t}), \mu_K(Q_{\mathbf{X}|Y_t})\right\rangle_{\mathbb{H}_K} \nonumber\\
&=& \int  K(\mathbf{X}_1,\mathbf{X}_2) d(P_{\mathbf{X}_1|\Delta N_t,Y_t} \times P_{\mathbf{X}_2|\Delta N_t,Y_t})+
\int  K(\mathbf{X}_1,\mathbf{X}_2) d(Q_{\mathbf{X}_1|Y_t} \times Q_{\mathbf{X}_2|Y_t})  \\&&-2\int  K(\mathbf{X}_1,\mathbf{X}_2) d(P_{\mathbf{X}_1|\Delta N_t,Y_t} \times Q_{\mathbf{X}_2|Y_t})\\
&=& E \{K(\mathbf{X}_1,\mathbf{X}_2) |\Delta N_1(t)=1,Y_1(t)=1,\Delta N_2(t)=1,Y_2(t)=1 \} \\
  && + E \{K(\mathbf{X}_1,\mathbf{X}_2) | Y_1(t)=1, Y_2(t)=1  \}-  2E \{ K(\mathbf{X}_1,\mathbf{X}_2)|\Delta N_1(t)=1,Y_1(t)=1, Y_2(t)=1  \}.
\end{eqnarray*}
By similar arguments of \eqref{eq5:prob-proof} and \eqref{eq6:char-proof}, we obtain that
\begin{eqnarray*}
 \lim_{\Delta t\rightarrow 0^+}   \Vert \mu_K(P_{\mathbf{X}|\Delta N_t,Y_t})-\mu_K(Q_{\mathbf{X}|Y_t})\Vert^2_{\mathcal{H}_K}
 &=&E \{ K(\mathbf{X}_1,\mathbf{X}_2)  \mid  Y_1=t, \delta_1=1,   Y_2=t, \delta_2=1\}
  \\&& -2  E\{K(\mathbf{X}_1,\mathbf{X}_2)   | Y_1=t, \delta_1=1, Y_2(t)=1\}   \\&& +  E\{K(\mathbf{X}_1,\mathbf{X}_2)   | Y_1(t)=1, Y_2(t)=1\},
\end{eqnarray*}
for any $t\in[0,\tau)$. This immediately shows part~(i).

(ii) $\operatorname{SID}_{\rho}(T,\mathbf{X})\geq 0$ is straightforward. If $\operatorname{SID}_{K}(T,\mathbf{X})=0$, we have that $  \lim_{ \Delta t\rightarrow 0^+}\mu_K(P_{\mathbf{X}|\Delta N_t,Y_t})=\mu_K(Q_{\mathbf{X}|Y_t}).$ Thus, when $K(\cdot,\cdot)$ is characteristic,
we have that $\lim_{ \Delta t\rightarrow 0^+}  P_{\mathbf{X}|\Delta N_t,Y_t}=Q_{\mathbf{X}|Y_t}$. Then, Theorem~\ref{theorem1.1} shows that $T$ and $\mathbf{X}$ are  independent.
Conversely, if  $T\indep \textbf{X}$, we have that $ \lim_{ \Delta t\rightarrow 0^+}P_{\mathbf{X}|\Delta N_t,Y_t}=Q_{\mathbf{X}|Y_t}$ and $\operatorname{SID}_{K}(T,\mathbf{X})=0$.
\end{proof}

\begin{proof}[\textbf{Proof of Theorem~\ref{Theorem3}.}]
Note that
\begin{eqnarray*}
&&- \int\rho(\mathbf{X},\mathbf{X}') d([P_{\mathbf{X}|\Delta N_t,Y_t}-Q_{\mathbf{X}|Y_t}]\times[P_{\mathbf{X}'|\Delta N_t,Y_t}-Q_{\mathbf{X}'|Y_t}]) \nonumber \\
&=& -\int \rho(\mathbf{X},\mathbf{X}') d(P_{\mathbf{X}|\Delta N_t,Y_t} \times P_{\mathbf{X}'|\Delta N_t,Y_t})
-\int \rho(\mathbf{X},\mathbf{X}') d(Q_{\mathbf{X}|Y_t} \times Q_{\mathbf{X}'|Y_t})\\
  &&+2\int\rho(\mathbf{X},\mathbf{X}') d(P_{\mathbf{X}|\Delta N_t,Y_t} \times Q_{\mathbf{X}'|Y_t})\\
&=& -E \{\rho(\mathbf{X}_1,\mathbf{X}_2) |\Delta N_1(t)=1,Y_1(t)=1,\Delta N_2(t)=1,Y_2(t)=1 \} - E \{\rho(\mathbf{X}_1,\mathbf{X}_2) | Y_1(t)=1, Y_2(t)=1  \}\\
  &&+  2E \{ \rho(\mathbf{X}_1,\mathbf{X}_2)|\Delta N_1(t)=1,Y_1(t)=1, Y_2(t)=1  \}.
\end{eqnarray*}
By similar arguments of \eqref{eq5:prob-proof} and \eqref{eq6:char-proof}, we obtain that
\begin{eqnarray*}
&&\lim_{\Delta t\rightarrow 0^+} - \int_{\mathcal{X}\times \mathcal{X}}\rho(\mathbf{X},\mathbf{X}') d([P_{\mathbf{X}|\Delta N_t,Y_t}-Q_{\mathbf{X}|Y_t}]\times[P_{\mathbf{X}'|\Delta N_t,Y_t}-Q_{\mathbf{X}'|Y_t}]) \nonumber
\\
 &=&E \{ \rho(\mathbf{X}_1,\mathbf{X}_2)  \mid  Y_1=t, \delta_1=1,   Y_2=t, \delta_2=1\}
   -2  E\{\rho(\mathbf{X}_1,\mathbf{X}_2)   | Y_1=t, \delta_1=1, Y_2(t)=1\} \\&& +  E\{\rho(\mathbf{X}_1,\mathbf{X}_2)   | Y_1(t)=1, Y_2(t)=1\},
\end{eqnarray*}
for any $t\in[0,\tau)$. This immediately shows this theorem.\qedhere

\end{proof}

\begin{proof}[\textbf{Proof  of Theorem~ \ref{Relationship}.}]
Note that $\rho(\mathbf{x},\mathbf{x}')=K(\mathbf{x},\mathbf{x})+K(\mathbf{x}',\mathbf{x}')-2K(\mathbf{x},\mathbf{x}').$ Then, we have that
\begin{eqnarray*}
&&  E\{\rho(\mathbf{X}_1,\mathbf{X}_2)  \mid Y_1=t, \delta_1=1,   Y_2=t, \delta_2=1\}  \nonumber\\
&=&2  E\{ K(\mathbf{X}_1,\mathbf{X}_1)  \mid Y_1=t, \delta_1=1 \}
 -2  E\{ K(\mathbf{X}_1,\mathbf{X}_2)  \mid Y_1=t, \delta_1=1,   Y_2=t, \delta_2=1\},\\
&&  E\{\rho(\mathbf{X}_1,\mathbf{X}_2)  \mid Y_1=t, \delta_1=1, Y_2(t)=1\}  \nonumber\\
&=&  E\{ K(\mathbf{X}_1,\mathbf{X}_1)  \mid Y_1=t, \delta_1=1 \}   +
 E\{ K(\mathbf{X}_2,\mathbf{X}_2)  \mid Y_2(t)=1 \}  \\
&&-2E\{ K(\mathbf{X}_1,\mathbf{X}_2)   \mid Y_1=t, \delta_1=1, Y_2(t)=1\},\\
&&  E\{ \rho(\mathbf{X}_1,\mathbf{X}_2)   \mid Y_1(t)=1, Y_2(t)=1\} \\
&=& 2E\{ K(\mathbf{X}_1,\mathbf{X}_1)   \mid Y_1(t)=1 \}
 -2 E\{K(\mathbf{X}_1,\mathbf{X}_2)   \mid Y_1(t)=1, Y_2(t)=1\}.
\end{eqnarray*}
By Theorem~\ref{Theorem3}, we have that
\begin{align*}
 &\operatorname{SID}_{\rho}(T,\mathbf{X})\\
 &=\!-\!\!\int_0^\tau\!\!\! \Big[2  E\{ K(\mathbf{X}_1,\mathbf{X}_1) \mid Y_1=t, \delta_1=1 \}\!\!  - \!\!\! 2E\{ K(\mathbf{X}_1,\mathbf{X}_2) \mid Y_1=t, \delta_1=1, Y_2=t, \delta_2=1\}\Big]a(t) \dx{}\nu(t) \\
 & \quad + \ 2\int_0^\tau\Big[E\{ K(\mathbf{X}_1,\mathbf{X}_1) \mid Y_1=t, \delta_1=1 \} +
 E\{ K(\mathbf{X}_2,\mathbf{X}_2) \mid Y_2(t)=1 \}  \\
& \quad - \ 2E\{ K(\mathbf{X}_1,\mathbf{X}_2) \mid Y_1=t, \delta_1=1, Y_2(t)=1\}\Big]a(t) \dx{}\nu(t)   -\int_0^\tau\Big[2E\{ K(\mathbf{X}_1,\mathbf{X}_1) \mid Y_1(t)=1 \} \\
& \quad - \ 2 E\{K(\mathbf{X}_1,\mathbf{X}_2) \mid Y_1(t)=1, Y_2(t)=1\}\Big]a(t) \dx{}\nu(t) \\&= 2\operatorname{SID}_{K}(T,\mathbf{X}).
\end{align*}
\end{proof}

\subsection{Proofs of results in Section~\ref{chapter-asymptotic}}

\begin{proof}[\textbf{Proof of Theorem~\ref{Theorem-consistency}.}]

We first show the convergence of $\operatorname{\widehat{SID}}_{K}(T,\mathbf{X})$ by the theory of $U$-statistics \citep{lee1990u}.
Let $\mathbf{Z}_i = (Y_i, \delta_i, \mathbf{X}_i)$ and
 $$P_{n}(\mathbf{Z}_{i_{1}}, \dots, \mathbf{Z}_{i_{5}})
  = {B}_{i_{1} i_{3} i_{5}} {K}_{i_{1}i_{2}} {B}_{i_{2}i_{4}i_{5}}\delta_{i_{5}},$$
 where
$K_{i_{1}i_{2}}=K(\mathbf{X}_{i_{1}},\mathbf{X}_{i_{2}})$ and
${B}_{i_{1} i_{3} i_{5}}= b_{i_{1} i_{3} i_{5}}- b_{ i_{3}i_{1} i_{5}}$
with $b_{i_{1} i_{3} i_{5}}=\delta_{i_{1}} W_h(Y_{i_{1}}-Y_{i_{5}})I(Y_{i_{3}}\geq Y_{i_{5}}).$
Then, by~\eqref{SK-U-est}, we have that
\begin{align*}
\operatorname{\widehat{SID}}_{K}(T,\mathbf{X}) &=
\begin{pmatrix} n \\ 5 \end{pmatrix}^{-1}
\sum_{i_{1}<i_{2}<\dots<i_{5}} P_{n}(\mathbf{Z}_{i_{1}}, \dots, \mathbf{Z}_{i_{5}}) \\
& =
\begin{pmatrix} n \\ 5 \end{pmatrix}^{-1}
\sum_{i_{1}<i_{2}<\dots<i_{5}} h_{n}(\mathbf{Z}_{i_{1}}, \dots, \mathbf{Z}_{i_{5}}),
\end{align*}
where
$$ h_n(\mathbf{Z}_{i_{1}}, \dots, \mathbf{Z}_{i_{5}} ) = \frac{1}{5 !} \sum_{\pi} P_{n} (\mathbf{Z}_{i'_{1}}, \ldots, \mathbf{Z}_{i'_{5}} ),$$ and $\sum_{\pi}$ denotes summation over all the $5!$ permutations $(i'_{1}, \ldots, i'_{5})$ of $(i_{1}, \ldots, i_{5})$.

Let
\begin{align*}
{h}_{n c}(\mathbf{z}_{1},\ldots, \mathbf{z}_{c}) &= E\{ {h}_{n}(\mathbf{Z}_{1}, \ldots, \mathbf{Z}_{5}) \mid \mathbf{Z}_{1}=\mathbf{z}_{1}, \ldots, \mathbf{Z}_{c}=\mathbf{z}_{c}\}, \\
P_{n c, j_{1},\ldots, j_{c}}(\mathbf{z}_{1},\ldots, \mathbf{z}_{c}) &=  E\{P_{n}( \mathbf{Z}_{1}, \ldots, \mathbf{Z}_{5}) \mid \mathbf{Z}_{j_{1}}= \mathbf{z}_{1},\ldots, \mathbf{Z}_{j_{c}}=\mathbf{z}_{c}\},
\end{align*}
 where $j_{1},\ldots, j_{c}\in\{1,2,3,4,5\}.$ Note that the subscripts $(j_{1},\ldots, j_{c})$ mark the position of the arguments of $P_{n}( \mathbf{Z}_{1}, \ldots, \mathbf{Z}_{5})$ due to the asymmetry.

Note that
$
\operatorname{\widehat{SID}}_{K}(T,\mathbf{X})
=E\{\operatorname{\widehat{SID}}_{K}(T,\mathbf{X})\}+O_{p}(\operatorname{Var}^{1/2}(\operatorname{\widehat{SID}}_{K}(T,\mathbf{X}))).
$
We establish the convergence of each term via the following two steps.

 Step 1: Establish the convergence of $\operatorname{Var}(\operatorname{\widehat{SID}}_{K}(T,\mathbf{X}))$.
By Theorem~3 in \citet{lee1990u}, we have that
\begin{equation}\label{variance-Un}
\begin{aligned}
 \operatorname{Var}(\operatorname{\widehat{SID}}_{K}(T,\mathbf{X})) &=
\begin{pmatrix} n \\ 5 \end{pmatrix}^{-1}
\sum_{c=1}^{5}
\begin{pmatrix} 5   \\ c   \end{pmatrix}
\begin{pmatrix} n-5 \\ k-c \end{pmatrix}
\sigma^2_{nc} \\
&= \sum_{c=1}^{5}
\begin{pmatrix} 5   \\ c   \end{pmatrix}
\frac{5!}{(5-c)!}
[n^{-c}+O(n^{-c-1})]\sigma^2_{nc},
\end{aligned}
\end{equation}
where $\sigma^2_{nc}=\operatorname{Var}(h_{n c} (\mathbf{Z}_{1}, \ldots, \mathbf{Z}_{c} )),$
  $c=1,2,\dots,5.$

We first consider $\sigma^2_{n1}=\operatorname{Var}(h_{n 1} (\mathbf{Z}_{1}))$.
After some algebra, we obtain that
\begin{align*}
E\{h_{n 1}^{2}(\mathbf{Z}_{1})\} &= \frac{1}{(5 !)^2}E\Big\{ \Big(E\Big\{ \sum_{\pi'} P_{n} (\mathbf{Z}_{1}, \mathbf{Z}_{i'_{2}},\mathbf{Z}_{i'_{3}},\mathbf{Z}_{i'_{4}}, \mathbf{Z}_{i'_{5}} )   + \ \sum_{\pi'} P_{n} (\mathbf{Z}_{i'_{2}},\mathbf{Z}_{1}, \mathbf{Z}_{i'_{3}},\mathbf{Z}_{i'_{4}},\mathbf{Z}_{i'_{5}}) \\
 & \quad +\dots+
 \sum_{\pi'} P_{n} (\mathbf{Z}_{i'_{2}},\mathbf{Z}_{i'_{3}}, \mathbf{Z}_{i'_{3}},\mathbf{Z}_{i'_{4}}, \mathbf{Z}_{1})\Big|\mathbf{Z}_{1}
\Big\}\Big)^2\Big\} \\
&= \frac{1}{25}E\Big\{[P_{n1,1}(\mathbf{Z}_{1})+P_{n1,2}(\mathbf{Z}_{1})
+\ldots+P_{n1,5}(\mathbf{Z}_{1})]^2\Big\} \\
&= \frac{1}{25 } \sum_{j=1}^5 E\{P_{n 1,j}^2(\mathbf{Z}_{1})\}+\frac{4}{5} {\operatorname{SID}}_{K}(T,\mathbf{X})^2,
\end{align*}
where $\sum_{\pi'}$ denotes summation
over all the $4 !$ permutations $(i'_{2},i'_{3}, i'_{4}, i'_{5})$ of $(2,3, 4, 5)$.
The second equality follows because $(\mathbf{Z}_{1}, \mathbf{Z}_{2},\mathbf{Z}_{3}, \mathbf{Z}_{4}, \mathbf{Z}_{5})$ are i.i.d. The third equality is from
$E\{P_{n}(\mathbf{Z}_{1},\ldots, \mathbf{Z}_{5})\}={\operatorname{SID}}_{K}(T,\mathbf{X})$.

We  show that $E\{P_{n 1,j}^2(\mathbf{Z}_{1})\}=O(1)$, $j=1, \ldots,5$.
We present only the proof that $E\{P_{n 1,1}^2(\mathbf{Z}_{1})\}=O(1)$. The other proofs can be obtained similarly. By the definition of
 $P_{n}(\mathbf{Z}_{i_{1}}, \dots, \mathbf{Z}_{i_{5}}),$
 we have that
$$P_{n}(\mathbf{Z}_{1}, \ldots, \mathbf{Z}_{5}) = {B}_{135} {K}_{12} {B}_{245}\delta_5= {K}_{12}[b_{1 3 5}b_{2 4 5}- 2b_{1 3 5}b_{4 2 5}
  +b_{ 3 1 5} b_{ 42 5}]\delta_5.
  $$
Then, we obtain that
\begin{align*}
  E\{P_{n 1,1}^{2}(\mathbf{Z}_{1} )\}
&= E\{E^2\{K_{12} b_{135}b_{245}\delta_5 \mid \mathbf{Z}_{1}\}\}+4E\{E^2\{K_{12} b_{135}b_{425}\delta_5 \mid \mathbf{Z}_{1}\}\} \\
& \quad + \ E \{E^2\{K_{12} b_{315}b_{425}\delta_5 \mid \mathbf{Z}_{1}\}\} \\
& \quad - \ 4E\{E\{K_{12} b_{135}b_{245}\delta_5 \mid \mathbf{Z}_{1}\}E\{K_{12} b_{135}b_{425}\delta_5 \mid \mathbf{Z}_{1}\}\} \\
& \quad - \ 4E\{E\{K_{12} b_{315}b_{425}\delta_5 \mid \mathbf{Z}_{1}\}E\{K_{12} b_{135}b_{425}\delta_5 \mid \mathbf{Z}_{1}\}\} \\
& \quad + \ E\{E\{K_{12} b_{135}b_{245}\delta_5 \mid \mathbf{Z}_{1}\}E\{K_{12} b_{315}b_{425}\delta_5 \mid \mathbf{Z}_{1}\}\} \\
&= E_{11}+4E_{12}+E_{13}-4E_{14}-4E_{15}+E_{16}.
\end{align*}
Consider the order of $E_{11}$.
Note that
\begin{equation}\label{kern-bound}
 K(\mathbf{x}_1,\mathbf{x}_2)\leq K^{1/2}(\mathbf{x}_1,\mathbf{x}_1) K^{1/2}(\mathbf{x}_2,\mathbf{x}_2)<\infty
\end{equation}
and
\begin{equation*}
 E\{K_{12} b_{135}b_{245}\delta_5 \mid \mathbf{Z}_{1}\}
=E\{K_{12} \delta_1 W_h(y_1-Y_5)\delta_2 W_h(Y_2-Y_5)[1-F_Y(Y_5)]^2\delta_5  \mid \mathbf{Z}_{1}\},
\end{equation*}
where $F_{Y}(y)$ is the marginal density distribution of $Y$.
By changing variables to $y_5=y_1+hu$ and $y_2=y_1+hv$, we have
\begin{align*}
E_{11} &= \int\Bigg[\int K(\mathbf{x}_1,\mathbf{x}_2)   W_h(y_1-y_5)  W_h(y_2-y_5)[1-F_Y(y_5)]^2f_{\mathbf{X},Y,1}(\mathbf{x}_2,y_2) \\&\times  f_{Y,\delta}(y_5,1) \dx{}\mathbf{x}_2  \dx{}y_2 \dx{}y_5\Bigg]^2f_{\mathbf{X},Y,\delta}(\mathbf{x}_1,y_1,1) \dx{}\mathbf{x}_1 \dx{}y_1\\
 &= \int\Bigg[\int K(\mathbf{x}_1,\mathbf{x}_2) W(u) W(v-u)[1-F_Y(y_1+hu)]^2 f_{X,Y,1}(\mathbf{x}_2,y_1+hv) \\
& \quad \times f_{Y,\delta}(y_1+hu,1) \dx{}\mathbf{x}_2\dx{}u\dx{}v\Bigg]^2f_{\mathbf{X},Y,\delta}(\mathbf{x}_1,y_1,1) \dx{}\mathbf{x}_1 \dx{}y_1= O(1).
\end{align*}
Similarly, we can obtain $E_{1j}=O(1)$, $j=2, \ldots,6$. Thus, $E\{P_{n 1,1}^{2}(\mathbf{Z}_{1} )\}=O(1)$ and $\sigma^2_{n1}=O(1).$

We can obtain that $\sigma^2_{n2}=\operatorname{Var}(h_{n 2} (\mathbf{Z}_{1},\mathbf{Z}_{2}))=O(1/h)$. Specifically, in this case, we need to calculate the analogues to $E_{1j}$, denoted by $E'_{1j}$.
We here present the proof of $E^{'}_{11}=E\{E^2\{K_{12} b_{135}b_{245}\delta_5 \mid \mathbf{Z}_{1},\mathbf{Z}_{2}\}\}\}$. The other proofs can be similarly obtained.
By changing variables to $y_5=y_1+hu$ and $y_2=y_1+hv$, we have
\begin{align*}
E^{'}_{11}  &= \int\Bigg[\int K(\mathbf{x}_1,\mathbf{x}_2)   W_h(y_1-y_5)  W_h(y_2-y_5)[1-F_Y(y_5)]^2 \\&\times f_{Y,\delta}(y_5,1) \dx{}y_5\Bigg]^2   f_{\mathbf{X},Y,\delta}(\mathbf{x}_1,y_1,1)f_{\mathbf{X},Y,\delta}(\mathbf{x}_2,y_2,1) \dx{}\mathbf{x}_1 \dx{}\mathbf{x}_2 \dx{}y_1 \dx{}y_2 \\
&= \int\Bigg[\int W(u)W(v-u)[1-F_Y(y_1+hu)]^2 f_{Y, \delta}(y_1+hu,1)\dx{}u \Bigg]^2 K^2(x_1,x_2) \\
& \quad \times \frac{1}{h}f_{\mathbf{X},Y,\delta}(\mathbf{x}_1,y_1,1)f_{\mathbf{X},Y,\delta}(\mathbf{x}_2,y_1+hv,1) \dx{}\mathbf{x}_1 \dx{}\mathbf{x}_2 \dx{}y_1\dx{}v = O(1/h).
\end{align*}
Then, by similar arguments for $E_{11}$, we obtain that $\sigma^2_{n2}=O(1/h)$.

Similarly, we can show $\sigma^2_{nc}=O(1/h^{c-1})$.
These, together with~\eqref{variance-Un}, yield that
\begin{equation}\label{re-step1}
\operatorname{Var}(\operatorname{\widehat{SID}}_{K}(T,\mathbf{X}))=O(1/n), \text{ as } nh\rightarrow \infty.
\end{equation}

Step 2: Establish the convergence of $E\{\widehat{\operatorname{SID}}_{K}(T,\mathbf{X})\}$.
Note that $P_{n}(\mathbf{Z}_{1}, \ldots, \mathbf{Z}_{5})= {K}_{12}[b_{1 3 5}b_{2 4 5} - 2b_{1 3 5}b_{4 2 5} + b_{ 3 1 5} b_{ 42 5}]\delta_5$. Thus,  we have that
\begin{eqnarray*}
  E\{h_{n}(\mathbf{Z}_{1},  \ldots, \mathbf{Z}_{5})\} &=&
   \frac{1}{5 !} \sum_{\pi}E \{ P_{n}(\mathbf{Z}_{i'_{1}},  \ldots, \mathbf{Z}_{i'_{5}} )\}\\
 &=& E\{ P_{n}(\mathbf{Z}_{1}, \ldots, \mathbf{Z}_{5})\}\\
 &=& E \{ K_{12} \delta_1 W_{h}( Y_1-Y_5) I(Y_3>Y_5) \delta_2 W_{h}( Y_2-Y_5) I(Y_4>Y_5)\delta_5 \}\\
&&- 2 E \{ K_{12}\delta_1 W_{h}( Y_1-Y_5) I(Y_3>Y_5)  \delta_4 W_{h}( Y_4-Y_5) I(Y_2>Y_5)\delta_5 \}\\
&&+  E \{ K_{12}\delta_3 W_{h}( Y_3-Y_5) I(Y_1>Y_5)  \delta_4 W_{h}( Y_4-Y_5) I(Y_2>Y_5)\delta_5 \} \\
 &=& \theta_{n 1}-2\theta_{n 2}+\theta_{n 3}.
\end{eqnarray*}
The second equation follows because $(\mathbf{Z}_{1}, \mathbf{Z}_{2},\mathbf{Z}_{3}, \mathbf{Z}_{4}, \mathbf{Z}_{5})$ are i.i.d.

Consider the first term. For given $t\in [0,\tau),$ we define
\begin{align*}
\theta_{n 1}(t) = E \{ K_{12} W_{h}( Y_1-t) W_{h}( Y_2-t)\delta_1\delta_2\}.
\end{align*}
Then, we have
\begin{align*}
\theta_{n 1}(t) &= E \Big\{ E\{K_{12}  \mid Y_1, \delta_1,Y_2, \delta_2\} W_{h}( Y_1-t)\delta_1 W_{h}( Y_2-t)\delta_2\Big\} \\
&= \int E\{K_{12}  \mid Y_1=y_1, \delta_1=1,Y_2=y_2, \delta_2=1  \}\frac{1}{h^2}  W\left(\frac{y_1-t}{h}\right) W\left(\frac{y_2-t}{h}\right) \\
  & \quad \times f_{Y, \delta}(y_1,1)f_{Y, \delta}(y_2,1) \dx{}y_1 \dx{}y_2 \\
&=   E\{K_{12}   \mid Y_1=t, \delta_1=1,Y_2=t, \delta_2=1  \} f_{Y, \delta}(t,1) f_{Y, \delta}(t,1) +O(h^{2}).
\end{align*}
Thus, we have
\begin{align*}
\theta_{n 1} &= E\Big\{\int_0^\tau \theta_{n 1}(t)  E^2 \{  I(Y>t)\} \dx{}N(t)\Big\} \\
&= E\Big\{\!\! \int_0^\tau \!\!\!\! E\{K_{12}   \mid Y_1=t, \delta_1=1,Y_2=t, \delta_2=1  \} f^2 _{Y, \delta}(t,1)   E^2 \{  I(Y>t)\} \dx{}N(t)\Big\} + \ O(h^{2}) \\
&= \int_0^\tau E\{ K_{12} \mid Y_1=t, \delta_1=1, Y_2=t, \delta_2=1\} a^*(t) \dx{}\nu(t)+O(h^{2}).
\end{align*}

Similarly, we can obtain that
\begin{align*}
\theta_{n2}(t) &= E \{ K_{12} W_{h}( Y_1-t)\delta_1 I(Y_2>t)\}, \\
\theta_{n2}    &= E\Big\{\int_0^\tau \theta_{n 2}(t)  E \{ W_{h}( Y_1-t)\delta_1  I(Y_2>t)\} \dx{}N(t)\Big\} \\
&= \int_0^\tau E\{ K_{12} \mid Y_1=t, \delta_1=1, Y_2(t)=1\} a^*(t) \dx{}\nu(t)+O(h^{2}),
\end{align*}
and
\begin{align*}
\theta_{n3}(t) &= E \{ K_{12}  I(Y_1>t)   I(Y_2>t)\}.\\
\theta_{n3}    &= E \left\{\int_0^\tau \theta_{n 3}(t) E \{   W_{h}( Y_1-t) W_{h}( Y_2-t)\delta_1\delta_2 )\} \dx{}N(t)\right\} \\
   &= \int_0^\tau E\{ K_{12} \mid Y_1(t)=1, Y_2(t)=1\} a^*(t) \dx{}\nu(t)+O(h^{2}).
\end{align*}
Then, by Theorem~\ref{Theorem4}, we have
\begin{equation}\label{re-step2}
E\{\operatorname{\widehat{SID}}_{K}(T,\mathbf{X})\}=E\{h_{n}(\mathbf{Z}_{1}, \ldots, \mathbf{Z}_{5})\}=\operatorname{SID}_{K}(T,\mathbf{X})+O(h^{2}).
\end{equation}
By~\eqref{re-step1} and~\eqref{re-step2}, we have that
 $\operatorname{\widehat{SID}}_{K}(T,\mathbf{X})=\operatorname{SKIC}_{K}(T,\mathbf{X})+O(h^{2}).$

By the relation between the $V$-statistic and $U$-statistic \citep[Section~4.2, Theorem~1]{lee1990u}, we have that $\widetilde{\operatorname{SID}}_{K}(T,\mathbf{X})=\operatorname{SID}_{K}(T,\mathbf{X})+O(h^{2})$.
Thus, we complete the proof of the theorem.\qedhere
\end{proof}

\begin{proof}[\textbf{Proof of Lemma~\ref{Hn2-degenation}.}]
We first calculate ${h}_{n 1}(z;t)$.
By the definition of ${h}_{n 1}(z;t)$, we have that
\begin{equation}\label{h1-0}
\begin{aligned}
{h}_{n 1}(\mathbf{z};t) &= \frac{1}{4}E\Big\{[ K(\mathbf{x}, {\bf X}_3)- K(\mathbf{x}, {\bf X}_4)- K(\mathbf{X}_2, \mathbf{X}_3)+ K(\mathbf{X}_2, \mathbf{X}_4)] \\
& \quad \times [\tilde{\delta} W_h(y-t)I(Y_3\geq t)- \delta_3 W_h(Y_3-t)I(y\geq t)] \\
& \quad \times[\delta_2 W_h(Y_2-t)I(Y_4\geq t)- \delta_4 W_h(Y_4-t)I(Y_2\geq t)] \Big\} \\
&= \frac{1}{4}E\Big\{[E \{K(\mathbf{x}, \mathbf{X}_3) \mid  Y_3(t)=1\} - E \{K(\mathbf{x}, {\bf X}_4) \mid  Y_4(t)=1\} \\
& \quad - \ E \{K(\mathbf{X}_2, \mathbf{X}_3) \mid  Y_2(t)=1, Y_3(t)=1\} \\
& \quad + \ E \{K(\mathbf{X}_2, \mathbf{X}_4) \mid  Y_2(t)=1, Y_4(t)=1\}] \\
& \quad \times [\tilde{\delta} W_h(y-t)I(Y_3\geq t)- \delta_3 W_h(Y_3-t)I(y\geq t)] \\
& \quad \times[\delta_2 W_h(Y_2-t)I(Y_4\geq t)- \delta_4 W_h(Y_4-t)I(Y_2\geq t)] \Big\}.
\end{aligned}
\end{equation}
The last equation holds due to the property of the conditional expectation and Theorem~\ref{theorem1.1}, which implies
 $P\{\mathbf{X}\leq \mathbf{x} \mid Y=t, \delta=1, Y(t)=1\}=P\{\mathbf{X}\leq \mathbf{x} \mid Y(t)=1\}.$

By the properties of conditional expectation, we have that
\begin{eqnarray*}
&&E \{K(x, {\bf X}_3)|  Y_3(t)=1\}
-E \{K(x, {\bf X}_4)| Y_4(t)=1\} \\
&&-E \{K({\bf X}_2, {\bf X}_3)|  Y_2(t)=1,  Y_3(t)=1\}
 +  E \{K({\bf X}_2, {\bf X}_4)|  Y_2(t)=1,  Y_4(t)=1\}]\\
&=& \frac{E \{K(x, {\bf X}_3)I(Y_3\geq t)\}}{P\{Y_3\geq t\}} -
\frac{E \{K(x, {\bf X}_4)I(Y_4\geq t)\}}{P\{Y_4\geq t\}} \\
&&
-\frac{E \{K({\bf X}_2, {\bf X}_3)I(Y_2\geq t)I(Y_3\geq t)\}}{P\{Y_2\geq t,Y_3\geq t\}}+
\frac{E \{K({\bf X}_2, {\bf X}_4)I(Y_2\geq t)I(Y_4\geq t)\}}{P\{Y_2\geq t,Y_4\geq t\}}=0.
\end{eqnarray*}
The last equation holds since $(Y_{i}, \delta_{i}, \mathbf{X}_{i}),i=1,2,3,4$, are i.i.d. This, together with~\eqref{h1-0}, yields that ${h}_{n 1}(\mathbf{z};t)=0.$

We next calculate ${h}_{n 2}(\mathbf{z},\mathbf{z}';t)$.
By the definition of ${h}_{n 2}(\mathbf{z},\mathbf{z}';t)$, we have that
\begin{align*}
 {h}_{n 2}(\mathbf{z},\mathbf{z}';t) &= \frac{1}{6}E\Big\{ [ K(\mathbf{x}, \mathbf{x}')- K(\mathbf{x}, \mathbf{X}_4)- K(\mathbf{x}', \mathbf{X}_3)+ K(\mathbf{X}_3, \mathbf{X}_4)] \\
& \quad \times [\tilde{\delta} W_h(y-t)I(Y_3\geq t)- \delta_3 W_h(Y_3-t)I(y\geq t)] \\
& \quad \times[\tilde{\delta}' W_h(y'-t)I(Y_4\geq t)- \delta_4 W_h(Y_4-t)I(y'\geq t)] \Big\} \\
&= \frac{1}{6}E\Big\{[K(\mathbf{x}, \mathbf{x}') - E \{K(\mathbf{x}, \mathbf{X}_4) \mid Y_4(t)=1\} \\
& \quad - \ E \{K(\mathbf{x}', \mathbf{X}_3) \mid Y_3(t)=1\} + E \{K(\mathbf{X}_2, \mathbf{X}_4) \mid Y_2(t)=1, Y_4(t)=1\}] \\
& \quad \times [\tilde{\delta} W_h(y-t)I(Y_3\geq t)- \delta_3 W_h(Y_3-t)I(y\geq t)] \\
& \quad \times[\tilde{\delta}' W_h(y'-t)I(Y_4\geq t)- \delta_4 W_h(Y_4-t)I(y'\geq t)] \Big\} \\
&= \frac{1}{6}  U(\mathbf{x}, \mathbf{x}^{\prime};t) V(y, \tilde{\delta}; y^{\prime}, \tilde{\delta}^{\prime};t).
\end{align*}
The proof is completed.\qedhere
\end{proof}

\begin{proof}[\textbf{Proof of Theorem~\ref{Theorem-asymptotic-null}.}]
Note that
\begin{eqnarray}\label{discomposition}
 &&n h^{1/ 2}\widehat{\mathrm{SID}}_{K}(T,\mathbf{X}) \nonumber\\
&=&n h^{1/ 2}\widehat{\mathrm{SID}}_{K}^\nu(T,\mathbf{X})
 +
n h^{1/ 2}\left(\begin{array}{l}
n \\ 4
\end{array}\right)^{-1} \sum_{i<j<k<l}\int^\tau_0 h_{n}(\mathbf{Z}_{i},  \mathbf{Z}_{j}, \mathbf{Z}_{k}, \mathbf{Z}_{l};t)d [\overline{N}(t)-\nu(t)], \nonumber\\
\end{eqnarray}
where $\operatorname{\widehat{SID}}_{K}^\nu (T,\mathbf{X})$ and $h_{n}(\mathbf{Z}_{i}, \mathbf{Z}_{j}, \mathbf{Z}_{k}, \mathbf{Z}_{l};t)$ are defined in~\eqref{sid-k-nu} and \eqref{sid-k-kernel}.

We next use Lemma~B.4 in \citet{Li96Fan} to establish the null limiting distribution of $nh^{1/ 2} \operatorname{\widehat{SID}}^\nu_{K}(T,\mathbf{X})$. Note that
\begin{align*}
\operatorname{\widehat{SID}}_{K}^\nu(T,\mathbf{X}) &=
\begin{pmatrix} n \\ 4 \end{pmatrix}^{-1}
\!\!\!\!\!\!\sum_{i<j<k<l}\int^\tau_0 h_{n}(\mathbf{Z}_{i}, \mathbf{Z}_{j}, \mathbf{Z}_{k}, \mathbf{Z}_{l};t) \dx{}\nu(t)  \triangleq
\begin{pmatrix} n \\ 4 \end{pmatrix}^{-1}
\!\!\!\!\!\!\sum_{i<j<k<l} H_{n}(\mathbf{Z}_{i}, \mathbf{Z}_{j}, \mathbf{Z}_{k}, \mathbf{Z}_{l}),
\end{align*}
where
$H_{n}(\mathbf{Z}_{i}, \mathbf{Z}_{j}, \mathbf{Z}_{k}, \mathbf{Z}_{l})
=\int^\tau_0 h_{n}(\mathbf{Z}_{i}, \mathbf{Z}_{j}, \mathbf{Z}_{k}, \mathbf{Z}_{l};t) \dx{}\nu(t).$
Define the $c$-order projection as
\begin{equation*}
{H}_{n c}(z_{1},\ldots, z_{c}) = E\{ {H}_{n}(\mathbf{Z}_{1},\mathbf{Z}_{2},\mathbf{Z}_{3},\mathbf{Z}_{4})  \mid  \mathbf{Z}_{1}=z_{1}, \ldots, \mathbf{Z}_{c}=z_{c}\},
\end{equation*}
for $c=1,2,3,4.$ Let $\sigma^2_{nc}=\operatorname{Var}(H_{n c} (\mathbf{Z}_{1}, \ldots, \mathbf{Z}_{c} )).$

By Lemma~B.4 in \citet{Li96Fan}, we need to verify
the conditions: \\
\textbf{Condition 1:} Under   $H_0$,
 ${H}_{n 1}(z_{1})=0$ and $E\{H^2_n(\mathbf{Z}_{1},\mathbf{Z}_{2},\mathbf{Z}_{3},\mathbf{Z}_{4})\}<\infty.$\\
\textbf{Condition 2:} As $n \rightarrow \infty$,
$$
\frac{E\{{G}_{n}^{2}(\mathbf{Z}_{1}, \mathbf{Z}_{2})\}+n^{-1} E\{H_{n 2}^{4}(\mathbf{Z}_{1}, \mathbf{Z}_{2})\}}{E^{2}\{H_{n 2}^{2} (\mathbf{Z}_{1}, \mathbf{Z}_{2} )\}} \rightarrow 0,
$$
where
$
{G}_{n}(\mathbf{Z}_{1}, \mathbf{Z}_{2})=E\{H_{n 2}(\mathbf{Z}_{1}, \mathbf{Z}_{3}) H_{n 2}(\mathbf{Z}_{2}, \mathbf{Z}_{3})  \mid  \mathbf{Z}_{1}, \mathbf{Z}_{2}\}.
$\\
\textbf{Condition 3:} As $n \rightarrow \infty$, $\sigma^2_{nc}/\sigma^2_{n2}=o(n^{(c-2)}),$ for $c=3,4.$

Under the above Condition~1--3, the proof of Lemma~B.4 in \citet{Li96Fan} suggests that
$\operatorname{\widehat{SID}}^{\nu}_{K}(T,\mathbf{X})$ can be approximated by the $U$-statistic based on the second-order projection. That is,
\begin{equation}
\begin{aligned}
n \operatorname{\widehat{SID}}^{\nu}_{K}(T,\mathbf{X}) &= n
\begin{pmatrix} 4 \\ 2 \end{pmatrix}
\begin{pmatrix} n \\ 2 \end{pmatrix}^{-1}
\sum_{i< j}  {H}_{n 2}(\mathbf{Z}_i,\mathbf{Z}_j)[1 +o_p(1)]  \stackrel{d}{\rightarrow} N\left(0, \frac{4^{2}(4-1)^{2}}{2} \sigma^2_{n2}\right).
\end{aligned}
\end{equation}
We verify Condition~1--3 via the following steps.

\textbf{Step 1: Verify Condition 1.}
By Lemma~\ref{Hn2-degenation}, we have that
\begin{equation*}
{H}_{n 1}(z_{1}) = \int^\tau_0 E\{{h}_{n}(\mathbf{Z}_{1},\mathbf{Z}_{2},\mathbf{Z}_{3},\mathbf{Z}_{4};t)  \mid  \mathbf{Z}_{1}=\mathbf{z}_{1}\} \dx{}\nu(t) = \int^\tau_0h_{n1}(\mathbf{z}_{1};t)\nu(t)=0.
 \end{equation*}
Since $K(\mathbf{x}_{1},\mathbf{x}_{2})$ and $W(u)$ are bounded, we  obtain that
$
E\{H^2_n(\mathbf{Z}_{1}, \ldots,\mathbf{Z}_{4})\}<\infty.
$

\textbf{Step 2: Verify Condition 2.}
By Lemma~\ref{var-degenation}, we have that
\begin{eqnarray}\label{Exp-Hn2-1}
&&E\{H_{n 2}^{2} (\mathbf{Z}_{1}, \mathbf{Z}_{2} )\}\nonumber\\
         &=& \frac{1}{36 h}\iint U^2(\mathbf{x}_{1}, \mathbf{x}_{2};s) S^4(s) f_{\mathbf{X},Y,\delta}(\mathbf{x}_{1},s,1)f_{\mathbf{X},Y,\delta}(\mathbf{x}_{2},s,1)  f_{Y,\delta}(s,1) \dx{}\mathbf{x}_1 \dx{}\mathbf{x}_2 \dx{}\nu(s) \nonumber\\
         &&\quad \times \iint W(a)W(a+b) W(c)W(c+ b) \dx{}a \dx{}b  \dx{}c
         +O(1).
\end{eqnarray}
 By Lemma~\ref{Gn-degenation}, we have that
\begin{equation}\label{Exp-Hn2-2}
 E\{H_{n 2}^{4}(\mathbf{Z}_{1}, \mathbf{Z}_{2})\}=O(h^{-3}), \quad
 E\{{G}_{n}^{2}(\mathbf{Z}_{1}, \mathbf{Z}_{2})\}=O(h^{-1}).
\end{equation}
By
\eqref{Exp-Hn2-1} and~\eqref{Exp-Hn2-2}, we have that
 $$
\frac{E\{{G}_{n}^{2}(\mathbf{Z}_{1}, \mathbf{Z}_{2})\}+n^{-1} E\{H_{n 2}^{4}(\mathbf{Z}_{1}, \mathbf{Z}_{2})\}}{E^{2}\{H_{n 2}^{2} (\mathbf{Z}_{1}, \mathbf{Z}_{2} )\}} \rightarrow 0,
$$
as $nh\rightarrow\infty$.

\textbf{Step 3: Verify Condition 3.}
By Lemma~\ref{Hn3-degenation} and the argument of Lemma~\ref{var-degenation}, we can easily show that
\[
\sigma^2_{n3}=O(1/h^2), \quad \sigma^2_{n4}=O(1/h^3).
\]
Then, we have that $\sigma^2_{n3}/\sigma^2_{n2}=O(1/h)$ and $\sigma^2_{n4}/\sigma^2_{n2}=O(1/h^2),$ which implies that
  $\sigma^2_{nc}/\sigma^2_{n2}=o(n^{(c-2)}),$
as $nh \rightarrow \infty$.

According to Lemma~B.4 in \citet{Li96Fan}, it follows that
\begin{equation}\label{asymptotic-null-oracal}
nh^{1/ 2} \operatorname{\widehat{SID}}^{\nu}_{K}(T,\mathbf{X}) \stackrel{d}{\rightarrow} N\left(0, 2 \sigma^2\right),
\end{equation}
where
 \begin{eqnarray*}
 \sigma^2 &=& \iint U^2(x_{1}, x_{2};s) S^4(s) f_{\mathbf{X},Y,\delta}(x_{1},s,1)f_{\mathbf{X},Y,\delta}(x_{2},s,1)f_{Y,\delta}(s,1)dx_1dx_2  d\nu(s)\\
 &&\times
           \iint W(a)W(a+b) W(c)W(c+ b) dadb  dc.
\end{eqnarray*}

We next show that
\begin{equation*}
n h^{1/ 2}
\begin{pmatrix} n \\ 4 \end{pmatrix}^{-1}
\sum_{i<j<k<l}\int^\tau_0 h_{n}(\mathbf{Z}_{i}, \mathbf{Z}_{j}, \mathbf{Z}_{k}, \mathbf{Z}_{l};t) \dx [\overline{N}(t)-\nu(t)]=o_p(1).
\end{equation*}
By the theory of the empirical process, we have that
\begin{equation}\label{nt-diff-1}
 \sup_{t\in[0,\tau]} \left| \overline{N}(t)-\nu(t) \right| = O_p(1/\sqrt{n}).
\end{equation}
Note that
\begin{equation*}
 nh^{1/ 2}
\begin{pmatrix} n \\ 4 \end{pmatrix}^{-1}
\sum_{i<j<k<l} h_{n}(\mathbf{Z}_{i}, \mathbf{Z}_{j}, \mathbf{Z}_{k}, \mathbf{Z}_{l};t), \quad t\in[0,\tau),
\end{equation*}
is a degenerate $U$-process. By the maximal inequalities for degenerate
$U$-processes \citep{sherman1994maximal}, we have that
\begin{equation}\label{maximal-u-process}
\begin{pmatrix} n \\ 4 \end{pmatrix}^{-1}
\sum_{i<j<k<l} h_{n}(\mathbf{Z}_{i}, \mathbf{Z}_{j}, \mathbf{Z}_{k}, \mathbf{Z}_{l};t)=\frac{2}{n(n-1)}  \sum_{i< j} {h}_{n 2}(\mathbf{Z}_i,\mathbf{Z}_j;t)[1+o_p(1)],
\end{equation}
uniformly for $t\in[0,\tau)$. Let
$$
T_{n}(\mathbf{x},t)= \frac{1}{n}\sum_{ j=1}^n {U}(\mathbf{x}, \mathbf{X}_j;t) [ {\delta}_j  W_h(Y_j-t)S(t)-I(Y_j\geq t)F_h(t)].
$$
By a similar argument in Lemma~\ref{Hn2-degenation}, we obtain that
 $E\{T_{n}(\mathbf{x},t)\}=0$. Then, by the continuity argument used by \citet{mack1982weak} or Lemma~6.6 in \citet{10.1214/009053607000000352}, we have that
\begin{equation*}
T_{n}(\mathbf{x},t) =E\{T_{n}(\mathbf{x},t)\}+ O_p\left(  \sqrt{\frac{\ln (1 / h)}{n h}} \right)=  O_p\left(  \sqrt{\frac{\ln (1 / h)}{n h}} \right),
\end{equation*}
uniformly for $t\in[0,\tau)$ and $\mathbf{x}\in \mathbb{R}^p,$ and
\begin{multline*}
\begin{aligned}
\frac{1}{n^2}  \sum_{i, j=1}^n  {h}_{n 2}(\mathbf{Z}_i,\mathbf{Z}_j;t)
&= \frac{1}{n}\sum_{ i=1}^nT_{n}(\mathbf{X}_i,t) [ {\delta}_i W_h(Y_i-t)S(t)-I(Y_i\geq t)F_h(t)] \\
  &= O_p\left(  \sqrt{\frac{\ln (1 / h)}{n h}} \right) \frac{1}{n}\sum_{ i=1}^n [ {\delta}_i W_h(Y_i-t)S(t)-I(Y_i\geq t)F_h(t)] \\
  &= O_p\left(  \frac{\ln (1 / h)}{n h} \right).
\end{aligned}
\end{multline*}
By~\eqref{maximal-u-process} and the relation between $V$- and $U$-statistics, we have that
\begin{equation}\label{maximal-u-process-1}
\begin{aligned}
 \sup_{t\in[0,\tau]} \left| n h^{1/ 2} \begin{pmatrix} n \\ 4 \end{pmatrix}^{-1}\sum_{i<j<k<l} h_{n}(\mathbf{Z}_{i}, \mathbf{Z}_{j}, \mathbf{Z}_{k}, \mathbf{Z}_{l};t) \right| &= n h^{1/ 2} O_p\left( \frac{\ln (1 / h)}{n h} \right)  = O_p\left( \frac{\ln (1 / h)}{h^{1/ 2}} \right).
\end{aligned}
\end{equation}
By~\eqref{nt-diff-1} and~\eqref{maximal-u-process-1}, we have that
\begin{equation}\label{term-2}
\begin{aligned}
n h^{1/ 2} \begin{pmatrix} n \\ 4 \end{pmatrix}^{-1} \!\!\!\!\!\!\sum_{i<j<k<l}\int^\tau_0 h_{n}(\mathbf{Z}_{i}, \mathbf{Z}_{j}, \mathbf{Z}_{k}, \mathbf{Z}_{l};t) \dx{} [\overline{N}(t)-\nu(t)] &= O_p\left( \frac{\ln (1 / h)}{\sqrt{nh} } \right)  = o_p(1).
\end{aligned}
\end{equation}

Note that
\begin{multline*}
\begin{aligned}
 \sigma^2
&= C \dot\iint U^2(\mathbf{x}_{1}, \mathbf{x}_{2};s) S^4(s) \frac{f_{\mathbf{X},Y,\delta}(\mathbf{x}_{1},s,1)}{f_{Y,\delta}(s,1)}
         \frac{f_{\mathbf{X},Y,\delta}(\mathbf{x}_{2},s,1)}{f_{Y,\delta}(s,1)}f^4_{Y,\delta}(s,1) \dx{}\mathbf{x}_1 \dx{}\mathbf{x}_2 \dx{}s \\
&= C \times E\{E\{U^2(\mathbf{X}_{1}, \mathbf{X}_{2};Y)|Y,\delta=1\}S^4(Y)f^3_{Y,\delta}(Y,1)\}.
\end{aligned}
\end{multline*}
This, together with~\eqref{discomposition},~\eqref{asymptotic-null-oracal}, and~\eqref{term-2}, leads to
$
n h^{1/ 2} \operatorname{\widehat{SID}}_{K}(T,\mathbf{X}) \stackrel{d}{\rightarrow}  N\left(0, 2\sigma^2\right). $
\end{proof}

\begin{proof}[\textbf{Proof  of Theorem~\ref{Theorem-asymptotic-alternatives}.}]
By the proof of Theorem~\ref{Theorem-consistency}, we can show that
\begin{equation*}
\operatorname{Var}(\operatorname{\widehat{SID}}_{K}(T,\mathbf{X}) \mid H_{1})=O(1/n), \text{ as } nh\rightarrow \infty,
\end{equation*}
and thus
$\operatorname{\widehat{SID}}_{K}(T,\mathbf{X})
=E\{\operatorname{\widehat{SID}}_{K}(T,\mathbf{X})\}
+O_{p}(1/n),$ under $H_{1}.$
This implies that
\begin{eqnarray*}
nh^{1/ 2}\operatorname{\widehat{SID}}_{K}(T,\mathbf{X}) &=&  nh^{1/ 2}E\{\operatorname{\widehat{SID}}_{K}(T,\mathbf{X})\}
+O_{p}(h^{1/ 2})\\  &=&  nh^{1/ 2} \{ {\operatorname{SID}}_{K}(T,\mathbf{X})+h^2\}
+O_{p}(h^{1/ 2}).
\end{eqnarray*}
This completes the proof of the theorem.\qedhere
\end{proof}

\subsection{Proofs of results in Section~\ref{chapter-bootstrap}}

\begin{proof}[\textbf{Proof  of Theorem~\ref{Theorem-asymptotic-boostrap}.}]
We first need to show that
\begin{equation}\label{SID-nu-app}
\operatorname{\widehat{SID}}^*_{K}(T,\mathbf{X})= \frac{12}{n(n-1)}  \sum_{i< j}e_i e_j \int^\tau_0  {h}_{n 2}(\mathbf{Z}_i,\mathbf{Z}_j;t) \dx{}\nu(t)+o_{p}^* (1/ \{n h^{1/ 2}\}).
 \end{equation}

Note that
\begin{eqnarray*}
&& \widehat{\mathrm{SID}}^*_{K}(T,\mathbf{X})-\frac{12}{n(n-1)}  \sum_{i< j}e_i e_j \int^\tau_0  {h}_{n 2}(\mathbf{Z}_i,\mathbf{Z}_j;t)d\nu(t)\\
&=&\frac{2}{n(n-1)}  \sum_{i< j}e_i e_j \int^\tau_0  \widehat{ U}(\mathbf{X}_i,  \mathbf{X}_j;t) \widehat{V}(Y_i,  \delta_i; Y_j,  \delta_j;t)d [\overline{N}(t)-\nu(t)]\\
  &&
 +\frac{2}{n(n-1)}  \sum_{i< j}e_i e_j \int^\tau_0 \Big[ \widehat{ U}(\mathbf{X}_i,  \mathbf{X}_j;t) \widehat{V}(Y_i,  \delta_i; Y_j,  \delta_j;t)-6{h}_{n 2}(\mathbf{Z}_i,\mathbf{Z}_j;t)
 \Big]d  \nu(t) \\
  &\triangleq&D_{n1}+D_{n2}.
 \end{eqnarray*}
We need to study the orders of $n h^{1/ 2}D_{n1}$ and $n h^{1/ 2}D_{n2}$.

Consider the term $n h^{1/ 2}D_{n1}$.
By the theory of the empirical process, we have that
 $$\sup_{t\in[0,\tau]} \left| \overline{N}(t)-\nu(t) \right| = O_p(1/\sqrt{n}).$$ This, together with the boundedness of $K(\mathbf{x}_{1},\mathbf{x}_{2})$ and $W(u)$, yields that
\begin{multline}\label{D-n1-var}
\begin{aligned}
&\operatorname{Var}^*(n h^{1/ 2}D_{n1})\\
 &= \frac{4n^2 h}{((n)_2)^2}\sum_{i< j}\Big[ \int^\tau_0 \widehat{ U}(\mathbf{X}_i, \mathbf{X}_j;t) \widehat{V}(Y_i, \delta_i; Y_j, \delta_j;t) \dx [\overline{N}(t)-\nu(t)]\Big]^2 \\
 &= \frac{4n^2 h}{((n)_2)^2} \sum_{i< j} \iint  \widehat{ U}(\mathbf{X}_i, \mathbf{X}_j;t) \widehat{V}(Y_i, \delta_i; Y_j, \delta_j;t) \\
 & \quad \times \widehat{ U}(\mathbf{X}_i, \mathbf{X}_j;s) \widehat{V}(Y_i, \delta_i; Y_j, \delta_j;s) \dx [\overline{N}(s)-\nu(s)] \dx [\overline{N}(t)-\nu(t)] \\
 &= O_p\Big(\frac{1}{nh}\Big).
\end{aligned}
\end{multline}

Consider the term $n h^{1/ 2}D_{n2}$. From the Cauchy--Schwarz inequality, we have that
 \begin{eqnarray}\label{var-Dn-2}
 \operatorname{Var}^*(n h^{1/ 2}D_{n2})
&=&\frac{4n^2 h}{((n)_2)^2}\sum_{i< j}\Big[\int^\tau_0 [ \widehat{ U}(\mathbf{X}_i,  \mathbf{X}_j;t) \widehat{V}(Y_i,  \delta_i; Y_j,  \delta_j;t)-6{h}_{n 2}(\mathbf{Z}_i,\mathbf{Z}_j;t)
  ]d  \nu(t)\Big]^2\nonumber\\
 &\leq &  \frac{4n^2 h}{((n)_2)^2}\sum_{i< j} \int^\tau_0 [ \widehat{ U}(\mathbf{X}_i,  \mathbf{X}_j;t) \widehat{V}(Y_i,  \delta_i; Y_j,  \delta_j;t)-6{h}_{n 2}(\mathbf{Z}_i,\mathbf{Z}_j;t)
  ]^2d  \nu(t).  \nonumber\\
 \end{eqnarray}
By Lemma~\ref{Hn2-degenation} and the rate of uniform convergence for kernel estimation in \citet{mack1982weak}, we have that
 \begin{eqnarray*}
\sup_{t\in [0,\tau)}\frac{1}{ (n)_2 }\sum_{i< j} [ \widehat{ U}(\mathbf{X}_i,  \mathbf{X}_j;t) \widehat{V}(Y_i,  \delta_i; Y_j,  \delta_j;t)-6{h}_{n 2}(\mathbf{Z}_i,\mathbf{Z}_j;t)
  ]^2 = O_p\left(  \sqrt{\frac{\ln (1 / h)}{n h}}   \right)=o_p( 1).
 \end{eqnarray*}
This, together with~\eqref{var-Dn-2}, leads to
\begin{equation}\label{var-Dn-2-1}
 \operatorname{Var}^*(n h^{1/ 2}D_{n2})=o_p(1).
\end{equation}
With~\eqref{D-n1-var} and~\eqref{var-Dn-2-1}, we prove~\eqref{SID-nu-app}.

With~\eqref{SID-nu-app}, we need to obtain the limiting distribution of the following statistic
\begin{equation*}
\mathcal{H}_{n}^{*(2)}\triangleq
\frac{12}{n(n-1)}    \sum_{i< j}e_i e_j \int^\tau_0  {h}_{n 2}(\mathbf{Z}_i,\mathbf{Z}_j;t) \dx{}\nu(t).
\end{equation*}
The limiting distribution of $\mathcal{H}_{n}^{*(2)}$ can be established by Theorem 1 of \citet{hall1984central}. Specifically,
by similar arguments in the proof of Theorem~\ref{Theorem-asymptotic-null}, we can verify that all the conditions of Theorem 1 of \citet{hall1984central} are satisfied with respective to $\{e_i\}_{i=1}^n$.
In addition, by Lemma~\ref{var-degenation}, we have that
\begin{align*}
 \operatorname{Var}^*(\mathcal{H}_{n}^{*(2)}) &= \frac{12^2}{n^2(n-1)^2} \sum_{i< j}  \Big[\int^\tau_0{h}_{n 2}(\mathbf{Z}_i,\mathbf{Z}_j;t) \dx{}\nu(t)\Big]^2\\
&= \frac{72}{n(n-1)}  \operatorname{Var}\Big( \int^\tau_0{h}_{n 2}(\mathbf{Z}_i,\mathbf{Z}_j;t) \dx{}\nu(t)\Big)[1+o_p(1)] \\
&= \frac{72}{n(n-1)}\frac{1}{36h}\sigma^{2}[1+o_p(1)] \\
&= \frac{2}{n(n-1)}\frac{1}{h}\sigma^{2}[1+o_p(1)].
\end{align*}
This, together with Theorem 1 of \citet{hall1984central}, yields that
\begin{equation*}
n h^{1/ 2}\frac{12}{n(n-1)}      \sum_{i< j}e_i e_j \int^\tau_0  {h}_{n 2}(\mathbf{Z}_i,\mathbf{Z}_j;t) \dx{}\nu(t)\stackrel{d^*}{\rightarrow} N\left(0,2\sigma^{2}\right),
\end{equation*}
 which immediately proves the theorem.\qedhere
\end{proof}

\begin{proof}[\textbf{Proof of Theorem~\ref{Theorem-power-boostrap}.}]
We first need to bound the critical value $\widehat{Q}_{n,(1-\alpha)}^*$ above.
By Markov's inequality, we have that
\begin{equation}\label{second-term}
\begin{aligned}
{P} \left\{ \widehat{Q}_{n,(1-\alpha)}^*>\gamma \right\} &\leq
 {P}\Big\{ \cup_{b \in\{1, \ldots, B\}}\{\gamma< n h^{1/ 2} \operatorname{\widehat{SID}}^{*(b)}_{K}(T,\mathbf{X}) \}\Big\} \\
   &\leq B\times {P} \left\{  \gamma< n h^{1/ 2} \operatorname{\widehat{SID}}^{*}_{K}(T,\mathbf{X}) \right\} \\
   &\leq B\times{E \left\{ n^{2} h \operatorname{\widehat{SID}}^{*2}_{K}(T,\mathbf{X}) \right\} }/{\gamma^2},
\end{aligned}
\end{equation}
for any $\gamma>0$. From the Cauchy--Schwarz inequality, we obtain that
\begin{align*}
 E^* \left\{ n^{2} h \operatorname{\widehat{SID}}^{*2}_{K}(T,\mathbf{X}) \right\}
&= \frac{n^{2}h}{(n)^2_4}\sum_{i\neq j}^{n} \left[ \sum_{
  \stackrel{k\neq l}{k,l\neq\{i,j\}}}^{n}
 \int^\tau_0 h_{n}(\mathbf{Z}_{i}, \mathbf{Z}_{j}, \mathbf{Z}_{k}, \mathbf{Z}_{l};t) \dx \overline{N}(t) \right]^2 \\
 &\leq \frac{n^2 h}{((n)_4)^2} (n-2)(n-3)\sum_{i\neq j\neq k\neq l}^{n}  \iint h_{n}(\mathbf{Z}_{i}, \mathbf{Z}_{j}, \mathbf{Z}_{k}, \mathbf{Z}_{l};s)\nonumber\\
 & \quad \times h_{n}(\mathbf{Z}_{i}, \mathbf{Z}_{j}, \mathbf{Z}_{k}, \mathbf{Z}_{l};t) \dx \overline{N}(s) \dx \overline{N}(t) \\
 &= \frac{n }{n-1} T_n,
\end{align*}
where
 $$T_n =\frac{h}{ (n)_4 }\sum_{i\neq j\neq k\neq l}^{n}  \iint h_{n}(\mathbf{Z}_{i}, \mathbf{Z}_{j}, \mathbf{Z}_{k}, \mathbf{Z}_{l};s) h_{n}(\mathbf{Z}_{i}, \mathbf{Z}_{j}, \mathbf{Z}_{k}, \mathbf{Z}_{l};t) \dx \overline{N}(s) \dx \overline{N}(t).$$
By similar arguments in the proof of Lemma~\ref{var-degenation},
we have that $E\{T_n\}\leq C_0,$ where $C_0>0$ is a constant (independent of $h$). This, together with~\eqref{second-term}, indicates that
\begin{equation*}\label{Markov-inequality}
 {P} \left\{ \widehat{Q}_{n,(1-\alpha)}^*>\gamma \right\} \leq \frac{n }{n-1}B\times  E \{ T_n\}\leq B  \frac{ C'_0 }{\gamma^2},
\end{equation*}
for constant $C'_0$ [$>nC_0/(n-1)$]. Choosing $\gamma=\{B C'_0/\varepsilon\}^{1/ 2}$ for  $\varepsilon>0,$ we have that
 \begin{equation}\label{term2-power-3}
 {P} \left\{ \widehat{Q}_{n,(1-\alpha)}^*>\gamma \right\} \leq \varepsilon.
\end{equation}
Under $H_1$, we have that ${\operatorname{SID}}_{K}(T,\mathbf{X})>0$. By~\eqref{term2-power-3}, we have that, as $n h\rightarrow \infty,$
$$(n \sqrt{h} )^{-1} \widehat{Q}_{n,(1-\alpha)}<\frac{1}{2} {\operatorname{SID}}_{K}(T,\mathbf{X}),$$
with probability at least $1-\varepsilon$. By the proof of Theorem~\ref{Theorem-asymptotic-alternatives}, we obtain that
\[
\operatorname{Var} \left\{ \operatorname{\widehat{SID}}_{K}(T,\mathbf{X}) \mid H_1 \right\} = O(1/n).
\]
These imply that
\begin{eqnarray*}
&& \limsup _{n \rightarrow \infty}  {P}\{n h^{1/ 2}\widehat{\mathrm{SID}}_{K}(T,\mathbf{X}) < \widehat{Q}_{n,(1-\alpha)}^* | H_1 \} \nonumber\\
&\leq& \limsup _{n \rightarrow \infty}{P}\{n h^{1/ 2}\widehat{\mathrm{SID}}_{K}(T,\mathbf{X}) < \widehat{Q}_{n,(1-\alpha)}^*,\widehat{Q}_{n,(1-\alpha)}^*\leq\gamma| H_1 \}\nonumber\\&&+
\limsup _{n \rightarrow \infty}{P}\{n h^{1/ 2}\widehat{\mathrm{SID}}_{K}(T,\mathbf{X}) < \widehat{Q}_{n,(1-\alpha)}^*,\widehat{Q}_{n,(1-\alpha)}^*>\gamma| H_1  \}\nonumber\\
&=& \limsup _{n \rightarrow \infty}  {P}\Big\{
   {\mathrm{SID}}_{K}(T,\mathbf{X})- \widehat{\mathrm{SID}}_{K}(T,\mathbf{X})
   \geq{\mathrm{SID}}_{K}(T,\mathbf{X}) \! -\! (n \sqrt{h} )^{-1} \widehat{Q}_{n,(1-\alpha)}^*
     | H_1\Big\}
     \\&&+\limsup _{n \rightarrow \infty}{P}\{ \widehat{Q}_{n,(1-\alpha)}^*>\gamma| H_1  \}\nonumber\\
   &\leq&
   \limsup _{n \rightarrow \infty}  {P}\Big\{
   {\mathrm{SID}}_{K}(T,\mathbf{X}) - \widehat{\mathrm{SID}}_{K}(T,\mathbf{X})
   \geq \frac{1}{2} {\mathrm{SID}}_{K}(T,\mathbf{X})
     | H_1\Big\}  +\varepsilon\nonumber\\
     &\leq&\limsup _{n \rightarrow \infty}
     \frac{4\mathrm{Var}\{\widehat{\mathrm{SID}}_{K}(T,\mathbf{X})| H_1\}}
   { {\mathrm{SID}}^2_{K}(T,\mathbf{X})} +\varepsilon\nonumber\\
     &\leq&\lim _{m, n \rightarrow \infty}
     \frac{4 }{n{\mathrm{SID}}^2_{K}(T,\mathbf{X})}+\varepsilon\leq 2\varepsilon.
\end{eqnarray*}
Thus, we have that
$
\lim_{n \rightarrow \infty} {P} \left\{  h^{1/ 2}\operatorname{\widehat{SID}}_{K}(T,\mathbf{X})\geq \widehat{Q}_{n,(1-\alpha)}^* \mid H_1 \right\} = 1.
$
\end{proof}

\section{Lemmas for the proof of Theorem~\ref{Theorem-asymptotic-null} } \label{Appendix-B}
\renewcommand{\theequation}{B.\arabic{equation}}
\renewcommand{\thelemma}{B.\arabic{lemma}}

In the section, we provide the following lemmas to prove Theorem~\ref{Theorem-asymptotic-null}.

\begin{lemma}\label{Hn3-degenation}
Let $\mathbf{z}=(y, \tilde{\delta}, \mathbf{x})$, $\mathbf{z}'=(y', \tilde{\delta}', \mathbf{x}')$,
$\mathbf{z}''=(y'', \tilde{\delta}'', \mathbf{x}'')$, and $\mathbf{z}'''=(y''', \tilde{\delta}''', \mathbf{x}''')$.
Under $H_0$, we have that
\begin{align*}
{h}_{n 3}(\mathbf{z},\mathbf{z}',\mathbf{z}'';t) &= \frac{1}{12} \Big[ [ U(\mathbf{x}, \mathbf{x}')-U(\mathbf{x}', \mathbf{x}'')] V_1(y,\tilde{\delta}; y'', \tilde{\delta}'';t)
V_0(y',\tilde{\delta}'; t) \\
  & \quad + \ [ U(\mathbf{x}, \mathbf{x}')-U(\mathbf{x}, \mathbf{x}'')]V_1(y',\tilde{\delta}'; y'', \tilde{\delta}'';t) V_0(y,\tilde{\delta}; t) \\
  & \quad + \ [ U(\mathbf{x}, \mathbf{x}')-U(\mathbf{x}', \mathbf{x}'')+E \{K(\mathbf{x}', \mathbf{X}_4) \mid Y_4(t)=1\} \\
  & \quad - \ E \{K(\mathbf{x}'',\mathbf{X}_4) \mid Y_4(t)=1\}]  V_1(y,\tilde{\delta}; y', \tilde{\delta}';t)V_0(y'',\tilde{\delta}'';t)\Big]
\end{align*}
and
\begin{align*}
&{h}_{n 4}(\mathbf{z},\mathbf{z}',\mathbf{z}'',\mathbf{z}''';t)\\
&= \frac{1}{12}\Big[  [ K(\mathbf{x}, \mathbf{x}')- K(\mathbf{x}, \mathbf{x}''')- K(\mathbf{x}', \mathbf{x}'') + K(\mathbf{x}'', \mathbf{x}''')]  V_1(y,\tilde{\delta}; y'', \tilde{\delta}'';t) V_1(y',\tilde{\delta}'; y''', \tilde{\delta}''';t)
  \\
& \quad + \ [ K(\mathbf{x}, \mathbf{x}')- K(\mathbf{x}',\mathbf{x}''')- K(\mathbf{x}, \mathbf{x}'')+ K(\mathbf{x}'', \mathbf{x}''')]  V_1(y,\tilde{\delta}; y''', \tilde{\delta}''';t) V_1(y',\tilde{\delta}'; y'', \tilde{\delta}'';t)
 \\
& \quad + \ [ K(\mathbf{x}, \mathbf{x}'')- K(\mathbf{x}, \mathbf{x}''')- K(\mathbf{x}', \mathbf{x}'')+ K(\mathbf{x}', \mathbf{x}''')]  V_1(y,\tilde{\delta}; y', \tilde{\delta}';t) V_1(y'',\tilde{\delta}''; y''', \tilde{\delta}''';t)
  \Big],
\end{align*}
where
$V_0(y,\tilde{\delta};t)=
 [\tilde{\delta} W_h(y-t)S(t)-I(y\geq t)F_h(t)]$ and $
 V_1(y,\tilde{\delta}; y^{\prime}, \tilde{\delta}^{\prime};t)=
 [\tilde{\delta} W_h(y-t)I(y'\geq t)- \tilde{\delta}' W_h(y'-t)I(y\geq t)]$.
\end{lemma}
\begin{proof}
By~\eqref{symmetric-kernel-detail} and the definition of ${h}_{n 3}(\mathbf{z},\mathbf{z}',\mathbf{z}'';t)$, some calculations  can yield  this lemma.\qedhere
\end{proof}

\begin{lemma}\label{var-degenation}
Under $H_0$, we have that
\begin{eqnarray*}
\operatorname{Var}(H_{n 2}(\mathbf{Z}_{1}, \mathbf{Z}_{2})) &=&
\frac{1}{36h}\iint U^2(\mathbf{x}_{1}, \mathbf{x}_{2};s) S^4(s) f_{\mathbf{X},Y,\delta}(\mathbf{x}_{1},s,1)f_{\mathbf{X},Y,\delta}(\mathbf{x}_{2},s,1)f_{Y,\delta}(s,1) \\
&&\times \dx{}\mathbf{x}_1 \dx{}\mathbf{x}_2 \dx{}\nu(s) \iint W(a)W(a+b) W(c)W(c+ b) \dx{}a \dx{}b \dx{}c+O(1).
\end{eqnarray*}
\end{lemma}

\begin{proof}
Note that $\operatorname{Var}(H_{n 2}(\mathbf{Z}_{1}, \mathbf{Z}_{2}))=E\{H_{n 2}^{2}(\mathbf{Z}_{1}, \mathbf{Z}_{2})\}.$  After some algebra, we have
\begin{eqnarray*}\label{Exp-Hn2}
&&E\{H_{n 2}^{2}(\mathbf{Z}_{1}, \mathbf{Z}_{2})\}\\
 &=& \frac{1}{36} E\Big\{ \Big[\int^\tau_0  U(\mathbf{X}_{1}, \mathbf{X}_{2};t) V(Y_1, \delta_1; Y_2, \delta_2;t) \dx{}\nu(t)\Big]^2\Big\} \\
&=& \frac{1}{36} E\Big\{ \iint U(\mathbf{X}_{1}, \mathbf{X}_{2};s)  U(\mathbf{X}_{1}, \mathbf{X}_{2};t)V(Y_1, \delta_1; Y_2, \delta_2;s)  V(Y_1, \delta_1; Y_2, \delta_2;t) \dx{}\nu(s) \dx{}\nu(t) \Big\} \\
&=& \frac{1}{36}  \iint U(\mathbf{x}_{1}, \mathbf{x}_{2};s)  U(\mathbf{x}_{1}, \mathbf{x}_{2};t) [ W_h(y_1-t)W_h(y_2-t)S^2(t)-2W_h(y_1-t)I(y_2\geq t)F_h(t) \\
&& \quad  + \ I(y_1\geq t)I(y_2\geq t)F^2_h(t)]  [ W_h(y_1-s)W_h(y_2-s)S^2(s)-2W_h(y_1-s)I(y_2\geq s)F_h(s) \\
&& \quad  + \ I(y_1\geq s)I(y_2\geq s)F^2_h(s)] f_{\mathbf{X},Y,\delta}(\mathbf{x}_{1},y_1,1)f_{\mathbf{X},Y,\delta}(\mathbf{x}_{2},y_2,1) \dx{}\mathbf{x}_1 \dx{}\mathbf{x}_2 \dx{}y_1 \dx{}y_2 \dx{}\nu(s) \dx{}\nu(t) \\
&\triangleq& \frac{1}{36} [E_1-2E_2+E_3-2E_4+4E_5-2E_6+E_7-2E_8+E_9],
\end{eqnarray*}
  where
  \begin{align*}
 E_1 &= \iint U(\mathbf{x}_{1}, \mathbf{x}_{2};s)  U(\mathbf{x}_{1}, \mathbf{x}_{2};t) W_h(y_1-t) W_h(y_2-t)S^2(t)W_h(y_1-s)W_h(y_2-s)S^2(s) \\
 & \quad \times f_{\mathbf{X},Y,\delta}(\mathbf{x}_{1},y_1,1)f_{\mathbf{X},Y,\delta}(\mathbf{x}_{2},y_2,1) \dx{}\mathbf{x}_1 \dx{}\mathbf{x}_2 \dx{}y_1 \dx{}y_2 \dx{}\nu(s) \dx{}\nu(t),
 \end{align*}
and  $E_j, j=2\ldots,9$ are defined similarly.

Using the change of variables and $\nu(t)=P\{Y\leq t,\delta=1 \}$, we have that
\begin{align*}
E_1
       &= \frac{1}{h}\iint U^2(\mathbf{x}_{1}, \mathbf{x}_{2};s) S^4(s) f_{\mathbf{X},Y,\delta}(\mathbf{x}_{1},s,1)f_{\mathbf{X},Y,\delta}(\mathbf{x}_{2},s,1)f_{Y,\delta}(s,1)\dx{}\mathbf{x}_1 \dx{}\mathbf{x}_2  \dx{}\nu(s) \\
       & \quad \times \iint W(a)W(a+b) W(c)W(c+ b) \dx{}a \dx{}b  \dx{}c
         +O(h)
\end{align*}
and
\begin{align*}
E_2    &= \iint U^2(\mathbf{x}_{1}, \mathbf{x}_{2};s)W(a)W(a+b)S^2(s)W(c)I(y_2\geq s)
f_{\mathbf{X},Y,\delta}(\mathbf{x}_{1},s,1) \\
   & \quad \times f_{\mathbf{X},Y,\delta}(\mathbf{x}_{2},s,1)f^2_{Y,\delta}(s,1) \dx{}\mathbf{x}_1 \dx{}\mathbf{x}_2 \dx{}a \dx{}b \dx{}\nu(s) \dx{}c.
\end{align*}
Thus, we have that $E_1=O(1/h)$ and $E_2=O(1)$. By the same argument for $E_2$, we obtain that $E_j=O(1),$ $j=3,\ldots, 8.$ This immediately proves this lemma.\qedhere
\end{proof}

\begin{lemma}\label{Gn-degenation}
Under $H_0$, we have that
\begin{equation*}
E\{H_{n 2}^{4}(\mathbf{Z}_{1}, \mathbf{Z}_{2})\}=O(h^{-3}), \quad
E\{{G}_{n}^{2}(\mathbf{Z}_{1}, \mathbf{Z}_{2})\}=O(h^{-1}).
\end{equation*}
\end{lemma}

\begin{proof}
We first show that $E\{H_{n 2}^{4}(\mathbf{Z}_{1}, \mathbf{Z}_{2})\}=O(h^{-3}).$
By Lemma~\ref{Hn2-degenation}, we have that
\begin{multline*}
E\{H_{n 2}^{4}(\mathbf{Z}_{1}, \mathbf{Z}_{2})\} \\
\begin{aligned}
&= \frac{1}{6^4} E\Big\{ \Big[\int^\tau_0  U(\mathbf{X}_{1}, \mathbf{X}_{2};t) V(Y_1, \delta_1; Y_2, \delta_2;t) \dx{}\nu(t)\Big]^4\Big\} \\
&= \frac{1}{6^4}\iint U(\mathbf{x}_{1}, \mathbf{x}_{2};t_1)U(\mathbf{x}_{1}, \mathbf{x}_{2};t_2)U(\mathbf{x}_{1}, \mathbf{x}_{2};t_3)U(\mathbf{x}_{1}, \mathbf{x}_{2};t_4) \\
& \quad \times [W_h(y_1-t_1)S(t_1)-I(y_1\geq t_1)F_h(t_1)][W_h(y_2-t_1)S(t)-I(y_2\geq t_1)F_h(t_1)] \\
& \quad \times [W_h(y_1-t_2)S(t_2)-I(y_1\geq t_2)F_h(t_2)][W_h(y_2-t_2)S(t)-I(y_2\geq t_2)F_h(t_2)] \\
& \quad \times [W_h(y_1-t_3)S(t_3)-I(y_1\geq t_3)F_h(t_3)][W_h(y_2-t_3)S(t)-I(y_2\geq t_3)F_h(t_3)] \\
& \quad \times [W_h(y_1-t_4)S(t_4)-I(y_1\geq t_4)F_h(t_4)][W_h(y_2-t_4)S(t)-I(y_2\geq t_4)F_h(t_4)] \\
& \quad \times f_{\mathbf{X},Y,\delta}(\mathbf{x}_{1},y_1,1)f_{\mathbf{X},Y,\delta}(\mathbf{x}_{2},y_2,1) \dx{}\mathbf{x}_1 \dx{}\mathbf{x}_2 \dx{}y_1 \dx{}y_2 \dx{}\nu(t_1) \dx{}\nu(t_2) \dx{}\nu(t_3) \dx{}\nu(t_4) \\
& = O(h^{-3}).
\end{aligned}
\end{multline*}

We next calculate the order of $E\{{G}_{n}^{2}(\mathbf{Z}_{1}, \mathbf{Z}_{2})\}.$
By Lemma~\ref{Hn2-degenation}, we have that
\begin{align*}
{G}_{n}( \mathbf{z}, \mathbf{z}') &= E\{H_{n 2}(\mathbf{z}, \mathbf{Z}_{3}) H_{n 2}( \mathbf{z}', \mathbf{Z}_{3})  \} \\
 &= \frac{1}{6^2} E\Big\{ \iint U(\mathbf{x}, \mathbf{X}_{3};s)  U(\mathbf{x}', \mathbf{X}_{3};t)V(y, \tilde{\delta}; Y_3, \delta_{3};s) V(y', \tilde{\delta}'; Y_{3}, \delta_{3};t)  \dx{}\nu(s) \dx{}\nu(t) \Big\},
\end{align*}
where $\mathbf{z}=(y, \tilde{\delta}, \mathbf{x})$ and $\mathbf{z}'=(y', \tilde{\delta}', \mathbf{x}')$.
Then, we have that
\begin{align*}
 &{G}_{n}^{2}(\mathbf{Z}_{1}, \mathbf{Z}_{2})\\
  &=  \frac{1}{6^4}
 E_{\mathbf{Z}_{3},\mathbf{Z}_{4} |\mathbf{Z}_{1},\mathbf{Z}_{2}}\Big\{\!\!\iint\!\!  U\!(\mathbf{X}_{1}, \mathbf{X}_{3};s)  U\!(\mathbf{X}_{2}, \mathbf{X}_{3};t)  V\!(Y_1, {\delta}_1; Y_3, \delta_{3};s) V(Y_2, {\delta}_2; Y_{3}, \delta_{3};t) \dx{}\nu(s) \dx{}\nu(t) \\
& \quad \times \iint U(\mathbf{X}_{1}, \mathbf{X}_{4};s)  U(\mathbf{X}_{2}, \mathbf{X}_{4};t)  V(Y_1, {\delta}_1; Y_4, \delta_{4};s) V(Y_2, {\delta}_2; Y_{4}, \delta_{4};t) \dx{}\nu(s) \dx{}\nu(t)  \Big\} \\
 &= \frac{1}{6^4} E_{\mathbf{Z}_{3},\mathbf{Z}_{4} |\mathbf{Z}_{1},\mathbf{Z}_{2}} \Big\{ \iint U(\mathbf{X}_{1}, \mathbf{X}_{3};s)  U(\mathbf{X}_{2}, \mathbf{X}_{3};t)U(\mathbf{X}_{1}, \mathbf{X}_{4};u)   U(\mathbf{X}_{2}, \mathbf{X}_{4};v) V(Y_1, {\delta}_1; Y_3, \delta_{3};s) \\
& \quad \times V(Y_2, {\delta}_2; Y_{3}, \delta_{3};t)V(Y_1, {\delta}_1; Y_4, \delta_{4};u) V(Y_2, {\delta}_2; Y_{4}, \delta_{4};v) \dx{}\nu(s) \dx{}\nu(t) \dx{}\nu(u) \dx{}\nu(v)  \Big\}
\end{align*}
and thus
\begin{multline*}
\begin{aligned}
E\{{G}_{n}^{2}(\mathbf{Z}_{1}, \mathbf{Z}_{2})\}
   &= \frac{1}{6^4} E \left\{  \iint U(\mathbf{X}_{1}, \mathbf{X}_{3};s)  U(\mathbf{X}_{2}, \mathbf{X}_{3};t)U(\mathbf{X}_{1}, \mathbf{X}_{4};u)  U(\mathbf{X}_{2}, \mathbf{X}_{4};v) \right. \\
  & \quad \times V(Y_1, {\delta}_1; Y_3, \delta_{3};s)  V(Y_2, {\delta}_2; Y_{3}, \delta_{3};t) V(Y_1, {\delta}_1; Y_4, \delta_{4};u) \\
  & \quad \times V(Y_2, {\delta}_2; Y_{4}, \delta_{4};v) \dx{}\nu(s) \dx{}\nu(t) \dx{}\nu(u) \dx{}\nu(v)  \Big\} = O(h^{-1}).
\end{aligned}
\end{multline*}
The proof is completed.\qedhere
\end{proof}

\section{Additional  simulation studies} \label{Appendix-C}
 \renewcommand{\theexample}{C.\arabic{example}}
 \renewcommand{\thetable}{C.\arabic{table}}
 \renewcommand{\thefigure }{C.\arabic{figure}}

In this section, we further investigate the performance of the proposed SID method under different types of covariates and in high-dimensional settings, as illustrated  in     Examples~\ref{exampl_7}-\ref{exampl_9}.  Moreover, we   evaluate the impact of the parameter  \(\beta\) on   \(\operatorname{SID}_\beta\) in Example~\ref{exampl_10}.

\begin{example}\label{exampl_7}(Different types of covariates)
This example evaluates the type-I error rates of the proposed method when covariates are discrete or heavy-tailed. The survival time $T$ and censoring time $C$ are generated as follows:
  $$ T  \sim\operatorname{Exp}(\lambda)
  ~~\text{  and  } ~~  C\sim\operatorname{Exp}( e^{X / 3}).$$
The covariates are set in the following two cases:
\[ \text{Case 1: } X \sim \text{Poisson}(3); \qquad \text{Case 2: } X \sim t(3), \]
where $\text{Poisson}(3)$ is the Poisson distribution with a parameter of 3, and $t(3)$ is the $t$-distribution with 3 degrees of freedom.
  \end{example}

Table \ref{tabapp1}  summarizes the empirical type-I error rates for various methods under   30\% censoring with $n=50$ and 100.   These results  demonstrate the KLR and our  $\operatorname{SID}$ methods maintain type-I error rates close to the nominal significance levels $\alpha = 0.01, 0.05$, and $0.10$. In contrast, the IPCW   shows  inflated empirical type-I error rates. For instance, when $\alpha = 0.05$ and $X \sim \text{Poisson}(3)$, the IPCW method exhibits  error rates of 0.066 for $n=50$ and 0.088 for $n=150$.  These results are   consistent with those observed in Example  \ref{exampl_1}.

 \begin{table}[http]
			\caption{Empirical type-I error rate under 30\% censoring for Example~\ref{exampl_7}.}
			\label{tabapp1}\scriptsize
			\resizebox{\linewidth}{!}{%
				\begin{tabular}{cccccccccccc}
					\hline
					&&  &  &   &  & &\multicolumn{2}{c}{$\operatorname{SID}_\beta$ } & &\multicolumn{2}{c}{$\operatorname{SID}_K$}    \\
					\cline{8-9}                   \cline{11-12}
					$\alpha$   &Case   &$n$ &CPH  &KLR  &IPCW  & &$\operatorname{SID}_1$ &$\operatorname{SID}_{0.5}$   &   & $\operatorname{SID}_{\mathrm{gauss}}$ &$\operatorname{SID}_{\mathrm{lap}}$   \\
					\hline
					0.01 &$X\sim \text{Poisson}(3)$        &50 &0.011&0.008 &0.014 &&0.008  &0.009 &&0.010 &0.011      \\
				         &                        &150&0.012&0.006 &0.018 &&0.011  &0.012 &&0.009 &0.009      \\[0.2cm]
					     &$X\sim t(3)$            &50 &0.015&0.006 &0.016 &&0.012  &0.014 &&0.012 &0.012      \\
					     &                        &150&0.014&0.008 &0.014 &&0.011  &0.010 &&0.012 &0.010      \\
                         \hline
					0.05 &$X\sim \text{Poisson}(3)$        &50 &0.048&0.054 &0.066 &&0.054  &0.054 &&0.044 &0.054      \\
                         &                        &150&0.054&0.058 &0.088 &&0.051  &0.048 &&0.048 &0.048      \\[0.2cm]
                         &$X\sim t(3)$            &50 &0.056&0.043 &0.064 &&0.043  &0.050 &&0.054 &0.058      \\
                         &                        &150&0.044&0.054 &0.047 &&0.052  &0.051 &&0.046 &0.051      \\
                         \hline
                    0.10 &$X\sim \text{Poisson}(3)$        &50 &0.109&0.112 &0.142 &&0.114  &0.113 &&0.104 &0.105      \\
                         &                        &150&0.107&0.104 &0.138 &&0.106  &0.105 &&0.094 &0.101      \\
                         &$X\sim t(3)$            &50 &0.093&0.088 &0.128 &&0.112  &0.104 &&0.102 &0.107      \\
                         &                        &150&0.104&0.104 &0.112 &&0.104  &0.101 &&0.098 &0.102      \\
                         \hline
			\end{tabular}}
		\end{table}

 \begin{example}\label{exampl_8}(High dimensional covariates)
This example investigates the type-I error rates of the proposed method when covariates are high dimensional. The data are generated from the models defined in Case 4 of Example~\ref{exampl_1}, with the exception that the dimension of   $\mathbf{X}$ is set to $p$. Specifically, $p$ is set to be 20, 40, 60, 80, 100, and 120.
 \end{example}

The empirical type-I error rates for Example \ref{tabapp2} under 30\% censoring with $n=150$ are summarized. The results indicate that the proposed $\operatorname{SID}$ methods, along with the KLR, can robustly control type-I error rates   as the dimensionality $p$ increases. Moreover, these results further confirm the finding from Example \ref{exampl_1} that the CPH   is unable to effectively control the type-I error when $p$ is large.  This issue arises because the   partial likelihood method used in the CPH may fail   in high-dimensional settings.

	\begin{table}[http]
        \caption{ Empirical type-I error rate under 30\% censoring with $n=150$ for Example~\ref{exampl_8}.}
		\label{tabapp2} \scriptsize
		\resizebox{\linewidth}{!}{%
			\begin{tabular}{cccccccccccc}
				\hline
				& &  &   &  & &\multicolumn{2}{c}{$\operatorname{SID}_\beta$ } & &\multicolumn{2}{c}{$\operatorname{SID}_K$}    \\
				\cline{7-8}                   \cline{10-11}
				$\alpha$     &$p$ &CPH  &KLR  &IPCW  & &$\operatorname{SID}_1$ &$\operatorname{SID}_{0.5}$   &   & $\operatorname{SID}_{\mathrm{gauss}}$ &$\operatorname{SID}_{\mathrm{lap}}$   \\
				\hline
				0.01    &20   &0.026&0.014 &0.006 &&0.016  &0.012 &&0.013 &0.014      \\
				           &40   &0.030 &0.012 &0.018 &&0.010   &0.010 &&0.010 &0.010      \\
				           &60   &0.068&0.008 &0.018 &&0.009  &0.008 &&0.009 &0.008      \\				
 				           &80   &0.146&0.007 &0.014 &&0.008  &0.008 &&0.008 &0.008      \\
				           &100 &0.270&0.008 &0.010 &&0.011   &0.009 &&0.011 &0.011      \\ 				
				           &120 &0.459&0.010  &0.013 &&0.012   &0.011 &&0.012 &0.012      \\ \hline
				0.05   &20   &0.078&0.058 &0.042 &&0.064  &0.062 &&0.060 &0.056      \\
                           &40   &0.110&0.054 &0.064 &&0.058  &0.042 &&0.042 &0.044      \\
                           &60   &0.184&0.046 &0.058 &&0.044  &0.045 &&0.042 &0.042      \\				
                           &80   &0.332&0.038 &0.050 &&0.050  &0.054 &&0.054 &0.054      \\
                           &100  &0.472&0.042 &0.034 &&0.045  &0.042 &&0.042 &0.044      \\ 				
                           &120  &0.692&0.046 &0.058 &&0.047  &0.048 &&0.050 &0.048      \\ \hline		
				0.1     &20   &0.134 &0.122  &0.084 &&0.106  &0.120 &&0.112 &0.0116     \\
                           &40   &0.178&0.104  &0.116  &&0.090  &0.092 &&0.090 &0.098      \\
                           &60   &0.300&0.102 &0.104  &&0.088  &0.092 &&0.092 &0.094     \\				
                           &80   &0.454&0.094&0.100  &&0.092  &0.100 &&0.104 &0.104      \\
                           &100  &0.628&0.084&0.086 &&0.100  &0.096 &&0.092 &0.092      \\ 				
                           &120  &0.818&0.096 &0.082 &&0.096  &0.102 &&0.098 &0.094      \\ \hline		
			             \end{tabular}}
	\end{table}

 \begin{example}\label{exampl_9}
This example assesses  the   power of our method  under different types of covariates and high-dimensional settings. Here,   \( T \) and   \( C \) are generated from the following model:	
		\begin{equation*}
			T \sim \operatorname{Exp}\left(e^{(\mathbf{1}_{p}^T\mathbf{X})^2/5}\right) \quad \text{and} \quad C \sim \operatorname{Exp}(\lambda).
		\end{equation*}		
The covariates are set in the following   cases
\[
\begin{array}{ll}
\text{Case 1:}~X \sim \text{Poisson}(3) \text{ with } p=1;\quad \quad \quad   & \text{Case 2:} ~ X \sim t(3) \text{ with } p=1; \\
\text{Case 3:} ~ X \sim N(0,1) \text{ with } p=1; \quad  \quad & \text{Case 4:} ~ \text{High dimensional } \mathbf{X}: ~ \mathbf{X} \sim N_{p}\left(\mathbf{0}, \boldsymbol{\Sigma}_{p}\right),
\end{array}
\] 		
where $\boldsymbol{\Sigma}_{p} = (0.5^{|j-k|})$.

\end{example}

\begin{figure}[http]
    \centering
		\begin{minipage}[t]{0.44\textwidth}
			\centering
			\includegraphics[width=\textwidth]{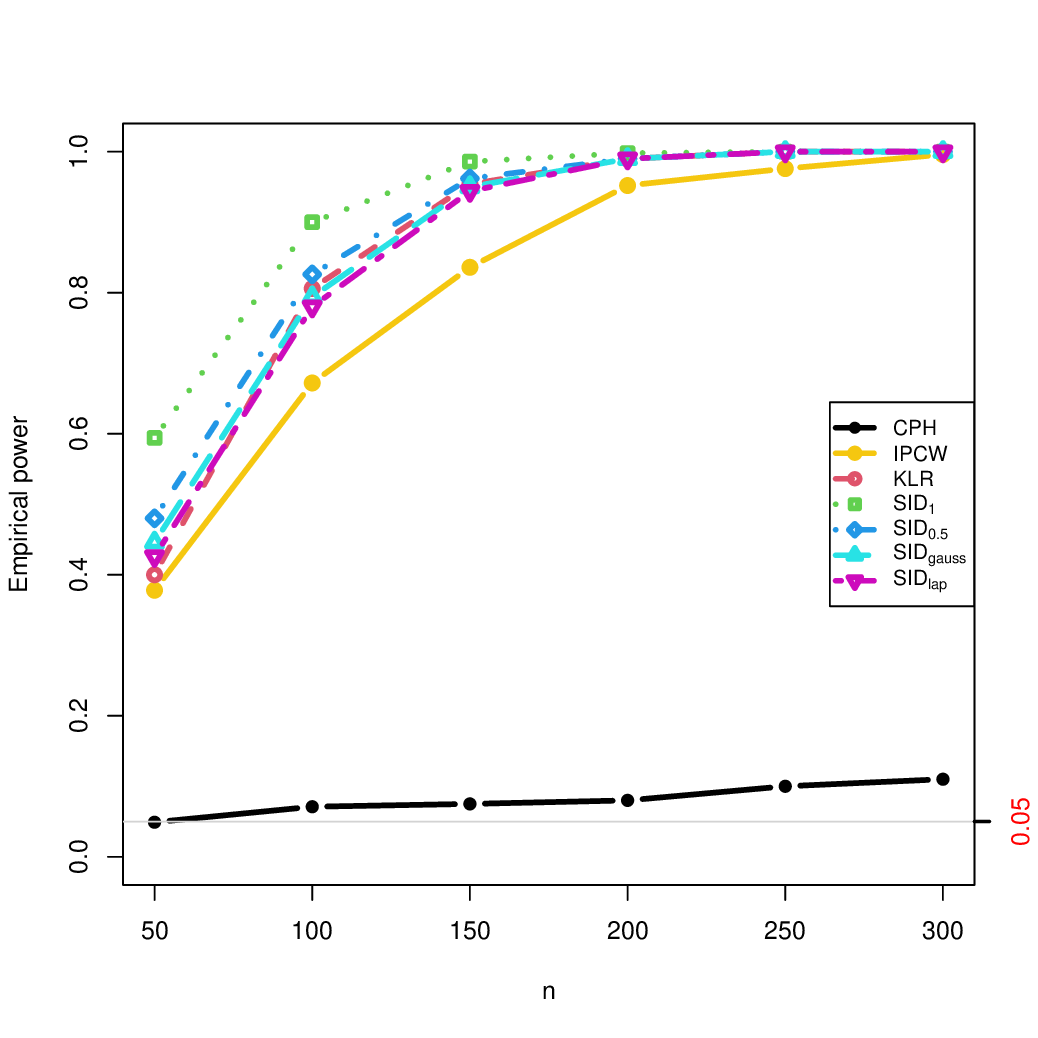}\vspace{-3mm}
			{\footnotesize (a)~ Case~1:~$X\sim \text{Poisson}(3)$  }
		\end{minipage}
		\begin{minipage}[t]{0.44\textwidth}
			\centering
			\includegraphics[width=\textwidth]{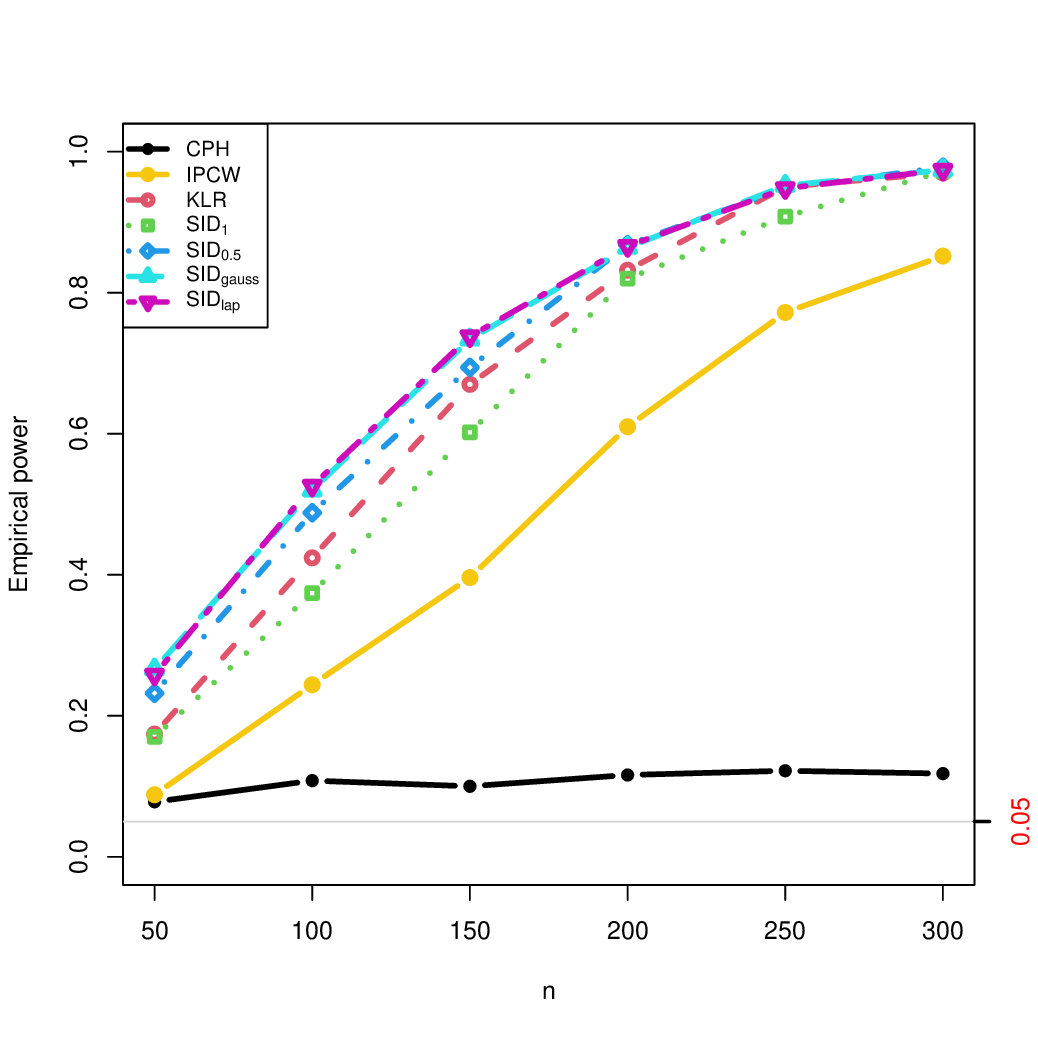}\vspace{-3mm}
			{\footnotesize (b)~ Case~2:~$X\sim t(3)$}
		\end{minipage}
		\begin{minipage}[t]{0.44\textwidth}
			\centering
			\includegraphics[width=\textwidth]{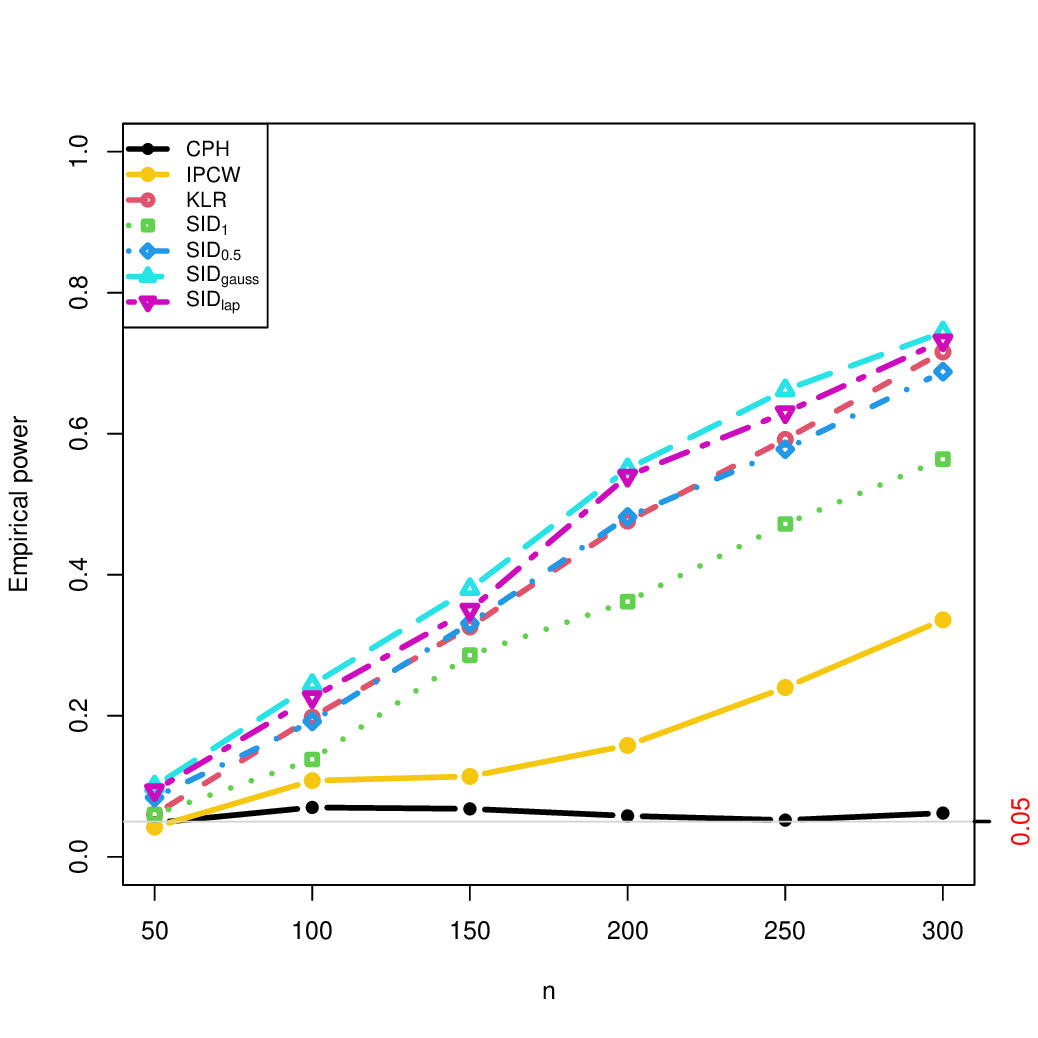}\vspace{-3mm}
			{\footnotesize (c) ~Case~3:~$X\sim N(0,1)$ }
		\end{minipage}
		\begin{minipage}[t]{0.44\textwidth}
			\centering
			\includegraphics[width=\textwidth]{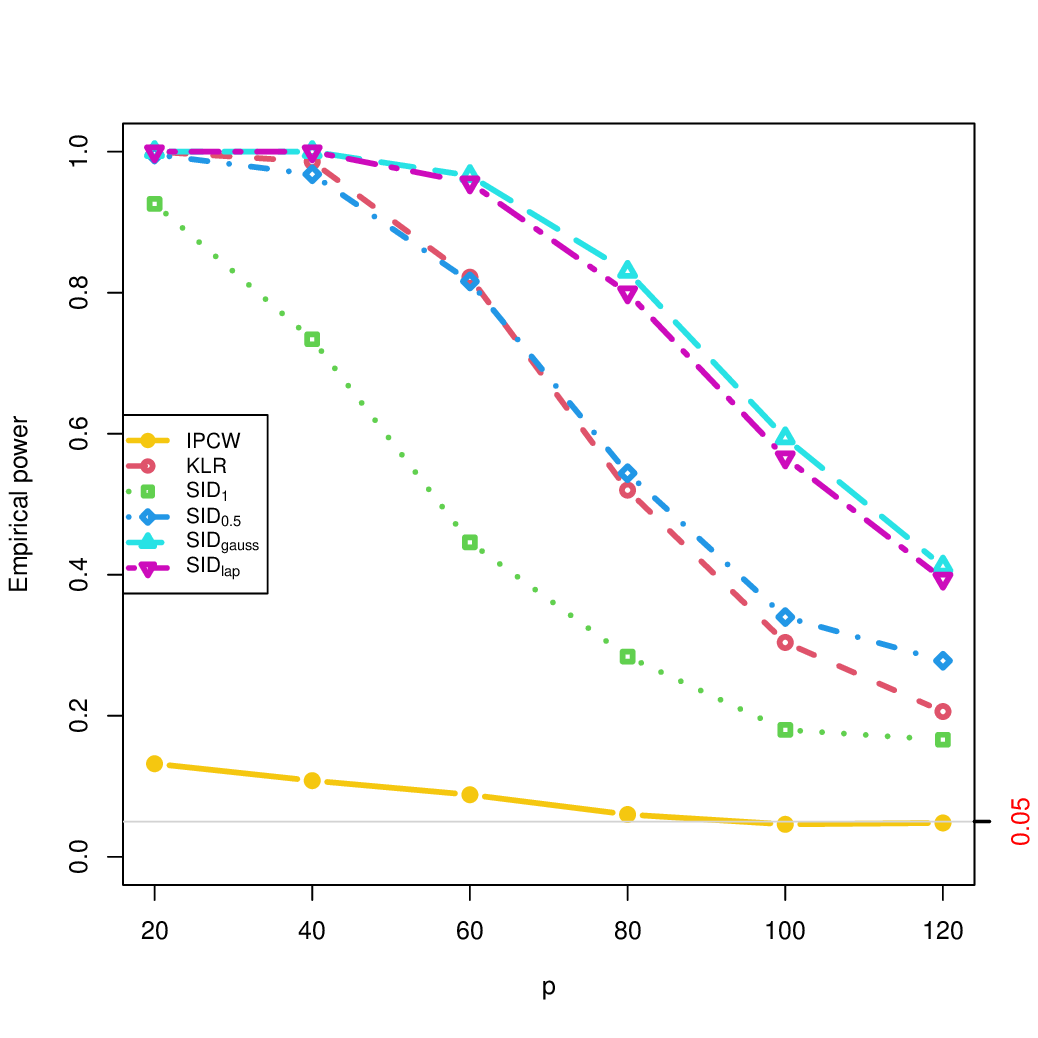}\vspace{-3mm}
			{\footnotesize (d)~ Case~4:~ $\mathbf{X} \sim N_{p}(\mathbf{0}, \boldsymbol{\Sigma}_{p})$ }
		\end{minipage}
		\caption{Comparisons of empirical power at $\alpha=0.05$ under 30\% censoring  for Example~\ref{exampl_9}.}
		\label{fig:exampl_9}
	\end{figure}

Figure~\ref{fig:exampl_9}(a)-(c) display power curves versus $n$ for Cases 1-3, while Figure~\ref{exampl_9}(d) shows power curves versus $p$  with  $n=150$   for Case 4. In Figure~\ref{fig:exampl_9}(d), the CPH method is excluded due to its failure to control type-I errors in high dimensions, see Example \ref{exampl_8}. From Figure~\ref{fig:exampl_9}(a)-(c), our SID method demonstrates higher empirical power across all covariate types. The results in Figure~\ref{fig:exampl_9}(d) further confirm that the SID method maintains superior power even with large $p$.  These findings highlight the robustness of the SID  in handling diverse and high-dimensional covariates.

\begin{example}\label{exampl_10}
This example evaluates the impact of the parameter \(\beta\) on \(\operatorname{SID}_\beta\). The survival time \(T\) is generated by the following linear and nonlinear Cox models:
\[
\text{(1)~Linear Cox model: } T \sim \operatorname{Exp}(e^{X/5});
\qquad
\text{(2)~Nonlinear Cox model: } T \sim \operatorname{Exp}(e^{X^2/2}),
\]
where \(X \sim N(0,1)\) and \(C \sim \operatorname{Exp}(\lambda)\).
\end{example}

\begin{figure}[http]
     \centering
		\begin{minipage}[t]{0.40\textwidth}
			\centering
			\includegraphics[width=\textwidth]{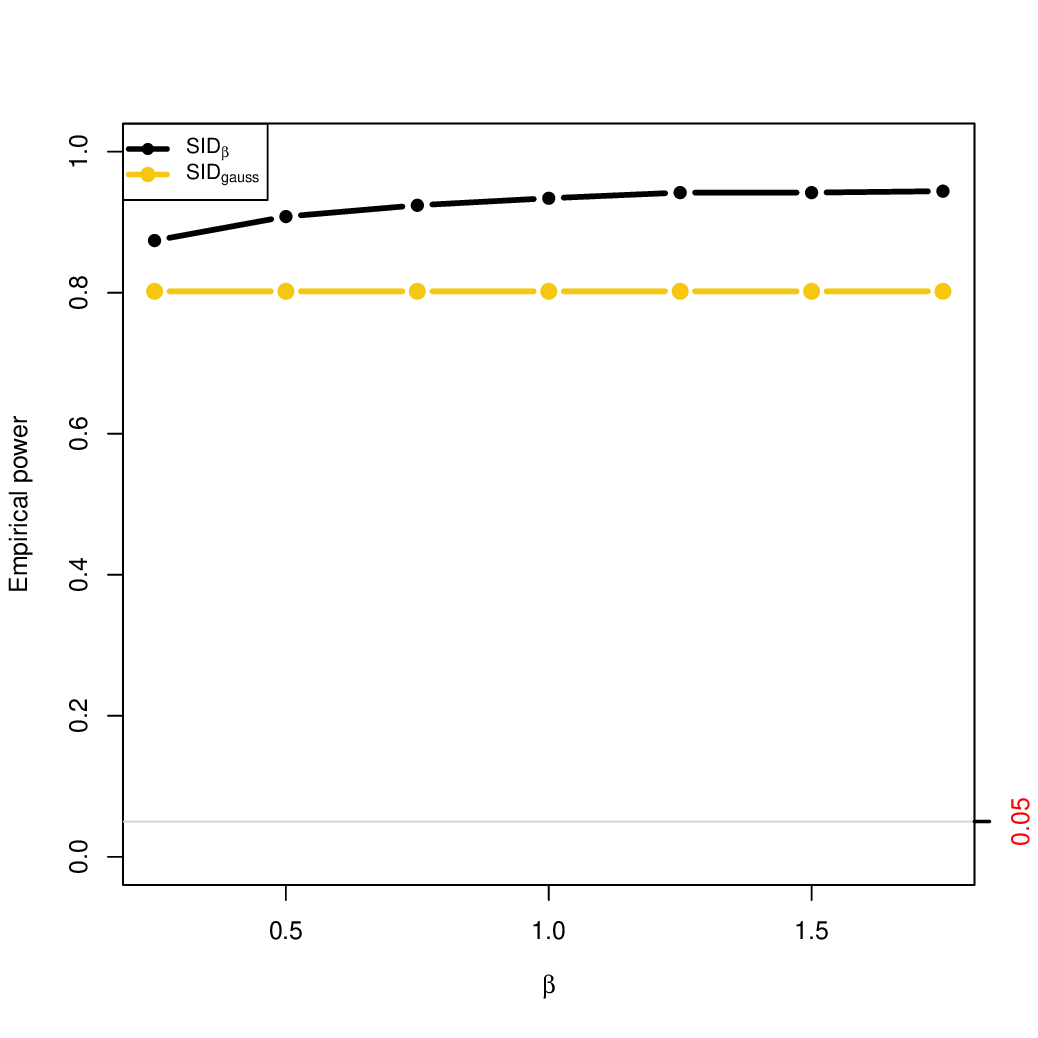}\vspace{-3mm}
			{\footnotesize (a)~ Linear Cox model}
		\end{minipage}
		\begin{minipage}[t]{0.40\textwidth}
			\centering
			\includegraphics[width=\textwidth]{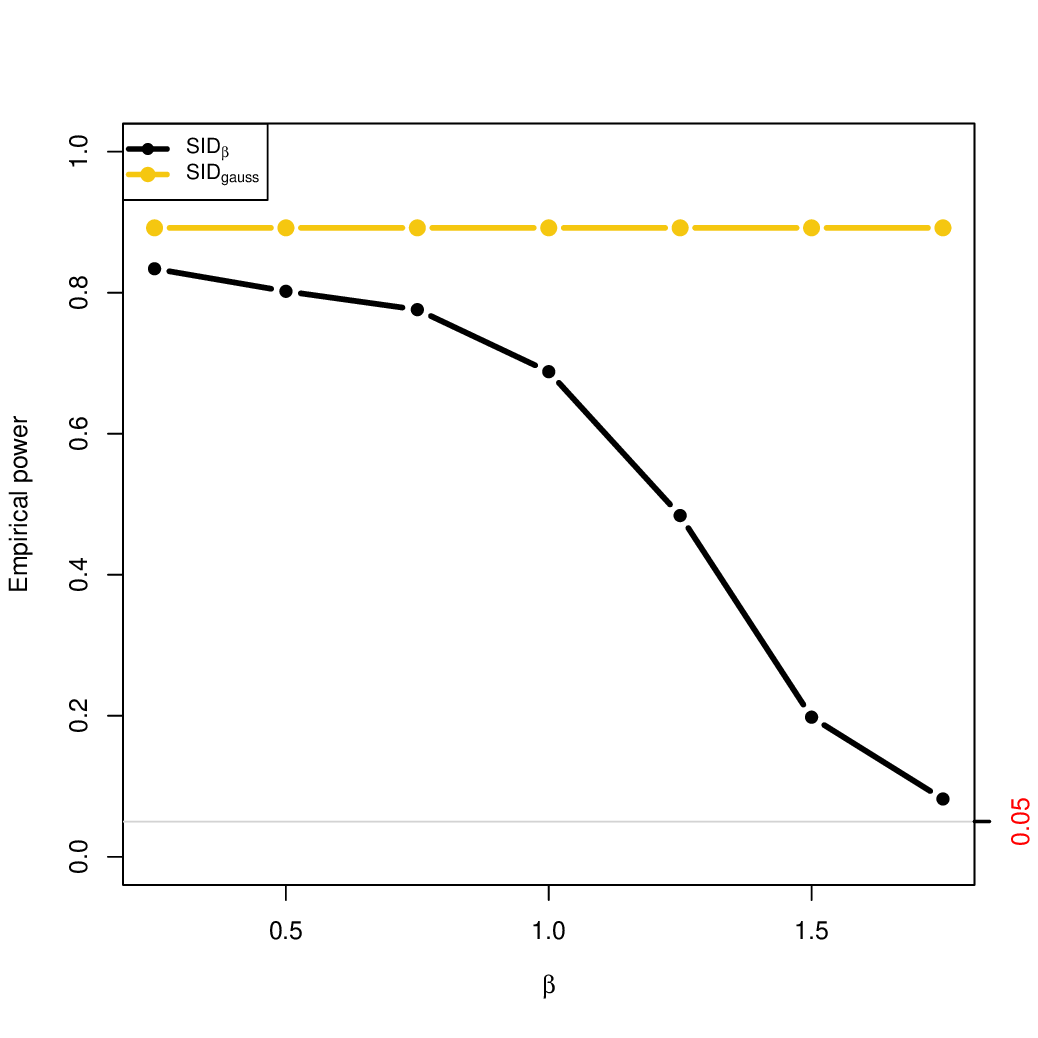}\vspace{-3mm}
			{\footnotesize  (b)~  Nonlinear Cox model}
		\end{minipage}
		\caption{Comparisons of empirical power at $\alpha=0.05$ under 30\% censoring with $n=150$ for Example~\ref{exampl_10}.}
		\label{fig:exampl_10}
	\end{figure}

In Figure~\ref{fig:exampl_10}, we display power curves versus \( \beta \) for \(\operatorname{SID}_\beta\), with \(\operatorname{SID}_{\mathrm{gauss}}\) as a benchmark.  From Figure~\ref{fig:exampl_10}(a), \(\operatorname{SID}_\beta\) shows higher empirical power as \(\beta\) increases in the linear dependence scenario, outperforming \(\operatorname{SID}_{\mathrm{gauss}}\). In contrast, Figure~\ref{fig:exampl_10}(b) reveals that in the nonlinear dependence scenario, smaller \(\beta\) values yield higher empirical power, though \(\operatorname{SID}_{\mathrm{gauss}}\) outperforms \(\operatorname{SID}_\beta\). These power curves demonstrate that different values of \(\beta\) affect the ability to capture nonlinear dependence.  Empirically,  higher levels of nonlinear dependence  require    smaller \(\beta\).

\end{appendix}


